\documentclass[11pt,a4paper]{article}
\usepackage{jheppub}

\usepackage{amsmath}
\usepackage{amsthm}
\usepackage{amssymb}
\usepackage{amsfonts}
\usepackage{physics}
\usepackage{graphicx}
\usepackage{caption}
\usepackage{subcaption}
\usepackage{multirow}
\usepackage{multicol}
\usepackage{booktabs}
\usepackage[noabbrev]{cleveref}
\usepackage[normalem]{ulem}

\makeatletter
\AddToHook{cmd/appendix/before}{\crefalias{section}{appendix}}
\makeatother

\title{Running of neutrino mass parameters in the Zee model}
\author[a]{Michael A. Schmidt,}
\author[a]{and James Vandeleur}
\affiliation[a]{Sydney Consortium for Particle Physics and Cosmology, School of Physics,\\
The University of New South Wales, Sydney, New South Wales 2052, Australia}
\emailAdd{m.schmidt@unsw.edu.au}
\emailAdd{james.vandeleur@unsw.edu.au}
\abstract{
We analyse the size of quantum corrections in the Zee model using effective field theory techniques. We derive the relevant 1-loop matching conditions and use them together with the existing renormalisation group equations in the two Higgs doublet model to calculate quantum corrections to the neutrino mass squared differences, mixing angles, and phases. Using four benchmark scenarios, we demonstrate when quantum corrections have to be included in studies of neutrino mass parameters in the Zee model.
}
\keywords{Renormalization Group, Neutrino Physics, Zee Model, Lepton Mixing}
\arxivnumber{2512.17235}

\begin{document}

\maketitle

\section{Introduction}
\label{sec:intro}
The Standard Model (SM) of particle physics can be used to make predictions with very high precision. 
Neutrino masses are one of the few physical phenomena beyond the SM (BSM) for which we have experimental evidence from lab-based experiments. Since the discovery of atmospheric neutrino oscillations by the Super-Kamiokande experiment in 1998~\cite{Super-Kamiokande:1998kpq} and solar neutrino oscillations in 2002 by SNO~\cite{SNO:2002tuh}, neutrino oscillation physics has seen tremendous progress in the last 30 years. Daya Bay measured the reactor mixing angle $\theta_{13}$~\cite{DayaBay:2012fng} and the next generation of neutrino oscillation experiments~\cite{JUNO:2022mxj,Hyper-Kamiokande:2018ofw,DUNE:2015lol} is expected to determine the mass ordering, measure the Dirac CP phase, and reach sub-percent level precision for some of the mixing parameters. A few weeks ago, the JUNO experiment reported the result of their first oscillation analysis~\cite{JUNO:2025gmd} improving the precision on the solar mixing angle $\sin^2\theta_{12} = 0.3092 \pm 0.0087$ and mass squared difference $\Delta m_{21}^2 = (7.50\pm0.12)\times 10^{-5}\,\mathrm{eV}^2$ by a factor 1.6 over the combination of previous measurements. JUNO is expected to measure $\sin^2(2\theta_{12})$ and the mass squared differences to at least 0.6\% precision within 6 years of data taking~\cite{JUNO:2021vlw}.
The increase in precision must be matched in theoretical predictions. 

There has been similar progress in the theoretical description of neutrino masses. 
One class of neutrino mass models is the radiative mass models, in which neutrino masses are generated at loop level. 
The Zee model was the first 1-loop neutrino mass model~\cite{Zee:1980ai,Wolfenstein:1980sy} and is the focus of this work; see~\cite{Herrero-Garcia:2017xdu} for a recent detailed phenomenological study. Other notable early radiative neutrino mass models include the Zee-Babu model~\cite{Zee:1985id,Babu:1988ki}, which generates neutrino masses at 2-loop order, the first 3-loop neutrino mass model~\cite{Krauss:2002px} and the scotogenic model~\cite{Tao:1996vb,Ma:2006km} which links neutrino masses with dark matter. For further details, see the review on radiative neutrino mass models~\cite{Cai:2017jrq}. 
In avoiding other constraints on new physics, neutrino mass models generally predict neutrino oscillation parameters at a scale much higher than the scale of neutrino oscillations, and thus renormalisation group (RG) corrections are required to link the two. 

The RG equations (RGEs) for the Weinberg operator~\cite{Weinberg:1979sa}, which describes Majorana neutrino masses in SM effective field theory (SMEFT), have been calculated at 1-loop order in ~\cite{Chankowski:1993tx,Babu:1993qv,Antusch:2001ck} and recently at 2-loop order in~\cite{Ibarra:2024tpt}; phenomenological implications have been studied in~\cite{Casas:1999tp,Casas:1999kc,Casas:1999tg,Chankowski:1999xc,Haba:1999ca,Antusch:2003kp,Gehrlein:2015ena,Chen:2025ruj}.
Beyond the Weinberg effective operator, renormalisation group studies of ultraviolet (UV) complete models have focused on seesaw models using 1-loop RGEs. The most well-studied of which is the type-I seesaw model~\cite{Minkowski:1977sc,Yanagida:1979as,Gell-Mann:1979vob,Mohapatra:1979ia,Glashow:1979nm}. RGEs for neutrino oscillation parameters within these models have been derived in~\cite{Mei:2005qp,Antusch:2005gp} (as well as in multi-Higgs doublet extensions of the seesaw model~\cite{Antusch:2001vn,Antusch:2002ek,Grimus:2004yh}) with different phenomenological aspects considered in~\cite{Ellis:1999my,Antusch:2002rr,Mei:2003gn,Mei:2005qp,Antusch:2005gp,Lindner:2005pk,Schmidt:2006rb}. 
In recent years, the 1-loop matching equations at the seesaw scale have also been derived~\cite{Zhang:2021tsq,Zhang:2021jdf,Ohlsson:2022hfl,Du:2022vso,Li:2022ipc,Li:2023ohq}, constructing a more complete EFT description of the type I seesaw model. 
RGEs in the type-II seesaw model~\cite{Magg:1980ut,Schechter:1980gr,Cheng:1980qt,Lazarides:1980nt,Wetterich:1981bx,Mohapatra:1980yp} with an electroweak triplet scalar field have been derived and studied in~\cite{Chao:2006ye,Schmidt:2007nq,Joaquim:2009vp,Zhang:2022osj,Wang:2023bdw,Zhang:2024weq}. Finally, the RGEs for the type-III seesaw model~\cite{Foot:1988aq} have been calculated in~\cite{Chakrabortty:2008zh}. In addition to the studies of the seesaw models, the RG corrections have been studied in the scotogenic model~\cite{Tao:1996vb,Ma:2006km} at 1-loop order~\cite{Bouchand:2012dx,Merle:2015ica} and the 1-loop matching to SMEFT was recently derived in~\cite{Liao:2022cwh}. These corrections in the scotogenic model were found to be potentially large, begging the question of whether similar behaviour is seen in other radiative models.

Motivated by upcoming precision experiments, we consider RG corrections in the Zee model. The Zee model introduces two additional scalar fields, a singly-charged scalar and a second electroweak doublet which together generate neutrino masses radiatively. The analysis requires us to introduce a sequence of effective field theories (EFTs), where each particle is integrated out at its respective mass scale.
Beginning with a degree of ambivalence as to the mass ordering of the two new particles, we will derive the matching conditions at 1-loop order for each case.
We will then focus on the scenario where the singly-charged scalar is heavier than the electroweak doublet scalars and proceed with a numerical solution of the RGEs.

This paper is organised as follows. \Cref{sec:models} introduces the Zee model with discussions of the scalar potential and the generation of neutrino masses. In \cref{sec:theory} we present the discussion of renormalisation group equations and the matching to the EFTs. In \cref{sec:results} we discuss numerical calculations of the neutrino mass parameters in three benchmark scenarios to illustrate the importance of quantum corrections. We present our conclusions and discuss the future outlook in \cref{sec:summ}. Five appendices provide technical details of the calculation.

\section{The Zee and two Higgs doublet models (2HDM)}
\label{sec:models}
In addition to the SM content, the Zee model~\cite{Zee:1980ai} includes a second Higgs-doublet $H_{2}$ and a singly charged scalar field $h$ with $SU(3)_{\rm C}\cross SU(2)_{\rm L} \cross U(1)_{\rm Y}$ SM gauge group charge assignments
\begin{align}
        H_{2} &\sim (1,2,-1/2) \qquad\text{and}\qquad  h \sim (1,1,1) \ .
\end{align}
The kinetic part of the Lagrangian is canonically normalised with the covariant derivative defined as
\begin{align}
        D_{\mu} &= \partial_{\mu} + ig_{3}G^{a}_{\mu}L_{a} + ig_2W^{a}_{\mu}T_{a} + ig_1B_{\mu}Y \ ,
        \label{eq:coD}
\end{align}
where $L_{a}$, $T_{a}$ and $Y$ are the generators of $SU(3)_{\rm C}$, $SU(2)_{\rm L}$ and $U(1)_{Y}$, respectively, with gauge couplings $g_3$, $g_2$, and $g_1$.
$H_{2}$ couples to the SM particle content via a set of Yukawa terms, exactly analogous to the Yukawa terms of the SM Higgs. The singly charged scalar couples to the lepton doublets via a lepton number violating term with antisymmetric Yukawa-like coupling $f$. Altogether, the Yukawa part of the Zee model Lagrangian is 
\begin{align}
        \mathcal{L}_{\rm Zee}^{\rm (Yuk)} &= \sum_{i=1}^{2} \left(\overline{Q}_{L}{Y_{d}^i}^{\dagger}H_{i}d_{R} + \overline{Q}_{L}{Y_u^i}^{\dagger}\tilde{H}_{i}u_{R} + \overline{L}{Y_e^i}^{\dagger}H_{i}e_{R}\right) + \bar{\tilde{L}}fLh + {\rm h.c.} 
        \ , \label{eq:ZeeYuk}
\end{align}
where $f$ is an anti-symmetric matrix in flavour space $(f_{\alpha\beta} = -f_{\beta\alpha})$, $Y_{e,u,d}^{1,2}$ are general complex matrices, and $\tilde{L}\equiv i\sigma_2L^c=i\sigma_2 C \overline{L}^{\rm T}$, with $\sigma_2$ being the second Pauli matrix, such that $\tilde{L}$ represents a charge-conjugated field that maintains the transformation properties of an $SU(2)$ doublet.

Finally, the most general scalar potential in the Zee model can be constructed from that in the two Higgs doublet model (2HDM) \cite{Davidson:2005cw,Herrero-Garcia:2017xdu}
 as 
\begin{align}
        V  &= \mu_1^2H_1^\dagger H_1 + \mu_2^2 H_2^\dagger H_2 - \left(\mu_3^2 H_1^\dagger H_2 + {\rm h.c.}\right)\nonumber\\
            &\phantom{=} + \frac{1}{2}\lambda_1\left(H_1^\dagger H_1\right)^2 + \frac{1}{2}\lambda_{2}\left(H_2^\dagger H_2\right)^2 + \lambda_3\left(H_1^\dagger H_1\right)\left(H_2^\dagger H_2\right) + \lambda_4\left(H_1^\dagger H_2\right)\left(H_2^\dagger H_1\right)\nonumber\\ 
            &\phantom{=} + \bigg\lbrace\frac{1}{2}\lambda_5\left(H_1^\dagger H_2\right)^2 + \left[\lambda_6\left(H_1^\dagger H_1\right) + \lambda_7\left(H_2^\dagger H_2\right)\right]H_1^\dagger H_2 + {\rm h.c.}\bigg\rbrace \nonumber\\
            &\phantom{=} + \mu_h^2 (h^*h) + \lambda_h \left(h^*h\right)^2 + \lambda_8 (h^*h) (H_1^\dagger H_1) + \lambda_9 (h^*h) (H_2^\dagger H_2) + (h^*h)\left(\lambda_{10}H_1^\dagger H_2 + {\rm h.c.}\right)\nonumber\\
            &\phantom{=} + \left(\mu_{\rm Zee}\varepsilon_{ij}H_{1}^{i}H_{2}^{j}h^* + {\rm h.c.} \right)
            \ . \label{eq:ZeePot}
\end{align}
From the hermiticity of the potential, all parameters of \cref{eq:ZeePot} must be real except for $\lambda_{5}$, $\lambda_{6}$, $\lambda_{7}$, $\lambda_{10}$, $\mu_{3}$, and $\mu_{\rm Zee}$.
Of these, we choose $\lambda_{5}$ to be real by redefining the phases of $H_{1}$ and $H_{2}$, and $\mu_{\rm Zee}$ to be real by redefining the phase of $h$.

In the following, we will refer to the Higgs basis, in which $H_2$ acquires no vacuum expectation value (vev). 
From the minimisation of the Higgs potential with respect to $H_1$ and $H_2$, and enforcing the vevs $\langle H_1\rangle  = (v/\sqrt{2},0)^T$, $\langle H_2\rangle =(0,0)^T$, we find the conditions 
 \begin{align}
     \mu_1^2 &= -\lambda_1 \frac{v^2}{2}\;,
     \qquad \mathrm{and}\qquad
     \mu_3^2  = \lambda_6 \frac{v^2}{2} \;. \label{eq:mu3tomu1}
 \end{align}

\subsection{Perturbativity and stability of the potential.}\label{sec:perstab}
To ensure perturbativity of the parameters in the scalar potential, it suffices to naively estimate $\lambda_{i} \leq \sqrt{4\pi}$~\cite{Herrero-Garcia:2017xdu}. A comprehensive analysis of the condition for boundedness from below is beyond the scope of this paper. We briefly summarize the relevant stability conditions of the potential as presented in~\cite{Herrero-Garcia:2017xdu}. It is straightforward to derive conditions on the couplings $\lambda_1$, $\lambda_2$, and $\lambda_h$ by considering large field values of the fields $H_1, H_2, h$, respectively. The resulting conditions are
 \begin{align}
         \lambda_{1}\geq 0, \qquad \lambda_{2}\geq 0, \ \ \text{ and }\ \ \lambda_{h} \geq 0 \ .
         \label{eq:lconstrainta}
 \end{align}
If any of the three couplings is equal to zero, the corresponding quadratic term has to be positive. In the numerical analysis in Sec.~\ref{sec:results} we use $\lambda_8=\lambda_9=\lambda_{10}=0$ to reduce the number of parameters, and thus it is sufficient to consider the scalar potential of a 2HDM. Within this framework, there are three additional conditions~\cite{Ivanov:2008er,Ferreira:2009jb}
\begin{align}\label{eq:lconstraintb}
        \lambda_{3} > - \sqrt{\lambda_{1}\lambda_{2}}\ , \qquad
        \lambda_{3}+\lambda_{4}-\abs{\lambda_{5}} > -\sqrt{\lambda_{1}\lambda_{2}} \ ,\nonumber\\
        \text{ and } \quad
        2\abs{\lambda_{6} + \lambda_{7}} < \frac{\lambda_{1} + \lambda_{2}}{2} + \lambda_{3} + \lambda_{4} + \lambda_{5} 
\end{align}
which must hold. 

\subsection{Heavy mass scales in the Higgs sector}
\label{sec:ZeeModelScales}
The BSM particle content of the Zee model is contained within the Higgs sector.
To implement an EFT description of the low-energy neutrino mass in this model, we must establish a separation of scales. We choose to work in the regime of $m_{h_{\rm SM}} \ll m_{H_2}, m_h$, where $m_{h_{\rm SM}}$ denotes the mass of the SM Higgs and $m_h$ and $m_{H_2}$ are the masses of the new scalar particles.
We leave the choice between the two hierarchies $m_h\gg m_{H_2}$ and $m_{H_2}\gg m_h$ of the heavy masses open for now and will discuss its implications on the matching result in the next section.
Because of the mixing between the Higgs states, we should be careful in how we enforce this scale separation as, naively, one expects the Lagrangian mass parameters to contribute to the physical mass states non-trivially.
Indeed, we want to impose a scale hierarchy on the physical masses as they will be the mass scales propagated by heavy mediator particles and thus the important quantities in our EFT expansion.

In the general basis, both Higgs doublets $H_{1}$ and $H_{2}$ acquire vevs $v_1$ and $v_2$. By identifying $\tan \beta = v_2/v_1$ we can rotate into the {\em Higgs basis} denoted as
\begin{align}
       \begin{pmatrix}
           H'_{1} \\ H'_{2}   
   \end{pmatrix}  &= 
   \begin{pmatrix}
           c_{\beta} & s_{\beta} \\
           -s_{\beta} & c_{\beta}
   \end{pmatrix}
   \begin{pmatrix}
       H_{1} \\ H_{2}
   \end{pmatrix} \, 
\end{align}
where $c_\beta \equiv \cos\beta$ and $s_{\beta}\equiv\sin\beta$ for some angle $\beta$. Under this rotation, the form of the scalar potential in \cref{eq:ZeePot} remains unchanged as long as the quartic couplings are redefined accordingly (see \cite{Davidson:2005cw} for these transformations). $H'_1$ acquires a vev $v=\sqrt{v_1^2+v_2^2}=(\sqrt{2}G_F)^{-1/2} \simeq 246$ GeV which we associate with the SM Higgs vev, while $H'_2$ does not acquire a vev.
We then expand about these minima in the Higgs basis,
\begin{align}
        H'_{1} = \begin{pmatrix}
                G^{+} \\ \frac{1}{\sqrt{2}}\left(v + h_{1} + iG^{0}\right)
        \end{pmatrix} \ , \
        H'_{2} =
        \begin{pmatrix}
                H^{+} \\ \frac{1}{\sqrt{2}}\left(h_{2} + iA\right)
        \end{pmatrix} \ . \label{eq:HiggsBasisExpand}
\end{align}
The charged and pseudo-scalar fields $G^{+}$ and $G^{0}$ are the would-be Goldstone bosons that become the SM $W^{+}$ and $Z$ bosons, leaving only the neutral scalars $h_{1}$ and $h_{2}$, the pseudo-scalar $A$ and the charged scalar $H^{+}$. Note that unlike the usual prescription in the 2HDM, we also have the singly charged $h^{+}$ from the Zee model.
We can find the mass-squared matrices for these states by substituting the expansion in \cref{eq:HiggsBasisExpand} into the scalar potential in \cref{eq:ZeePot} with couplings in the Higgs basis.

The CP even mass eigenstates $h_{\rm SM}$ and $H$ are found from a mixing of the neutral scalar components of $H_1$ and $H_2$ in the Higgs basis, conventionally written as
\begin{align}
    \begin{pmatrix}
        h_{\rm SM} \\ H
    \end{pmatrix} =
    \begin{pmatrix}
        s_{\beta-\alpha} & c_{\beta-\alpha}\\
        c_{\beta-\alpha} & -s_{\beta-\alpha}
    \end{pmatrix}
    \begin{pmatrix}
        h_1\\
        h_2
    \end{pmatrix} \ , \label{eq:neutralMassRot}
\end{align}
where $\alpha$ is defined as the mixing angle from the general basis to the mass basis, such that $\beta-\alpha$ represents the physical and therefore basis independent angle between the Higgs basis and the mass basis \cite{Davidson:2005cw}.
For our discussion, it is most instructive to write the masses in terms of this physical angle
\begin{align}
    m_{h_{\rm SM}}^2 &= \frac{c^2_{\beta-\alpha}}{2c^2_{\beta-\alpha}-1}\mu_2^2 + \frac{2(c^2_{\beta-\alpha}-1)\lambda_1 + c_{\beta-\alpha}^2(\lambda_3+\lambda_4+\lambda_5)}{4c^2_{\beta-\alpha}-2}v^2 \  \text{ and }\label{eq:mh}\\
    m_H^2 &= \frac{c_{\beta-\alpha}^2-1}{2c^2_{\beta-\alpha}-1}\mu_2^2 + \frac{2\lambda_1c^2_{\beta-\alpha}+(c^2_{\beta-\alpha}-1)(\lambda_3+\lambda_4+\lambda_5)}{4c^2_{\beta-\alpha}-2}v^2 \ .\label{eq:mH}
\end{align}
We now associate $h_{\rm SM}$ with the SM Higgs particle. We also have the mass of the pseudo-scalar
\begin{align}
m_A^2 &= \mu_2^2 + \frac{1}{2}v^2(\lambda_3+\lambda_4-\lambda_5) \ , \label{eq:mA}
\end{align}
which, like the masses of the neutral scalars, is the sum of the Lagrangian mass parameter $\mu_2^2$ and a term consisting of the quartic couplings $\lambda_i$.

In a similar manner, another mixing angle $\varphi$ rotates the charged scalars in the Higgs basis to the mass basis
\begin{align}
\label{eq:chargedMixing}
        \begin{pmatrix}
            h_{1}^{+} \\ h_{2}^{+}
    \end{pmatrix} &=
    \begin{pmatrix}
            s_{\varphi} & c_{\varphi} \\
            c_{\varphi} & -s_{\varphi}
    \end{pmatrix}
    \begin{pmatrix}
        h^{+}\\
        H^{+}
\end{pmatrix} \, ,
\, \text{with } \,
s_{2\varphi} = \frac{\sqrt{2} v\mu_{\rm Zee}}{m_{h_{2}^{+}}^{2} - m^{2}_{h_{1}^{+}}} \ .
\end{align}
    The masses are given by 
\begin{align}
    m_{h_1^+}^2
    &= \frac{c_\varphi^2-1}{c_\varphi^2}\mu_2^2 + \frac{1}{c_\varphi^2}\mu_h^2 + \frac{(c_\varphi^2-1)\lambda_2 + 2\lambda_8}{2c_\varphi^2}v^2 \ \text{ and }\label{eq:mh1}\\
    m_{h_2^+}^2
    &= \frac{1}{c_\varphi^2}\mu_2^2 + \frac{c_\varphi^2-1}{c_\varphi^2}\mu_h^2 + \frac{\lambda_3 + 2(c_\varphi^2-1)\lambda_8}{2c_\varphi^2}v^2 \ .\label{eq:mh2}
\end{align}
Exactly analogous to $\beta-\alpha$, $\varphi$ is a physical mixing angle. Furthermore, the mixing of the CP even states $h_{\rm SM}$ and $H$ is distinct from that of the CP odd states $h_1^+$ and $h_2^+$.

We now have a more precise identification of the mass scales in the Zee model. They are the SM Higgs mass $m_{h_{\rm SM}}$, the additional neutral Higgs states $m_H$ and $m_A$, and the two charged scalar states $m_{h_1^+}$ and $m_{h_2^+}$. 
As we will ultimately calculate the neutrino masses in the SMEFT, we must enforce a large separation $v\sim m_{h_{\rm SM}} \ll m_H,m_A,m_{h_1^+},m_{h_2^+}$ between these masses. 
If we enforce this scale separation while also insisting on the perturbativity of the quartic couplings $\lambda_i \lesssim \sqrt{4\pi}$, the neutral state masses given in \cref{eq:mh,eq:mH,eq:mA} require the Lagrangian mass parameter $\mu_2$ to be large, and the mixing angle $\beta-\alpha$ to be such that $c_{\beta-\alpha}\ll1$ \cite{Dawson:2023ebe}. Enforcing the separation of scales in the Lagrangian mass parameters is therefore enough to also enforce the separation of the physical mass scales. Furthermore, doing so automatically moves us close to the Higgs basis. 
Indeed, in the limit $v^2/\mu_2^2\to 0$, we approach the {\em alignment limit} where the mass basis aligns exactly with the Higgs basis. This alignment means that an inverse power series in the Lagrangian mass parameter $\mu_2$ is identical to `integrating out' the physical mass state~\cite{Dawson:2023ebe}.

In the case of the charged mass states, we are also free to take the Lagrangian mass parameter $\mu_h$ of the charged singlet as arbitrarily large. It is then clear that the charged mass states in \cref{eq:mh1,eq:mh2} will satisfy our earlier scale separation. The mass parameter of the charged singlet also allows another separation of scales, between $\mu_2$ and $\mu_h$, creating the hierarchy of three scales that we eluded to schematically at the start of this section. As seen from the expression for the mixing angle in \cref{eq:chargedMixing}, separating the charged mass scales will also move us close to another alignment limit between the mass states and the charged states in the Higgs basis. Although analogous to the alignment in the neutral case, $\varphi$ is an entirely independent parameter from $\beta-\alpha$, and thus these represent independent alignments.

From here, we will work in the Higgs basis.
For convenience, we will drop the notation $H'_{1,2}$ and simply refer to the two Higgs doublets (in the aligned Higgs basis) as $H_{1}$ and $H_{2}$. 
As the mixing of the Higgs states must be small, we can reasonably associate the bare propagating masses of each particle with the respective quadratic Lagrangian parameters. We identify $m_h^2 \equiv \mu_{h}^2 \simeq m_{h_1^+}^2$, $m_{H_2}^2 \equiv \mu_{2}^2 \simeq m_{H}^2, m_{A}^2, m_{h_2^+}^2$ and $m_{h_{\rm SM}}^2 \simeq \lambda_1 v^2$.

\subsection{Lepton sector and neutrino mass parameters}\label{sec:leptonsector}
In the Zee model, the Higgs mechanism endows the charged leptons with a mass matrix given by
\begin{align}
        m_{\rm E} 
        &= \frac{v}{\sqrt{2}}(c_\beta {Y_e^1}^{\dagger} + s_{\beta} {Y_e^2}^{\dagger}) \qquad 
        \underset{\text{Higgs\ basis}}{\longrightarrow}
        \qquad
        m_{\rm E}= \frac{v}{\sqrt{2}}{Y_e^1}^\dagger\ .
\end{align}
$m_{\rm E} = {\rm diag}(m_e,m_\mu,m_\tau)$ is a diagonal matrix with entries corresponding to the bare masses of the electron, muon and tau respectively.
The choice of the Higgs basis specifies $\beta\to 0$.
In this basis, $Y_e^1$ is clearly defined by the charged lepton masses and a priori, $Y_e^2$ is a general complex matrix.
However, $Y_e^2$ can be partially constrained from the phenomenology of charged lepton flavour violation (cLFV)~\cite{Petcov:1982en}. We discuss this point further in \cref{sec:results}.

\begin{figure}[!b]
    \centering
    \includegraphics[width=0.49\textwidth]{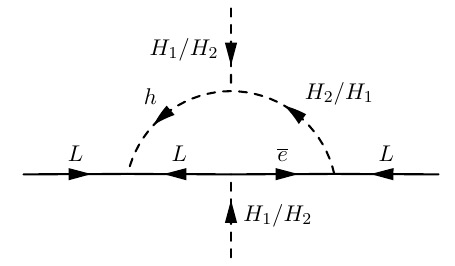}
    \includegraphics[width=0.49\textwidth]{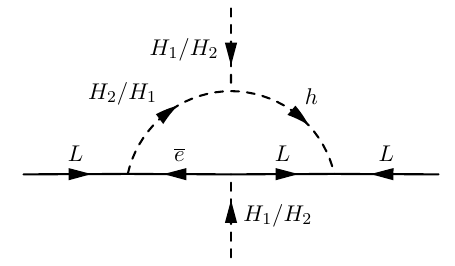}
    \caption{1-loop Feynman diagrams that generate neutrino masses in the Zee model. In the Higgs basis, only the diagrams with $H_1$ on the external legs survive with the $H_2$ internal propagator. The diagram on the right is the flavour-space transpose of the one on the left.}
    \label{fig:fulltheorymass}
\end{figure}

Importantly, the Zee model also generates an effective Majorana mass for the neutrinos via the 1-loop interaction shown in \cref{fig:fulltheorymass}.
This process violates lepton number by $\Delta L = 2$ and contributes to the dimension 5 Weinberg operator in the SMEFT.
More directly, applying the Higgs mechanism allows a two-point, $\Delta L = 2$ interaction containing only the left-handed neutrino fields to be extracted from the one loop, four-point diagrams in \cref{fig:fulltheorymass}. 
The resulting two-point function can be exactly reproduced from an effective Lagrangian below the electroweak scale containing a Majorana mass matrix for neutrinos,
\begin{align}
        \mathcal{L_{\rm Zee/\nu}} \supset -\frac{1}{2}\overline{\nu^{c}_{\rm L}} M_\nu\nu_{\rm L} \, , \label{eq:MajoranaMassTerm}
\end{align}
where the mass matrix $\mathcal{M}_\nu$ is calculated from the 1-loop diagram (\cref{fig:fulltheorymass}) in the general basis as~\cite{He:2011hs}
\begin{align}
        M_{\nu} &=
        \left(\frac{s_{2\varphi}t_{\beta}}{8\sqrt{2}\pi^{2}v}\ln\frac{m_{h_{2}^{+}}^{2}}{m_{h_{1}^{+}}^{2}}\right)
        \left[fm_{\rm E}^{2} + m_{\rm E}^{2}f^{\rm T} - \frac{v}{\sqrt{2}s_{\beta}}(fm_{\rm E}Y_e^2 + {Y_e^2}^{\rm T}m_{\rm E}f^{\rm T})\right] \ .
\end{align}
In the Higgs basis, where $\beta=0$, the mass matrix is simplified to
\begin{align}
        M_{\nu} &= -\frac{v^{2}}{16\pi^{2}}\frac{\mu_{\rm Zee}}{m_{h}^{2} - m_{H_2}^{2}}\log\left(\frac{m_{h}^{2}}{m_{H_2}^{2}}\right)\left(f{Y_e^1}^{\dagger}Y_e^2 + {Y_e^2}^{\rm T}{Y_e^1}^{*} f^{T}\right) \ . \label{eq:naiveMass}
\end{align}
Conventionally, the neutrino mass matrix is represented in terms of the neutrino mixing parameters. It can be diagonalised by the unitary matrix $U$ ($UU^\dagger = I$) using a Takagi factorization, $M_\nu = U^* D_\nu U^\dagger$ where $D_{\nu}={\rm diag}(m_{\nu_1},m_{\nu_2},m_{\nu_3})$ is a diagonal matrix with elements given by the neutrino masses. $U$ is the Pontecorvo-Maki-Nakagawa-Sakata (PMNS) mixing matrix parametrised as
\begin{align}
    U =
    \begin{pmatrix}
    c_{12}c_{13} &
    s_{12}c_{13} &
    s_{13}e^{-i\delta} \\
    -s_{12}c_{23} - c_{12}s_{23}s_{13}e^{i\delta} &
    c_{12}c_{23} - s_{12}s_{23}s_{13}e^{i\delta} &
    s_{23}c_{13} \\
    s_{12}s_{23} - c_{12}c_{23}s_{13}e^{i\delta} &
    -c_{12}s_{23} - s_{12}c_{23}s_{13}e^{i\delta} &
    c_{23}c_{13}
    \end{pmatrix}
    \cdot
    \begin{pmatrix}
    e^{i\phi_{1}/2} & 0 & 0 \\
    0 & e^{i\phi_{2}/2} & 0 \\
    0 & 0 & 0
    \end{pmatrix}\;, 
    \label{eq:UPMNS}
\end{align}
where unphysical phases have been removed. 
Here, $c_{ij}(s_{ij}) = \cos \theta_{ij} (\sin\theta_{ij})$, $\delta$ is the Dirac CP-violating phase, and $\phi_{1,2}$ are the Majorana CP-violating phases.
We also define the mass squared differences, $\Delta m_{21}^2 \equiv m_2^2 - m_1^2$ and $\Delta m_{3l}^2 \equiv m_3^2 - m_l^2$ where $\Delta m_{3l}^2 = \Delta m_{31}^2 > 0$ for the normal ordering (NO) of the neutrino masses and $\Delta m_{3l}^2 = \Delta m_{32}^2 < 0$ for the inverted ordering (IO).
Assuming either ordering, the two mass squared differences, the three mixing angles and the Dirac CP-violating phases have been experimentally constrained by neutrino oscillation experiments, $\delta$ being the least so. There are also upper limits on the absolute neutrino mass scale from tritium beta decay giving $\sqrt{\sum_i |U_{ei}|^2 m_i^2} < 0.45$ eV at 90\% C.L.~\cite{KATRIN:2024cdt}, neutrino-less double beta decay giving $m_{ee} = \sum_i U_{ei}^2 m_i < 28-122 $ meV at 90\% C.L.~\cite{KamLAND-Zen:2024eml}, an upper limit on the sum of neutrino masses obtained by the Planck collaboration giving $\sum_i m_i < 0.12 $ eV at 95\% C.L.~\cite{Planck:2018vyg}, and most recently the DESI collaboration found $\sum_i m_i < 0.0642 $ eV at 95\% C.L.~\cite{Elbers:2025vlz}.

In this way, it is said that the Zee model generates a Majorana mass for neutrinos at 1-loop. This understanding of radiative neutrino mass generation would be perfectly acceptable if not for the potentially large logarithm of the ratio of mass scales that appears in the expression of the neutrino mass matrix in \cref{eq:naiveMass}.
This term is not necessarily small and motivates turning to the machinery provided by an EFT framework, which is capable of accurately handling large separations of scales and the re-summing of large logarithms in particular.

\section{EFT calculation of neutrino mass matrix}
\label{sec:theory}
We have seen that the calculation of the neutrino mass matrix in the full theory suffers from a potentially large logarithm term that undermines the validity of perturbation theory.
The framework of EFT resolves this problem by explicitly splitting the ratio of the two mass scales.
We first calculate a matching condition between the high energy, complete theory and the low energy effective theory at the high scale, encountering only terms like $\ln\mu/m_{\rm large}$, which are well controlled in the high energy regime $\mu\simeq m_{\rm large}$.
The RGEs in the effective theory then allow us to re-sum the full series of large logarithm terms as we move to the low energy scale $\mu\simeq m_{\rm small}$.
We then calculate observables at the low scale $\mu\simeq m_{\rm small}$, encountering only logarithmic terms of the form $\ln \mu / m_{\rm small}$ which are now under control.
In particular, we carry out this prescription twice as we have two BSM mass scales, both the charged singlet and the additional Higgs doublet field that must be integrated out before reaching the SMEFT in which we calculate the neutrino mass matrix.
For a general discussion of the EFT framework, we refer to \cite{Manohar:2018aog}.

In this section, we will first in detail the matching procedure that allows us to define the effective, dimension $>4$ operators in the successive effective theories.
We will then move to the discussion of the RGEs.

We calculate each matching step and the RGEs all at 1-loop level, and never beyond dimension $5$ effective operators.
The best practice is to calculate the RG evolution or running at one loop order higher than the order of matching to ensure scheme independence~\cite{Bando:1992np,Ciuchini:1993fk,Ciuchini:1993ks}. 
2-loop RGEs are known in the SMEFT as well as in the 2HDM; see~\cite{Ibarra:2024tpt} for the recent calculation of the 2-loop RGEs of the Weinberg operator in SMEFT. However, to the best of our knowledge, there are no previously calculated 2-loop RGEs for the dimension $5$ Weinberg-like operators in the general 2HDM EFT. As the Weinberg operator only has one fermion bilinear, we do not expect any issues with scheme dependence and leave the calculation of 2-loop RGEs for future work. 

\subsection{EFT Matching}\label{sec:matching}
We discussed the mass scales that appear in the Zee model in \cref{sec:ZeeModelScales}. A priori, we have two mass hierarchies to choose from. The first is $m_h \gg m_{H_2} \gg m_{h_{\rm SM}} \sim v$, where we would first integrate out the charged singlet, producing a 2HDM EFT, then integrate out the remaining second Higgs doublet. Alternatively, we could have $m_{H_2} \gg m_h \gg m_{h_{\rm SM}}$, which would require first integrating out the second Higgs doublet followed by the charged singlet. 
In this paper, we will primarily be concerned with the first option, which is motivated by leptogenesis~\cite{Lackner:future}, in which we begin by matching the Zee model to the 2HDM EFT. However, we will show the matching for both and discuss their similarities in this section.

Let us first consider the mass hierarchy $m_h \gg m_{H_2} \gg m_{h_{\rm SM}}$ and begin by matching the Zee model to the 2HDM EFT.
Many non-renormalisable operators arise in the 2HDM EFT after integrating out the singly charged scalar, $h$. Important to our discussion of neutrino masses, are the Weinberg-like dimension 5 effective operators given by
\begin{align}
    \mathcal{L}_{\rm 2HDM\ EFT} \supset \kappa^{ij}_{\alpha\beta} (\bar{\tilde{L}}_\alpha H_i)(\tilde{H}_j^\dagger L_\beta) \ , \label{eq:WOP2HDM}
\end{align}
where there is an implicit sum over the Higgs flavour indices $i,j\in\{1,2\}$ and the lepton flavour indices $\alpha, \beta \in \{1,2,3\}$.
Note that these Weinberg-like coefficients maintain the symmetry $\kappa^{ij} = (\kappa^{ji})^{T}$ throughout all calculations of the matching and running.

The \texttt{Matchete} package~\cite{Fuentes-Martin:2022jrf} for Mathematica determines the matching conditions through the path integral formalism. Using this package, we find that the tree- and 1-loop level contributions to the Weinberg-like coefficient $\kappa^{ij}$ in the 2HDM EFT are given by\footnote{We verified the matching conditions in full by hand.}
\begin{subequations}    
\begin{align}
                                     \kappa^{ij}(\bar{\mu}) &= \kappa^{ij(0)}(\bar{\mu}) + \kappa^{ij(1)}(\bar{\mu})\;\;\text{with}  \\   
        \kappa^{ij(0)}(\bar{\mu}) &= -\frac{\mu_{\rm Zee}}{m_{h}^2}\epsilon_{ij}f \;\;\; \text{and } \label{eq:kappaZeeTree}\\
\kappa^{ij(1)}(\bar{\mu}) &= -\frac{1}{16\pi^{2}}\frac{\mu_{\rm Zee}}{m_{h}^2}\left(1+\log\left[\frac{\bar{\mu}^2}{m_h^2}\right]\right)\times \nonumber\\
                                     &\phantom{=}\, \left[\delta_{ik}\epsilon_{lj}f{Y_e^k}^{\dagger}Y_e^l + \delta_{lj}\epsilon_{ik}{Y_e^k}^{\rm T}{Y_e^l}^{*}f - \epsilon_{ij}f(\lambda_8+\lambda_9-4\lambda_h)\right] \;.\label{eq:kappaZeeLoop}
\end{align}\label{eq:kappaZee}%
\end{subequations}%
Note, that $\kappa^{ij}$ is always suppressed by the new physics scale $1/\Lambda = \mu_{\rm Zee}/m_h^2$. We introduced the scale $\bar{\mu}^2 = 4\pi\mu^2e^{-\gamma}$ in the $\overline{\rm MS}$ scheme, where $\gamma\approx0.577$ is the Euler-Mascheroni constant. 
It is precisely this scale that has taken the place of the small scale in the full theory. We can ensure the logarithmic term in \cref{eq:kappaZee} is not large by performing this matching step between the Zee model and 2HDM EFT at the scale $\bar{\mu} = m_h$. We will later use the RGEs in the 2HDM EFT to run $\kappa^{ij}(m_{h})$ down to the scale $\kappa^{ij}(m_{H_2})$.
We have some freedom in our choice of the high energy matching scale. For example, we could also have made the choice $\bar{\mu}=m_h e^{-1 /2}$, removing the 1-loop matching contribution entirely.
This is the first indication that in the EFT approach the neutrino masses are generated by 1-loop running effects and not directly from the matching. The discussion following \cref{eq:Cmh2_2} elaborates on this point.

\begin{figure}[tb!]
    \centering
    \begin{subfigure}[c]{0.4\textwidth}
        \centering
        \includegraphics[width=\textwidth]{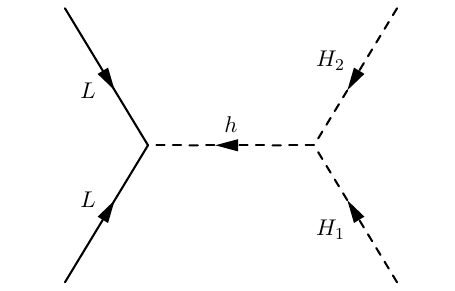}        
        \caption{}
        \label{fig:feynZeeA}
    \end{subfigure}
    \begin{subfigure}[c]{0.4\textwidth}
        \centering
        \includegraphics[width=\textwidth]{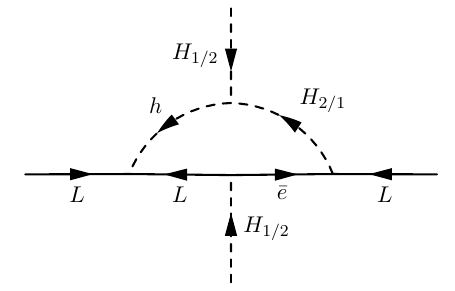}        
        \caption{}
        \label{fig:feynZeeB}
    \end{subfigure}
    \vspace{1cm}
    
    \begin{subfigure}[c]{0.4\textwidth}
        \centering
        \includegraphics[width=\textwidth]{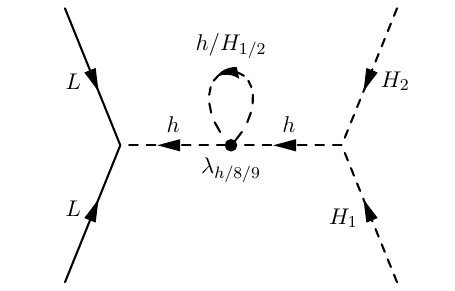}
        \caption{}
        \label{fig:feynZeeC}
    \end{subfigure}
    \begin{subfigure}[c]{0.4\textwidth}
        \centering
        \includegraphics[width=\textwidth]{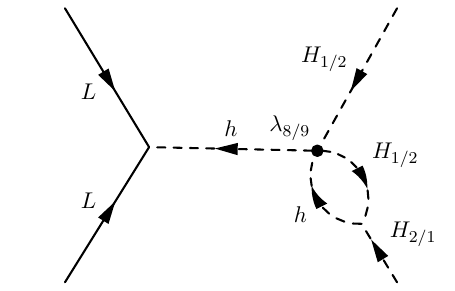}
        \caption{}
        \label{fig:feynZeeD}
    \end{subfigure}
    \caption{Tree and 1-loop level contributions from the Zee model to the Weinberg like $\kappa^{ij}$ in the 2HDM EFT. Different choices of fields of a diagram are indicated with slashes, e.g. there are three self energy insertions for diagram (c) with $h$, $H_1$ or $H_2$ in the loop and quartic couplings $\lambda_h$, $\lambda_8$ and $\lambda_9$, respectively.} 
    \label{fig:feynZee}
\end{figure}

We can understand the matching condition in \cref{eq:kappaZee} through an equivalent diagrammatic calculation.
The 2HDM EFT effective operator in \cref{eq:WOP2HDM} represents all diagrams up to one loop order in the full theory (Zee model) with the incoming fields $L$, $L$, $H_i$ and $H_j$ after the propagator of the heavy particle $h$ has been expanded in inverse powers of its mass.
By first calculating all relevant diagrams in the full (Zee) model and then expanding these amplitudes in powers of $p/m_{h}$, where it is assumed that the characteristic momenta of the incoming particles $p\ll m_{h}$, we can calculate expressions for the Wilson coefficients $\kappa^{ij}$ order by order and in terms of the parameters in the Zee model.
All such diagrams in the full Zee model are shown in \cref{fig:feynZee} with \cref{fig:feynZeeA} being the only contributing diagram at tree-level to $\kappa^{ij(0)}$ and \cref{fig:feynZeeB,fig:feynZeeC,fig:feynZeeD} all contributing at 1-loop level to $\kappa^{ij(1)}$. 
The diagram in \cref{fig:feynZeeC} provides zero contribution when the low scale masses $H_1$ or $H_2$ run in the loop as it leads to a scaleless integral in the matching conditions. 
Within the 1-loop matching condition (\cref{eq:kappaZeeLoop}), we can see that each term corresponds to one or more of the diagrams in \cref{fig:feynZee}.
Like the tree-level contribution, \cref{fig:feynZeeC,fig:feynZeeD} arise as terms anti-symmetric in the Higgs flavour $i,j$ due to the anti-symmetric coupling of the $hH_1H_2$ vertex.
The terms containing Yukawa couplings result from \cref{fig:feynZeeB} and its partner diagram mirrored left to right (corresponding to a transpose in the lepton flavour space).

\begin{figure}[htb!]
    \centering
    \includegraphics[width=0.4\textwidth]{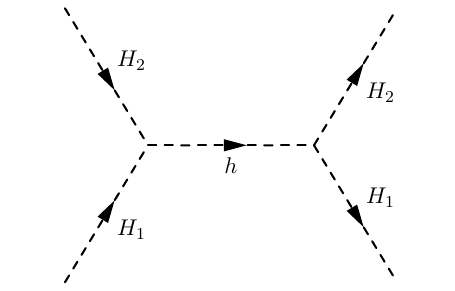}
    \caption{Tree level contribution in the Zee model to the quartic couplings $\lambda_3$ and $\lambda_4$ in the 2HDM EFT.}
    \label{fig:feynlambda}
\end{figure}

To finalise this matching step, we ought to consider the matching of the other parameters of the two theories.
$\kappa^{ij}(m_h)$ depends on the high energy parameters $f$, $\mu_{\rm Zee}$, $\lambda_8$, $\lambda_9$ and $\lambda_h$ of the Zee model as well as $Y_e^1$ and $Y_e^2$. Strictly, the Yukawa couplings are also defined in the Zee model, however unlike the other listed parameters, they remain as coupling parameters in the 2HDM EFT and thus cannot be treated as constant numerical parameters.
We should also consider the matching corrections to both these couplings and the quartic couplings (which do not enter directly in the matching of $\kappa^{ij}(m_h)$ but will be important in the RG running) in the renormalisable part of the 2HDM EFT. 
It is sufficient to consider these matching contributions at tree-level as they only contribute to the 1-loop component ${\kappa^{ij}}^{(1)}$; any 1-loop contribution to the matching of the Yukawa couplings in the 2HDM EFT would therefore contribute at 2-loop order or more to the final neutrino mass in the SMEFT. At tree level, only $\lambda_3$ and $\lambda_4$ acquire a matching contribution from the Zee model,
\begin{align}
    \lambda_3 = \lambda_3^{\rm (Zee)} - \frac{|\mu_{\rm Zee}|^2}{m_{h}^2} \ \text{ and } \
    \lambda_4 = \lambda_4^{\rm (Zee)} + \frac{|\mu_{\rm Zee}|^2}{m_{h}^2} \ . \label{eq:Zeelmatch}
\end{align}
These contributions can be seen in \cref{fig:feynlambda}.
The difference in sign between the two couplings in \cref{eq:Zeelmatch} is a result of how the contraction of $\rm SU(2)$ group indices is defined in their respective terms in the Lagrangian (c.f. \cref{eq:ZeePot}).
As this is our first matching step, we will avoid this complication by simply taking the values of $Y_e^2$ and the quartic couplings $\lambda_{1,\ldots,7}$ in the 2HDM EFT, at the scale of $m_{h}$, as inputs. 

\begin{figure}[tb!]
    \centering
    \begin{subfigure}[c]{0.32\textwidth}
        \centering
        \includegraphics[width=1.05\textwidth]{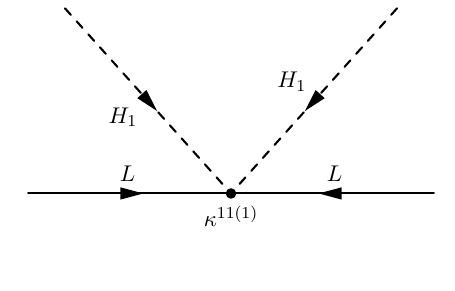}
        \caption{}
        \label{fig:feyn2HDMA}
    \end{subfigure}
    \hfill
    \begin{subfigure}[c]{0.32\textwidth}
        \includegraphics[width=1.05\textwidth]{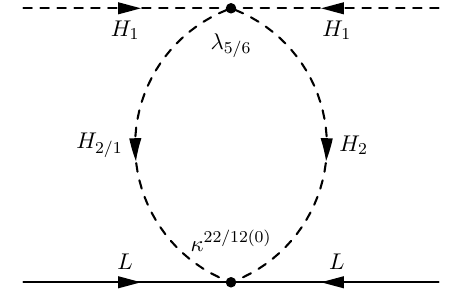}
        \caption{}
        \label{fig:feyn2HDMB}
    \end{subfigure}
    \hfill
    \begin{subfigure}[c]{0.32\textwidth}
        \centering
        \includegraphics[width=1.05\textwidth]{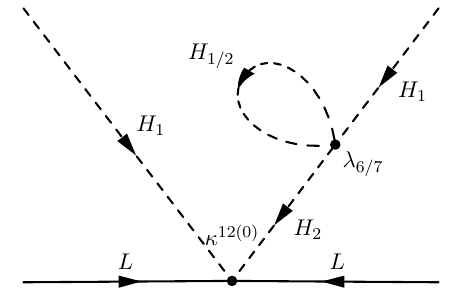}
        \caption{}
        \label{fig:feyn2HDMC}
    \end{subfigure}
    \hfill
    \vspace{1cm}
    \begin{subfigure}[c]{0.32\textwidth}
        \centering
        \includegraphics[width=1.05\textwidth]{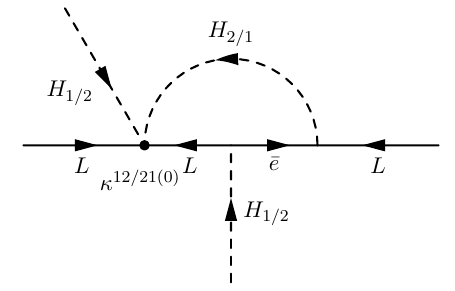}
        \caption{}
        \label{fig:feyn2HDMD}
    \end{subfigure}
    \hfill
    \begin{subfigure}[c]{0.32\textwidth}
        \centering
        \includegraphics[width=1.05\textwidth]{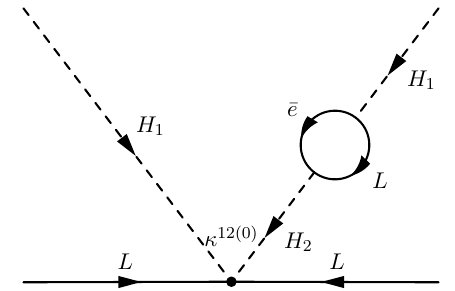}
        \caption{}
        \label{fig:feyn2HDME}
    \end{subfigure}
    \hfill
    \begin{subfigure}[c]{0.32\textwidth}
        \centering
        \includegraphics[width=1.05\textwidth]{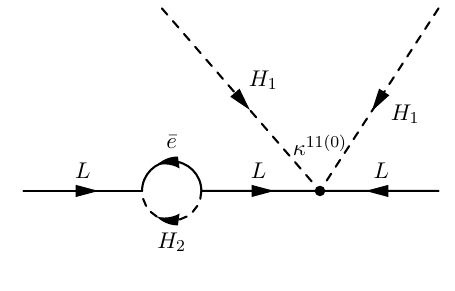}
        \caption{}
        \label{fig:feyn2HDMF}
    \end{subfigure}
    \hfill
    \caption{Tree and 1-loop level contributions from the 2HDM EFT to the SMEFT Weinberg operator $C$.
    }
    \label{fig:feyn2HDM}
\end{figure}

Now let us consider the matching of the 2HDM EFT to the SMEFT at the scale of the now `heavy' second Higgs doublet mass $m_{H_2}$. We emphasise that this requires a calculation of the renormalisation group running in the 2HDM EFT between these two scales, however, we save this discussion for the following section.
Integrating out the second Higgs doublet in the 2HDM EFT, assuming the running has been accounted for, will generate the SMEFT dimension 5 Weinberg operator
\begin{align}
\mathcal{L}_{\rm SMEFT} &\supset C_{\alpha\beta}(\bar{\tilde{L}}_{\alpha}H)(\tilde{H}^{\dagger}L_{\beta}) \ ,
\\
    C_{\alpha\beta}(\bar{\mu}) &= \kappa^{11} - \frac{1}{16\pi^{2}}\kappa^{22} \lambda_5 \log\left(\frac{\bar{\mu}^2}{m_{H_2}^2}\right)\nonumber\\
                    &\phantom{=}\ - \frac{1}{16\pi^{2}}\left(1+\log\left(\frac{\bar{\mu}^2}{m_{H_2}^2}\right)\right)
        \bigg[(\lambda_{6}^{*} - 3\lambda_{7}^{*})(\kappa^{12} + \kappa^{21}) 
                            - \kappa^{12}{Y_e^1}^{\dagger} {Y_e}^{2} - {{Y_e}^{2}}^{\rm T} {Y_e}^{1*}\kappa^{21} \bigg]\nonumber\\
                    &\phantom{=}\  -\frac{1}{16\pi^{2}}\frac{1}{8}\left(1+2\log\left(\frac{\bar{\mu}^{2}}{m_{H_2}^{2}}\right)\right)
                    \bigg[\kappa^{11}{Y_e^2}^\dagger Y_e^2
            + {Y_e^2}^{\rm T}Y_e^{2*}\kappa^{11}\bigg] \ . \label{eq:MatchWeinberg}
\end{align}
This is the operator that will generate the neutrino mass matrix after we enact the Higgs mechanism.
Again, all of the parameters in the renormalisable part of the SMEFT should also be modified with matching contributions that represent effective couplings from processes in the 2HDM EFT, with the second Higgs doublet integrated out. For brevity, we leave these remaining matching conditions for \cref{app:matching}. 
The Feynman diagrams in the 2HDM EFT that represent the contributions to the terms in \cref{eq:MatchWeinberg} are shown in \cref{fig:feyn2HDM}. Note that these diagrams include the $\kappa^{ij}$ vertices. The first diagram in \cref{fig:feyn2HDMA} corresponds to the first term in \cref{eq:MatchWeinberg} and is simply the lone $\kappa^{11}$ vertex.
Note that although this is a tree-level matching between the 2HDM EFT and the SMEFT, $\kappa^{11}$ is only generated at 1-loop level matching between the Zee model and the 2HDM EFT, as emphasised by the vertex label in \cref{fig:feyn2HDMA}.
All other diagrams in \cref{fig:feyn2HDM} are 1-loop contributions. Where these diagrams involve a $\kappa^{ij}$ vertex, we explicitly take only the tree level component to ensure that our full calculation of the final neutrino mass matrix is consistently at 1-loop order.

As before, we only have logarithmic terms of the form $\ln\bar{\mu}/m_{H_2}$ in \cref{eq:MatchWeinberg} and thus we ensure they are controlled by enforcing the matching condition at $\bar{\mu}=m_{H_2}$. 
Having defined the SMEFT Weinberg operator at $m_{H_2}$, we can then use the RGEs in the SMEFT to calculate the neutrino mass matrix at the electroweak scale.
The running is negligible below the EW scale, because neutrinos do not couple to the photon or gluon and all renormalizable interactions are flavour conserving. 

We now have the complete set of matching conditions for the $m_h\gg m_{H_2}\gg m_{h_{\rm SM}}$ hierarchy.
Let us briefly demonstrate the connection to the calculation in the full theory by considering the matching condition for $C(m_{H_2})$ (c.f. \cref{eq:MatchWeinberg}) where we also take the matching condition of the 2HDM EFT vertices $\kappa^{ij}(m_{H_2})$, having not yet taken running into account. Throwing away any accidental 2-loop effects with factors of $1/(4\pi)^4$, we find
\begin{align}
C^{\rm (1-loop)}(m_{H_2}) &= \kappa^{11}(m_{H_2}) + \frac{1}{16\pi^{2}}
    \bigg[\kappa^{12(0)}(m_{H_2}){Y_e^1}^\dagger Y_e^2 + {Y_e^2}^{\rm T}{Y_e^1}^{*}\kappa^{21(0)}(m_{H_2}) \bigg] \label{eq:Cmh2_1} \\
    &= \frac{1}{16\pi^2}\log\left(\frac{m_{H_2}^2}{m_h^2}\right)\frac{\mu_{\rm Zee}}{m_h^2}\big(
    f{Y_e^1}^\dagger Y_e^2 
    - {Y_e^2}^{\rm T}{Y_e^1}^* f
    \big) \ . \label{eq:Cmh2_2}
\end{align}
\Cref{eq:Cmh2_2} features an exact cancellation between $\kappa^{11}$ and the tree-level contribution to $\kappa^{12}$ and $\kappa^{21}$ that combines with the Yukawa couplings in \cref{eq:Cmh2_1}.
This result agrees with that in the full theory (c.f. \cref{eq:naiveMass}) except for the factor $(m_h^2-m_{H_2}^2)^{-1}$ in the full theory which in \cref{eq:Cmh2_2} is reduced to simply $m_h^{-2}$. These factors are consistent in the context of the EFT where we have assumed $m_h\gg m_{H_2}$.

Again, we could also have chosen to match at the similar scale $\bar{\mu} = m_{H_2}e^{-1/2}$. Such a choice would result in exactly the same final expression in \cref{eq:Cmh2_2}.
However, there would be no cancellation between $\kappa^{11}$ and $\kappa^{12(0)}$ in this instance.
The result in \cref{eq:Cmh2_2} would directly be the expression for $\kappa^{11}(m_{H_2}e^{-1/2})$, the only surviving term at $1$-loop in the SMEFT matching condition $C(m_{H_2}e^{-1/2})$ in \cref{eq:MatchWeinberg}.
For this modified matching scale, the cancellation is instead between the constants and the logarithmic scales in the individual matching conditions.

Regardless of the precise choice of matching scale, the EFT matching result in \cref{eq:Cmh2_2} contains the same overall logarithmic factor as in the full theory.
This is precisely because we have so far neglected the running, which would otherwise re-sum this logarithm.
Importantly, because \cref{eq:Cmh2_2} contains no terms other than the one proportional to this logarithm, it is clear that the entirety of the neutrino mass matrix is generated due to the running in the 2HDM EFT.
When we carefully match at the different scales appropriate for the different operator coefficients, these logarithm terms go to zero by design.
Therefore, unlike in the full theory case, there is no direct matching contribution from the Zee model to the neutrino mass matrix in the SMEFT at up to 1-loop order. 
Rather, it is generated via the spoiled cancellation due to the discrepancy in the running between $\kappa^{11}$ and the combination $\kappa^{12(0)}{Y_e^1}^\dagger Y_e^2$.

Although slightly different, the situation is fundamentally the same for the alternate choice of mass hierarchy, $m_{H_2}\gg m_h\gg m_{h_{\rm SM}}$. First integrating out the second Higgs doublet, we arrive at an EFT comprised of the SMEFT with the addition of the charged singlet h (hSMEFT). 
At tree level, we generate the dimension-$5$ operator
\begin{align}
\mathcal{L}_{\rm hSMEFT} \supset C^{(hHLe)}_{\alpha\beta} h^* H^i(\bar{e}_{\alpha}L^j_\beta)\varepsilon_{ij} \ ,\qquad
\text{where}\qquad \ C^{(hHLe)}_{\alpha\beta} = \frac{\mu_{\rm Zee}}{m_{H_2}^2}(Y_e^2)_{\alpha\beta} \ . \label{eq:inversematchZee}
\end{align}
After integrating out the charged singlet $h$, this results in an additional contribution to the Weinberg operator $C_{\alpha\beta}(\bar{\tilde{L}}_\alpha H)(\tilde{H}^\dagger L_\beta)$ in SMEFT at 1-loop,
\begin{align}
    C^{\rm SMEFT}_{\alpha\beta} &= C_{\alpha\beta}^{\rm hSMEFT} + \frac{1}{16\pi^2}\left(1+\log\left(\frac{\bar{\mu}^2}{m_h^2}\right)\right)\left(f{Y_e^1}^\dagger C^{(hHle)} - {C^{(hHle)}}^{\rm T}{Y_e^1}^* f\right)_{\alpha\beta}\;. \label{eq:CSMEFTalt}
\end{align}
The first term denotes the logarithmically enhanced contribution from RG mixing of the operator $h^* H^i(\bar e L^j) \epsilon_{ij}$ into the Weinberg operator within hSMEFT, while the contribution from 1-loop matching is not logarithmically enhanced when matching at $\bar \mu = m_h$. This again demonstrates that the logarithmically enhanced contribution originates from RG running.

Although through slightly different mechanisms, each hierarchy suggests a similar significance of the running in generating neutrino masses. 
Indeed, we should expect this to be the case as both EFT hierarchies stem from the full Zee model, which contains a large logarithm in the calculation of the neutrino mass matrix.
This logarithm of physical scales can only be produced via the running as the logarithm terms in the matching conditions always contain the physically arbitrary choice of scale $\bar{\mu}$\footnote{We stress that although arbitrary from the physical perspective, the choice of $\bar{\mu}$ is computationally important as it is intrinsically linked to the convergence of the series.}. 
We will restrict ourselves to studying the former hierarchy $m_h\gg m_{H_2} \gg m_{h_{\rm SM}}$ and anticipate that the qualitative behaviours we demonstrate are applicable to both hierarchies.

\subsection{Renormalisation group running}
\label{sec:running}
We take the hierarchy $m_h\gg m_{H_2}\gg m_{h_{\rm SM}}$ which requires two stages of matching, first at the high scale $m_h$ and then at the intermediate scale $m_{H_2}$.
As demonstrated in the previous section \cref{sec:theory}, it is vital that we calculate the RGEs in the 2HDM EFT to re-sum the large logarithm between these two scales.
We will also require the RGEs in the SMEFT to evolve the SMEFT matching condition at $\bar{\mu} = m_{H_2}$ down to the scale of neutrino mixing experiments and properly include the RG running effects.
The RGEs in each of these models can be found in the literature~\cite{Grimus:2004yh,Li:2016} or calculated with any of a wide selection of computational tools; we use the Mathematica package \texttt{RGBeta}~\cite{Thomsen:2021ncy} to calculate the RGEs of the renormalizable part of the Lagrangian.

In this section, we will focus on running in the 2HDM EFT up to dimension $5$ effective operators with a particular focus on the running of the Weinberg-like operators that generate neutrino mass. It is unnecessary for us to consider the running in the full Zee model as we will take the Zee model parameters as inputs at the scale of the first matching step, $m_h$.
The RGEs and the behaviour of the running in the SMEFT are already well understood~\cite{Antusch:2001ck,Antusch:2005gp,Jenkins:2013wua,Jenkins:2013zja,Alonso:2013hga,Ibarra:2024tpt} and so we shall not discuss it in detail here.
All $\beta$-functions in both the 2HDM EFT and the SMEFT that we use in the numeric calculation in \cref{sec:results}, including those not immediately discussed in this section, are collected in \cref{app:betas}.

The $\beta$-functions that most directly contribute to the running of the neutrino mass parameters in the 2HDM EFT are those of the Weinberg-like Wilson coefficients $\kappa^{ij}$. Where $t=\ln\bar{\mu}$ is the logarithm of the energy scale in the $\overline{\rm MS}$ scheme, these $\beta$-functions are~\cite{Grimus:2004yh,Li:2016}\footnote{The Honours thesis \cite{Li:2016} pointed out some corrections to the result in \cite{Grimus:2004yh}.} 
\begin{align}
        16\pi^{2}\frac{d}{dt}\kappa^{ij}(\mu) &=
-3g^2(2\kappa^{ij}-\kappa^{ji}) + 4\sum_{k,l=1}^2\kappa^{kl}\lambda_{kilj}\nonumber\\
                               &\phantom{=}\ + 2\sum_{k=1}^2\bigg[\kappa^{kj}{Y_e^{i\dag}}{Y_e}^k 
                                       - \left(\kappa^{ik} + \kappa^{ki}\right){Y_e^{j\dagger}}Y_e^{k} \nonumber\\
&\phantom{=}\qquad\qquad\qquad
+ {Y_e^k}^{\rm T}{Y_e^j}^*\kappa^{ik} 
- {Y_e^k}^{\rm T}{Y_i}^*(\kappa^{kj} + \kappa^{jk})\bigg]\nonumber\\
&\phantom{=}\ + \sum_{k=1}^2[T_{ki}\kappa^{kj} + T_{kj}\kappa^{ik}] + \kappa^{ij}S + (S^T)\kappa^{ij} \ , \label{eq:betaK_2HDM}\\
        \text{where, }\ S &= \frac{1}{2}\sum_{k=1}^2{Y_e^k}^{\dagger} {Y_e^k} \ \text{ and, } \ T_{ij} = \text{Tr}\left(Y_e^{i\dagger} Y_e^{j} + 3Y_u^{i\dagger} Y_u^{j} + 3Y_d^{i\dagger} Y_d^{j}\right) \ . \label{eq:STdefs}
\end{align}
It is the running of $\kappa^{11}$ and the tree-level contribution $\kappa^{12(0)}$ that are most interesting for the generation of the neutrino mass matrix, as these are the components that delicately cancel in the direct matching at $m_{H_2}$ (see \cref{eq:Cmh2_2}). Let us briefly discuss the key features of these $\beta$-functions now; more discussion will follow the analysis of the numeric results in \cref{sec:results}.

First, although it is not entirely clear how to maintain the separation of loop orders in the matching through the running, the $\beta$-function for $\kappa^{12(0)}$ can best be described by isolating the leading order contributions to the matching of terms in the running of $\kappa^{12}$. Doing so, we find for the running of the tree-level component $\kappa^{12(0)}$,
\begin{align}
    16\pi^{2}\frac{d}{dt}\kappa^{12(0)} &\simeq (2\lambda_3 + 2\lambda_4 - 9g_2^2 + T_{11} + T_{22})\kappa^{12(0)}\nonumber\\
    &\phantom{=} \qquad \qquad + 2\left(\kappa^{12(0)}Y_e^{1\dagger}Y_e^1 + {Y_e^2}^{\rm T}{Y_e^{2}}^*\kappa^{12(0)}\right) \ . \label{eq:betaK12(0)}
\end{align}
The dominant contributions in \cref{eq:betaK12(0)} come from the terms proportional to the gauge coupling $g_2$ and the traces $T_{11}$ and $T_{22}$ over the Yukawa matrices. Each of these coefficients are $\mathcal{O}(1)$ (recall that $T_{11}$ and $T_{22}$ include the quark Yukawa matrices, which contain an $\mathcal{O}(1)$ element for the top quark, as per \cref{eq:STdefs}).
All of these terms act as coefficients to $\kappa^{12(0)}$ itself and will thus only contribute to an overall exponential scaling of $\kappa^{12(0)}(t)$, in the sense that
\begin{align}
    \frac{d}{dt}\kappa^{12(0)} &= B\kappa^{12(0)} + f(\kappa^{12(0)},Y_e^{1,2},...)\\
    \implies \kappa^{12(0)} &= Ae^{Bt} + g(\kappa^{12(0)},Y_e^{1,2},...) \ , \label{eq:expscaling}
\end{align}
where $A$ and $B$ are constants and $f$ and $g$ are unspecified functions collecting all other terms in the $\beta$-function. If $f$ amounts to a small correction relative to the first term, then $g$ will amount to a small correction to the exponential scaling of $\kappa^{12(0)}$.

The only terms that will alter the shape of $\kappa^{12(0)}$ throughout the running are those in the second line of \cref{eq:betaK12(0)} which are suppressed by two factors of the charged-lepton Yukawa matrices and thus will be negligible.

A similar, although richer, structure is present in the running of $\kappa^{11}$, given by 
\begin{equation}
   \begin{aligned}
    16\pi^{2}\frac{d}{dt}\kappa^{11} &= (2\lambda_{1}-3g_2^{2} + 2T_{11})\kappa^{11} + (4\lambda_{6}^{*} + 2T_{21})\kappa^{12{\rm (sym)}} + 2\lambda_{5}^{*}\kappa^{22}
    \\
    &\phantom{=} \ + \kappa^{11}\left(-\frac{3}{2}{Y_e^1}^\dagger Y_e^1 + \frac{1}{2}{Y_e^2}^\dagger Y_e^2\right) + \left(-\frac{3}{2}{Y_e^1}^\dagger Y_e^1 + \frac{1}{2}{Y_e^2}^\dagger Y_e^2\right)^{\rm T}\kappa^{11}
    \\
    &\phantom{=} + 2\left(\kappa^{12}{Y_e^1}^\dagger Y_e^2 + {Y_e^2}^{\rm T}{Y_e^1}^{*}{\kappa^{12}}^{\rm T} \right) \ , \label{eq:BetaK11}
    \\
    \text{where } \kappa^{12{\rm (sym)}} &= \frac{1}{2}(\kappa^{12} + \kappa^{21}) \ .
\end{aligned}
\end{equation}
The first line of \cref{eq:BetaK11} contains three terms, one proportional to each of $\kappa^{11}$, $\kappa^{12{\rm (sym)}}$ and $\kappa^{22}$.
The first of these will result in an exponential scaling of $\kappa^{11}$, as before.
The other two terms will contribute to a linear dependence of $\kappa^{11}$ on $t$ for as long as the coefficients of these terms depend sufficiently little on $t$. 
The overall dependence of $\kappa^{11}$ on $t$ will depend, in part, on the relative size of these three terms.
Again, as the trace terms $T_{11}$ and $T_{21}$ are both $\mathcal{O}(1)$, we expect both the terms proportional to $\kappa^{11}$ and $\kappa^{12{\rm (sym)}}$ to be significant.
This implies that $\kappa^{11}(t)$ cannot be well approximated by a pure exponential, unlike the running of $\kappa^{12(0)}$ in \cref{eq:betaK12(0)}.

The second line of \cref{eq:BetaK11} contains mixing terms that contain $\kappa^{11}$ and carry an extra suppression of two charged-lepton Yukawa matrices.
Therefore, these terms will be small relative to the size of $\kappa^{11}$ and will not significantly contribute to the running which already has a contribution from the first term which is of the same order as $\kappa^{11}$.

The final line of \cref{eq:BetaK11} also contains two Yukawa couplings and thus we expect it to be suppressed. 
However, these terms contain the full $\kappa^{12}$ term and not only its symmetric component.
At the high scale, $\kappa^{12}(m_h)$ matches at tree level to the anti-symmetric coupling $f$ in the Zee model.
This matched term then combines with the Yukawa matrices in \cref{eq:BetaK11} to reproduce the same expression as in $\kappa^{11}(m_h)$.
Thus, near the UV scale, the terms in the third line of \cref{eq:BetaK11} will be significant, as they will effectively act as a term proportional to $\kappa^{11}$ with an $\mathcal{O}(1)$ coefficient.

We find that in our numeric calculations,
due to the tree-level contribution to the matching of $\kappa^{12}$, as well as contributions from terms like $\sim {Y_e^2}^\dagger Y_e^2$ that are not suppressed by the first Yukawa matrix $Y_e^1$, $\kappa^{12}$ is generally larger than $\kappa^{11}$ and therefore significantly contributes to the running of $\kappa^{11}$. This is not reciprocated by the contribution of $\kappa^{11}$ to the running of $\kappa^{12}$, leaving $\kappa^{12}$ to run very little.
Thus, as $\kappa^{11}$ runs through the 2HDM EFT, the terms containing $\kappa^{12}$ will remain as a significant linear term in $\kappa^{11}(t)$.

Comparing the running of $\kappa^{12(0)}$ and $\kappa^{11}$, it appears that the neutrino mixing parameters will be much more sensitive to the running of $\kappa^{11}$ due to the terms contributing a linear dependence to $\kappa^{11}(t)$ that are absent from the running of $\kappa^{12(0)}$ which predominately results in only an exponential scaling of $\kappa^{12(0)}$, altering the shape of the matrix much less.
We should note that it is not $\kappa^{12(0)}$ alone that cancels with $\kappa^{11}$ in \cref{eq:Cmh2_1}, rather $\kappa^{11}$ is cancelled by the combination
\begin{align}
    \frac{1}{16\pi^2}\left(\kappa^{12(0)}{Y_e^1}^\dagger Y_e^2 + {Y_e^2}^{\rm T}{Y_e^1}^*\kappa^{12(0)}\right) \ . \label{eq:cancelcombo}
\end{align}
The Yukawa matrices here have their own $\beta$-functions (see \cref{app:betas}) and thus the running of the full term in \cref{eq:cancelcombo} would also contribute to the mixing parameters.
The key distinction is that as the neutrino mass matrix will be generated by the discrepancy in the running between $\kappa^{11}$ and the combination of terms in \cref{eq:cancelcombo}, it is the size of the running of each expression relative to that expression's initial value that we should compare to determine which part of the cancellation contributes most to the running discrepancy.
As such, the exponential scaling in each expression is of little significance as clearly the originally cancelling expressions begin at the same value, and both will exponentially scale with an overall $\mathcal{O}(1)$ exponent\footnote{The $\beta$-functions for $Y_e^1$ and $Y_e^2$ in \cref{app:betas} show that the exponential scaling of these Yukawa couplings behaves similarly, dominated by the $T_{ij}$ term.}. 
Therefore, the terms contributing to the linear scaling will be those most responsible for the discrepancy in the running of the two expressions, and these terms are much stronger in the case of $\kappa^{11}$.
Indeed, the running of $\kappa^{11}$ in the 2HDM EFT is the dominant contribution to the neutrino mass matrix in the SMEFT.

To complete our discussion of the theory in the 2HDM EFT, let us make a crude large-logarithm approximation to demonstrate how the neutrino mass matrix is generated via the running. That is, we take the right-hand side of each $\beta$ function as a constant, such that the logarithmic dependence in $\kappa(\bar{\mu})$ (that is, the linear dependence in $\kappa(t)$) is dominant,
\begin{align}
\kappa^{ij}(\bar{\mu}) &\simeq \beta_{\kappa^{ij}}(m_{h})\log\left(\frac{\bar{\mu}}{m_{h}}\right) + \kappa^{ij}(m_{h}) \ . \label{eq:KLargeLog}
\end{align}
With this approximation, we can first use \cref{eq:kappaZee} to match $\kappa^{ij}$ between the Zee model and the 2HDM EFT at the scale $m_h$, then use \cref{eq:KLargeLog} to find the expressions for $\kappa^{ij}(m_{H_2})$ before substituting them into the matching condition for the SMEFT Weinberg operator $C(m_{H_2})$ (c.f. \cref{eq:MatchWeinberg}) matched at the scale of the second Higgs doublet $m_{H_2}$.
Keeping only the overall 1-loop level contributions, we retrieve the logarithm term from the direct matching in \cref{eq:Cmh2_2},
\begin{align}
    C_\text{large log}(m_{H_2}) &= \frac{1}{16\pi^{2}}\log\left(\frac{m_{H_2}^2}{m_{h}^2}\right)\frac{\mu_{\rm Zee}}{m_h^2}\left(f{Y_e^1}^{\dagger}Y_e^{2} - {Y_e^2}^{\rm T}{Y_e^1}^*f\right) + \mathcal{O}\left(\left(\frac{1}{16\pi^2}\right)^2\right)\label{eq:CLargeLog} \ .
\end{align}
The validity of this large log approximation remains to be seen from the numeric solution of the full set of coupled RGEs. However, it is already enough to demonstrate that the logarithm of the ratio of the mass scales is generated via the running.

The remaining $\beta$-functions involved in the running of the neutrino mass parameters in the 2HDM EFT are listed in \cref{app:betas} along with those of the other parameters in the SMEFT. Of note in the SMEFT is the running of the Weinberg operator coefficient,
\begin{align}
        16\pi^{2}\frac{d}{dt}C &= -\frac{3}{2}\left[C(Y_{e}^{\dagger}Y_{e}) + (Y_{e}^{\dagger}Y_{e})^{T}C\right] + 4\lambda C - 3g_{2}^{2}C + 2TC \label{eq:betaK} \ .
\end{align}
where, analogous to the case in the 2HDM EFT,
\begin{align}
    T = {\rm Tr}(3Y_{u}^{\dagger}Y_{u} + 3Y_{d}^{\dagger}Y_{d} + Y_{e}^{\dagger}Y_{e}) \ .
\end{align}
The running of the SMEFT Weinberg operator in \cref{eq:betaK} is similar to the running of $\kappa^{12(0)}$ in that it contains only terms resulting in an exponential scaling in $C(t)$, and terms otherwise suppressed by a factor of two charged-lepton Yukawa couplings. 
The running of the Weinberg operator in SMEFT is well understood~\cite{Chankowski:1993tx,Babu:1993qv,Antusch:2001ck,Ibarra:2024tpt}, and generally small. We include it in our analysis for completeness, although it is of comparatively little interest.
Note that we consider the running above the electroweak scale and before enacting the Higgs mechanism.

One might wonder whether the Zee-Wolfenstein model~\cite{Zee:1980ai,Wolfenstein:1980sy} with $Y_e^2(m_h)=0$ is excluded when taking into account RG corrections, as is the full-theory calculation~\cite{He:2003ih,Koide:2002uk,Frampton:2001eu}.
At leading-log order, the same conclusions hold as in the full theory, because only $\kappa^{12}$ is generated in the matching at the scale $m_h$, which is antisymmetric at tree level. Using the $\beta$-function of $\kappa^{11}$ we find that $\kappa^{11}$ also has vanishing diagonal elements in the leading-log approximation. However, non-zero $Y_e^2$ is generated if $T_{21}$ is non-zero which can induce non-zero diagonal couplings in $\kappa^{11}$, when solving the RG equations.

\section{Numeric Calculations}
\label{sec:results}
We are finally prepared to complete a full, numeric calculation of the neutrino mass parameters in the EFT framework.
In general, the goal of such a calculation would be to fit values for the UV parameters of the theory, in this case the Zee model, that correctly reproduce the experimental measurements at low energy (in the SMEFT). 
A typical statistical analysis of the space of all the UV parameters, such as that conducted in \cite{Herrero-Garcia:2017xdu}, is beyond the scope of this work.
Rather, our aim is to demonstrate the importance of the running effects in this calculation which are typically not taken into account.
As such, we will study the running under four, illustrative sets of benchmark parameters in the Zee model.

In this section, we first outline the framework that we use to choose our benchmark parameters and precisely how our calculation follows with them. We then discuss the interesting features of the running in the 2HDM EFT for each of the four benchmark cases, followed by an analysis of the resulting sensitivity of the observable neutrino mixing parameters to the choice of UV parameters. 

\subsection{Phenomenological constraints}\label{sec:constraints}
As eluded to in \cref{sec:theory}, the Zee model and its subsequent EFTs allow for cLFV processes through the second lepton Yukawa matrix, $Y_{e}^2$. Such processes are experimentally well constrained. In particular the tree-level decay $\mu\to eee$ and the loop-level radiative muon decay $\mu\to e\gamma$ will provide important constraints. Their branching ratios have been constrained to  
${\rm BR}(\mu \to eee) < 1.0\times 10^{-12}$ \cite{SINDRUM:1987nra} and
${\rm BR}(\mu\to e\gamma) < 1.5\times 10^{-13}$ \cite{MEGII:2025gzr}, respectively.
We calculate the approximate bound on the elements of $|Y_e^2|$ from each of these types of operators in \cref{app:pheno}. The strongest constraint is set by the decay $\mu\to eee$ with 
\begin{align}
    |(Y_e^2)_{ee} (Y_e^2)_{e\mu}| \lesssim 2.3\times 10^{-5}\left(\frac{m_{H_2}}{\rm TeV}\right)^2 \; . \label{eq:Ye2Constraint}
\end{align}
We impose it and the other constraints in our numerical analysis. It should be noted that although this constraint becomes weak for $m_{H_2} \gtrsim 100$ TeV, elements of the Yukawa matrix should always remain within the perturbative regime with $|(Y_e^2)_{\alpha\beta}|\lesssim \sqrt{4\pi}$. 
Constraints from $\tau$ cLFV processes are weaker due to their similarly sized branching ratios and weaker experimental constraints.

Secondly, we recall from \cref{sec:perstab} the discussion of perturbativity and the stability of the potential as it pertains to the quartic couplings $\lambda_i$. 
The constraints discussed there should remain true at all energy scales, however, it is not immediately obvious that if enforced at one scale, these relations will remain true after taking running into account.
For each set of benchmark we trial in the following, we have checked numerically that these relations continue to hold across the whole energy range we consider, and in particular between the scales $m_{H_2}$ and $m_h$ where the running is most significant.
If the conditions for stability hold in the 2HDM, they also hold in the Zee model because the tree level shift from the matching cancels in $\lambda_3 + \lambda_4$ and $\lambda_3$ in the Zee model is always larger than in the 2HDM, as shown in \cref{eq:Zeelmatch}.

This brings us to the discussion of mass scales.
Beyond enforcing our assumed hierarchy $m_{h_{\rm SM}}\ll m_{H_2}\ll m_h$ with enough separation that the EFT prescription holds, we only need to enforce that these additional scalar masses are above the TeV scale, lest they would have been detected at the Large Hadron Collider.
To provide a more tangible guide for these scales, we turn to the calculation of the neutrino mass matrix in the full Zee model, \cref{eq:naiveMass}.
This equation provides a natural upper bound as the neutrino mass matrix scales as $1 / m_h^2$ and for a large enough scale $m_h$, it will become too small to reproduce the scale of the measured atmospheric mass squared difference $\sqrt{\Delta m_{3l}^2}\sim 0.05$ eV.
Indeed, 
\begin{align}
    \mathcal{M}_{\nu} &\sim \frac{v^2}{16\pi^2}\frac{\mu_{\rm Zee}}{m_h^2}\log\left(\frac{m_{H_2}^2}{m_h^2}\right)fY_e^1Y_e^2\\\nonumber
    & \sim 0.05\, \mathrm{eV} 
    \times \frac{10^{11}\, \mathrm{GeV}}{m_h} \times \mathcal{O}(1) \times \mathcal{O}(1) \ ,
\end{align}
where we make the order of magnitude estimate $Y_e^1 \sim \sqrt{2} m_{\tau}/v$.
To maximise the possible scale of $m_h$, we choose $Y_e^{2}, f \sim\mathcal{O}(1)$ - the most generous Yukawa contributions that maintain perturbativity, and $\mu_{\rm Zee} = m_h$.
Explanation of the atmospheric mass scale $\sim 0.05$  eV thus requires $m_h\lesssim 10^{11}\, {\rm GeV}$ or, more realistically, $m_h\lesssim 10^9\, {\rm GeV}$ if $Y_e^{2}, f \lesssim\mathcal{O}(0.1)$. The logarithm would contribute at most an $\mathcal{O}(10)$ factor when comparing to $m_{H_2}>1\, {\rm TeV}$.
This provides us with $6$ orders of magnitude between $1$ and $10^6$ TeV to place reasonable heavy states with masses $m_{H_2}$ and $m_h$. We will see in the discussion of the calculation framework that the exact choice of scale will carry little significance in our calculations. We choose masses at the lower end of this range to emphasise the importance of these effects in even the `not-too-far-beyond' SM regime.

Finally, we check that our chosen hierarchy remains when running effects are accounted for.
The $\beta$-functions for the quadratic couplings, which are associated with the mass scales, are included in \cref{app:betas}.
In particular, $\beta_{\mu_{1}^{2}}$ includes a term proportional to $\mu_{2}^{2}$, suggesting that the scale dependence of $\mu_1$ may be very large in the context of $m_{h_{\rm SM}}\ll m_{H_2}$, which is a manifestation of the hierarchy problem.
To ensure that the running of the masses in the 2HDM EFT does not spoil our chosen hierarchy, we observe that more precisely than the dependence of each quadratic coupling being small, we require a small scale dependence for the ratio of the scales
\begin{align}
        \frac{d}{dt}\left(\frac{\mu_{1}^{2}}{\mu_{2}^{2}}\right) &=
        \frac{1}{\mu_{2}^{2}}\beta_{\mu_{1}^{2}} - \frac{\mu_{1}^{2}}{\mu_{2}^{4}}\beta_{\mu_{2}^{2}}\simeq \frac{1}{16\pi^{2}}\left[4\lambda_{3} + 2\lambda_{4} + \mathcal{O}\left(\frac{\mu_{1}^{2}}{\mu_{2}^{2}}\right) + \mathcal{O}\left(\frac{\mu_{3}^{2}}{\mu_{2}^{2}}\right) + \mathcal{O}\left(\frac{\mu_{1}^{2}}{\mu_{2}^{4}}\right)\right]  \ .
\end{align}
Indeed, with $\mu_2\gg\mu_1\sim\mu_3$ (see the condition in \cref{eq:mu3tomu1}), we see that the only contributions to the running of the mass hierarchy that are not suppressed by the mass scales are those from the quartic couplings $\lambda_{3}$ and $\lambda_{4}$. Due to the overall loop suppression, these couplings can be as large as perturbativity constraints allow, $\mathcal{O}(1)$, while limiting the overall running of $\mu_{1}^{2}/\mu_{2}^{2}$ to $\mathcal{O}(10^{-2})$. 
In principle, there is also a contribution from $\mu_h$ in the Zee model, induced by the quartic couplings $\lambda_{8,9,10}$ which we set to $0$ as their contributions are suppressed anyhow.
The running of the mass scales is therefore insufficient to invalidate the EFT calculation.

\subsection{Calculation framework}\label{sec:framework}
To numerically solve the system of RGEs, we must first specify a choice of the high energy (UV) model parameters.
The nature of EFTs is such that these UV parameters are not completely constrained by their counterparts in the low energy theory.
However, to ensure a reasonable choice of such parameters, it is useful to begin in the low energy theory (in our case, the SMEFT) and consider the running up to the high energy, introducing new high energy parameters along the way.
Once a complete set of high energy parameters has been chosen, we can then use the full set of RGEs to run the high energy parameters down to the experimental scale.

Beginning at the low scale in the SMEFT, we are fortunate that the running of the Yukawa matrices $Y_{e,u,d}$ and the quartic coupling $\lambda$ in the SMEFT is independent of the Weinberg coefficient $C$. 
The running of said couplings in the SMEFT is therefore entirely known from the measured values of the couplings at low energy (see \cref{tbl:inputs}) and their beta functions, \cref{eq:appBetaCSMEFT,eq:appBetaYeSMEFT,eq:appBetaYuSMEFT,eq:appBetaYdSMEFT,eq:appBetaLSMEFT,eq:appBetagSMEFT}.
After running these couplings in the SMEFT up to the scale of $m_{H_2}$, we are faced with the matching to the 2HDM EFT.

At $\bar{\mu}=m_{H_2}$, the matching condition for the quartic couplings in \cref{eq:appMatchL} depends only on $\lambda_1$, $\lambda_6$, and $\lambda_7$ from the 2HDM EFT. The coupling $\lambda_1$ is the dominant contribution to the quartic coupling $\lambda$ in the SM, while $\lambda_{6}$ and $\lambda_7$ retain more freedom, and we consider different benchmark values to illustrate their effects.

For the Yukawa matrices, first considering $Y_e$, it is possible to write the matching condition \cref{eq:appMatchYe} in the form 
\begin{align}
C &= AY_e^1 + Y_e^1B \label{eq:sylvester}
\end{align}
where $A$, $B$ and $C$ are all flavour space matrices depending on $Y_e^2$ with $C$ in particular also carrying the dependence on the SMEFT's $Y_e$. This matrix equation is known as the Sylvester equation and can be solved to find $Y_e^1$.
Although we are predominately interested in the electroweak sector, we will see that the trace of the quark Yukawas (in particular the contribution of the top quark) plays a vital role in the running of both the charged-lepton Yukawa matrices and the Weinberg coefficients.
This particular dependence allows us to reduce the parameter space we need to consider by taking the so called ``alignment limit'' in which $Y_{u,d}^{2} = \alpha Y_{u,d}^1$, for $\alpha \simeq \mathcal{O}(1)$.
The matching step for fitting the quark Yukawa matrices is then conducted iteratively with $Y_{u,d}^{1}$ originally set as the corresponding $Y_{u,d}$ in the SMEFT. This allows a $Y_{u,d}^{2}=\alpha Y_{u,d}^{1}$ to be defined so that $Y_{u,d}^{1}$ can be determined from \cref{eq:sylvester} before $Y_{u,d}^2$ is recalculated as $\alpha Y_{u,d}^1$.
Therefore, specifying only the general complex matrix $Y_e^2$ (with components bound as per the cLFV phenomenology of \cref{sec:constraints}) and the quartic couplings allows us to run all of the Yukawa and quartic couplings in the 2HDM EFT up to the high scale $\bar{\mu}=m_h$.

Finally, at $\bar{\mu}=m_h$, we are left with the choice of values for the singlet scalar Yukawa $f$ in the Zee model.
We set the additional quartic couplings $\lambda_{8,9,h}=0$ with strictly positive $0<\lambda_h\ll 1$ to ensure the stability of the potential, as these couplings contribute uninterestingly to the matching, and removing them simplifies our calculation. 
In particular, the extent of their contribution is effectively a loop-suppressed modification of the benchmark value of $f$ through their contribution to the matching of $\kappa^{ij}$, as in \cref{eq:kappaZee}. Therefore, for a fixed set of mass scales ($m_{h_{\rm SM}}$, $m_{H_2}$ and $m_{h}$) and $\alpha$, the only free parameters in this procedure are the quartic scalar couplings $\lambda_i$, the second charged-lepton Yukawa matrix $Y_e^2$ and the Yukawa-like coupling matrix $f$ of the Higgs singlet. 
Recall that $f$ is anti-symmetric and real, while $Y_e^2$ is a general complex matrix.

The most direct fitting approach is to solve the minimisation problem to find the choice(s) of $f$ and $Y_e^2$ that most accurately reproduce the measured mixing parameters of the neutrino mass matrix. 
Using the full EFT calculation, including both matching steps and the intermediate running of all relevant parameters, would make this minimisation numerically intractable.
Instead, within the fitting procedure, we will only numerically evaluate the running of the Yukawa and quartic couplings in the 2HDM EFT from $m_{H_2}$, where $Y_e^2(m_{H_2})$ is introduced, up to the scale $m_h$ of the Zee model.
At this high scale, we will choose $f(m_h)$ and then use the calculation of the neutrino mass matrix in the full theory to directly compare with the measured mixing parameters \cite{Esteban:2024eli}. 
We therefore seek to minimise the function
\begin{align}
        F\big(f(m_h),Y_e^{2}(m_{H_2});\beta_{Y_e^2}\big) &= \sum_{a\in P} \frac{(a_{\rm calc} - a_{\rm exp})^{2}}{(\Delta a_{\rm exp})^{2}} \label{eq:paramSearch}
\end{align}
where the sum is over the neutrino mass parameters $P = \lbrace \theta_{12}, \theta_{23}, \theta_{13}, \Delta m_{21}^2, \Delta m_{3l}^2, \delta_{\rm CP} \rbrace$ with $l=1(2)$ for normal(inverted) mass ordering and $\Delta a_{\rm exp}$ represents the $1\sigma$ experimental uncertainty in each parameter that weights the `importance' of each fit relative to the precision of its measurement.
$F$ depends explicitly on the choice of $f(m_h)$ and $Y_e^2(m_{H_2})$ and implicitly on the running $\beta_{Y_e^2}$ between $Y_e^2(m_{H_2})$ and $Y_e^2(m_h)$.
Of course, we argue that this will not reproduce quantitatively correct values for the UV parameters $Y_e^2$ and $f$, however, it will still allow us to qualitatively demonstrate the running effects and the significance of the full EFT calculation. 
The final neutrino mass scale in all of our numerical results is in slight tension with the bound from DESI assuming $\Lambda$CDM cosmology, but respects all other experimental constraints listed in \cref{sec:leptonsector}. 

The procedure for generating benchmark values for the UV model parameters, and in particular the coupling matrices $f$ and $Y_e^2$, is summarised as follows:
\begin{enumerate}
    \item Begin with SMEFT values for $Y_{e,u,d}(m_{H_2})$ and $\lambda_{1,\ldots,7}(m_{H_2})$.
    \item Generate a random, complex matrix $Y_{e}^2$ (subject to flavour physics constraints, \cref{sec:constraints}).
    \item Use $Y_e^2(m_{H_2})$ and $Y_e(m_{H_2})$ to determine $Y_e^1(m_{H_2})$ from the matching condition \cref{eq:appMatchYe}.
    \item Run $Y_e^2$ and $\lambda_{1,\ldots,7}$ in the 2HDM EFT up to $\bar{\mu}=m_h$.
    \item Generate random real parameters to construct $f(m_h)$.
    \item Use the now full set of UV parameters to calculate the neutrino mass matrix in the full Zee Model, \cref{eq:naiveMass}.
    \item Repeat steps 2-6 in a root-finding algorithm to determine values for $Y_e^2(m_{H_2})$ and $f(m_h)$ that optimally reproduce the experimentally determined neutrino mixing parameters \cite{Esteban:2024eli} in the full theory.
\end{enumerate}

Due to the high number of free parameters, especially in the general complex matrix $Y_e^2$, relative to the few measured neutrino mass parameters, the fitting of $f$ and $Y_e^2$ is not unique.
We account for this by considering four sets of benchmark parameters that we find most illustrative. 
Each benchmark adheres to the phenomenological constraints discussed in \cref{sec:constraints}.

As a final note, our chosen fitting procedure outlined above results in a peculiar independence of the qualitative behaviour of the running in the 2HDM EFT from the choice of mass scales.
Schematically, this occurs because the $\beta$-functions never depend explicitly on the scale $\bar{\mu}$ and likewise, the boundary values we enforce are essentially, though not entirely, independent of the mass scale. The values for the known couplings at the low scale $m_{H_2}$ are fixed by the running in the SMEFT, which is mostly negligible. As we will see, $Y_e^2$ generally runs relatively little in the 2HDM EFT, and therefore, the conditions for fitting $f$, dictated by the expression for the neutrino mass matrix in the full theory, differ only by an overall scale, which is absorbed by our choice of $f$. The scale of $f$ does not significantly alter the qualitative behaviour of the running. 
Consequently, we stress that the level of significance of the running that we explore in the following is applicable across all reasonable mass scales, including relatively light $\sim$TeV new physics, limited only by the requirement that the EFT expansions remain valid.
We demonstrate this by including one set of benchmark parameters at the lower $m_{H_2}=1$TeV scale.

\subsection{Benchmark cases}\label{sec:BMs}
For each set of benchmark parameters at the high scale, we show the running in the 2HDM EFT of the model parameters most relevant for generating the neutrino mass matrix in the eventual matching to the SMEFT. 
Unfortunately, representing the physical neutrino mixing parameters above this scale is ambiguous because the Weinberg operator, and therefore the neutrino mass matrix, is not well defined.

Our benchmark values are chosen to best illustrate the running behaviour and are always consistent with the various experimental and theoretical constraints we have discussed in the earlier sections. The benchmarks are differentiated by choices of $\alpha$, the quartic couplings $\lambda_i$ and the mass scales; each set of choices leads to different values for $f$, $Y_e^1$ and $Y_e^2$ in the fitting procedure described in \cref{sec:framework}. 
We always assert the benchmark values $f$, $Y_e^1$ and $Y_e^2$ (see \cref{tbl:benchmarks}) at the high scale $\bar{\mu}=m_h$, and we use the 2HDM EFT $\beta$-functions (see \cref{app:betas}) to run down to the lower scale $\bar{\mu}=m_{H_2}$.

\subsubsection{Benchmark 1: Minimally constrained case (\texorpdfstring{$Y_e^1\lesssim Y_e^2$, $\alpha=1$}{Ye1 < Ye2, alpha=1})}
\begin{figure}[!b]
  \centering
  \begin{subfigure}[b]{0.47\textwidth}
    \centering
    \includegraphics[width=\textwidth]{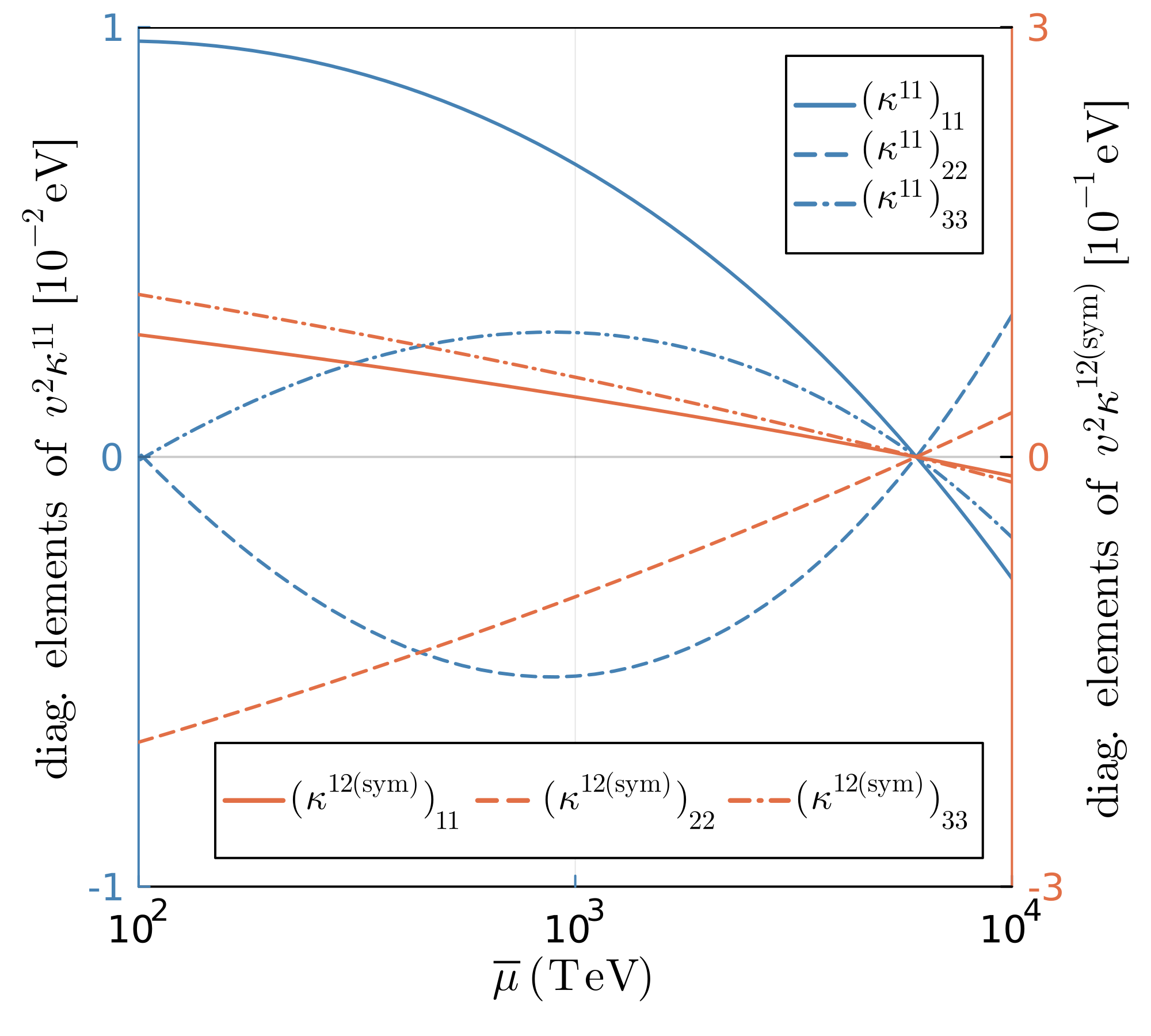}
    \caption{Running of $\kappa^{11}$ and $\kappa^{12{\rm(sym)}}$.}
    \label{fig:inspectK_B1}
  \end{subfigure}
  \begin{subfigure}[b]{0.47\textwidth}
    \centering
    \includegraphics[width=\textwidth]{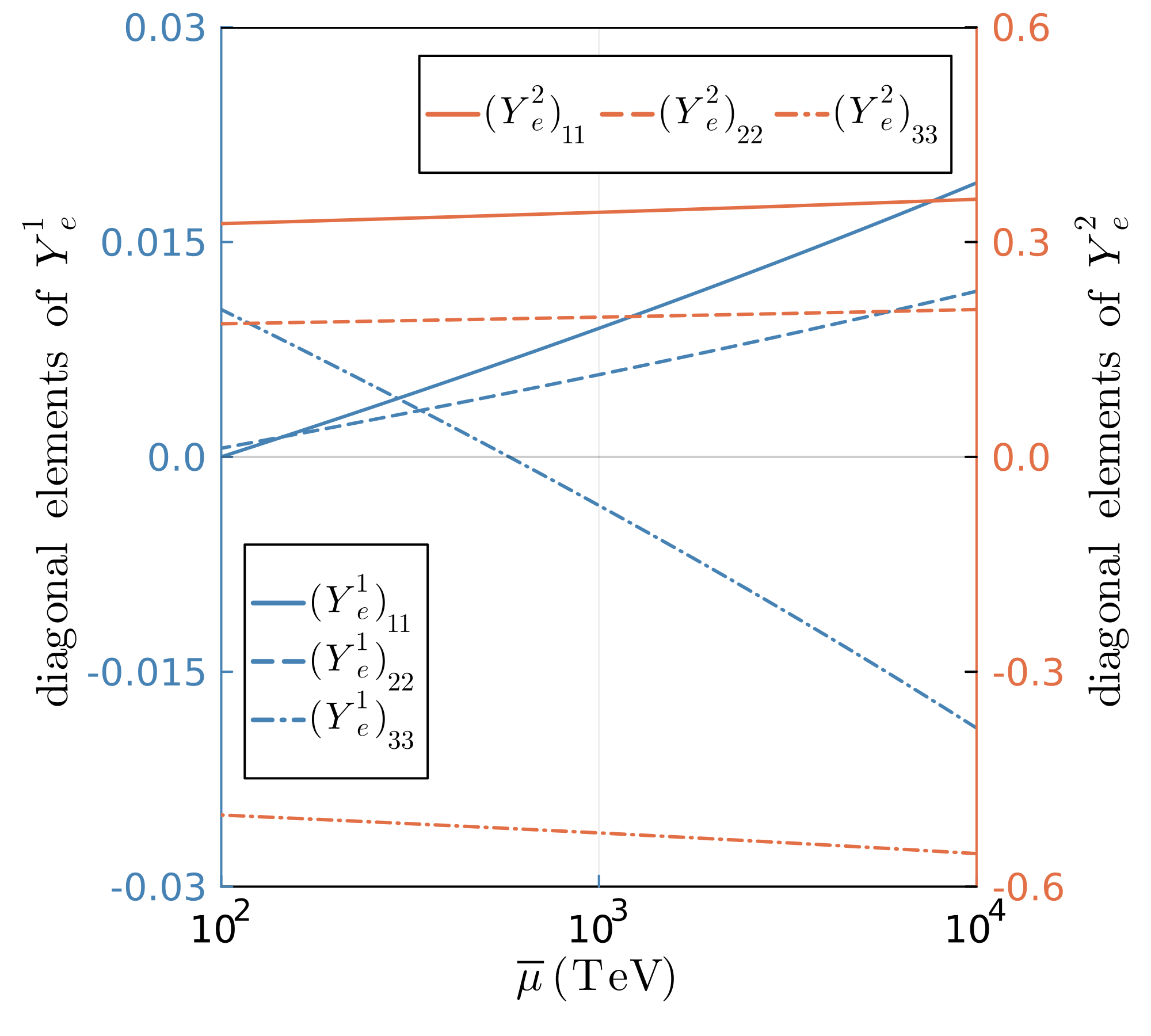}
    \caption{Running of $Y_e^1$ and $Y_e^2$.}
    \label{fig:inspectYe_B1}
  \end{subfigure}
  \caption{Running of $\kappa^{11}$, $\kappa^{12}$, $Y_e^1$ and $Y_e^2$ in the 2HDM EFT for the first set of high-scale benchmark parameters.
  The real component of the diagonal elements of each parameter matrix is plotted.
  The values of the Weinberg-like operator $\kappa$ are multiplied by $v^2$ to represent the scale of the resulting neutrino masses.
  }
  \label{fig:B1Running}
\end{figure}

Let us begin with the least engineered case, constraining the unknown Yukawa matrices only by the phenomenology discussed in \cref{sec:constraints} and the framework linking the UV parameters to the measured, low energy SMEFT parameters discussed in \cref{sec:framework}.
For this first set of benchmark parameters we choose $\lambda_1=\lambda_2=0.25$, $\lambda_{3,4,5,6,7}=0$ and $\alpha=1$ at the scale $\bar\mu = m_{H_2}$.
The running of the relevant parameters in the 2HDM EFT (between $\bar{\mu}=m_h$ and $m_{H_2}$) resulting from this first set of benchmark parameters is shown in \cref{fig:B1Running}. In particular, \cref{fig:inspectK_B1} shows the running of elements of $\kappa^{11}$, which dominantly contribute to the neutrino mass matrix as identified in \cref{sec:theory}, and that of the elements of $\kappa^{12{\rm (sym)}}$, which contribute to the running of $\kappa^{11}$. \Cref{fig:inspectYe_B1} shows the running of $Y_e^1$ and $Y_e^2$ which are also significant in the running of $\kappa^{11}$. In each case, we plot only the real component of the diagonal entries of the parameter matrices for clarity. These values provide an adequate qualitative summary of the behaviour of the matrices.

Prior to any quantitative comparison to the neutrino mass matrix in the full theory, \cref{fig:B1Running} demonstrates that the effect of the running is extremely qualitatively significant. Most importantly, it is clear that $\beta_{\kappa^{11}}$ cannot be approximated by a constant value over the chosen mass scales and thus a large log approximation will be invalid.

Recalling our discussion of \cref{eq:BetaK11} in \cref{sec:running}, there are two terms contributing significantly to $\beta_{\kappa^{11}}$ that do not contribute exponentially to the scaling of $\kappa^{11}$ in the sense of \cref{eq:expscaling}. One is dependent on $\kappa^{12{\rm (sym)}}$, and the other carries a tree-level contribution in the full $\kappa^{12}$. This tree level contribution is suppressed by the product of the two charged-lepton Yukawa couplings $\sim Y_e^1Y_e^2$ (throwing away here information that is not relevant to the overall scale). Considering \cref{fig:inspectYe_B1}, it appears that while $Y_e^2$ does not significantly change scale, $Y_e^1$ rapidly loses approximately one order of magnitude as it runs downward from $m_h$. This change of scale then propagates to $\beta_{\kappa^{11}}$, as the originally dominant $\kappa^{12}$ term becomes suppressed and relinquishes dominance over the running of $\kappa^{11}$ to the $\kappa^{12{\rm (sym)}}$ term. This behaviour is made apparent by the large curve in the running of $\kappa^{11}$ in \cref{fig:inspectK_B1} that results from a smooth change from one near-linear gradient (with respect to $t=\ln\bar{\mu}$) to another.

\subsubsection{Benchmark 2: Variations of the Yukawa couplings in the UV.}
\begin{figure}[htb!]
  \centering
  \begin{subfigure}[b]{0.45\textwidth}
    \centering
    \includegraphics[width=\textwidth]{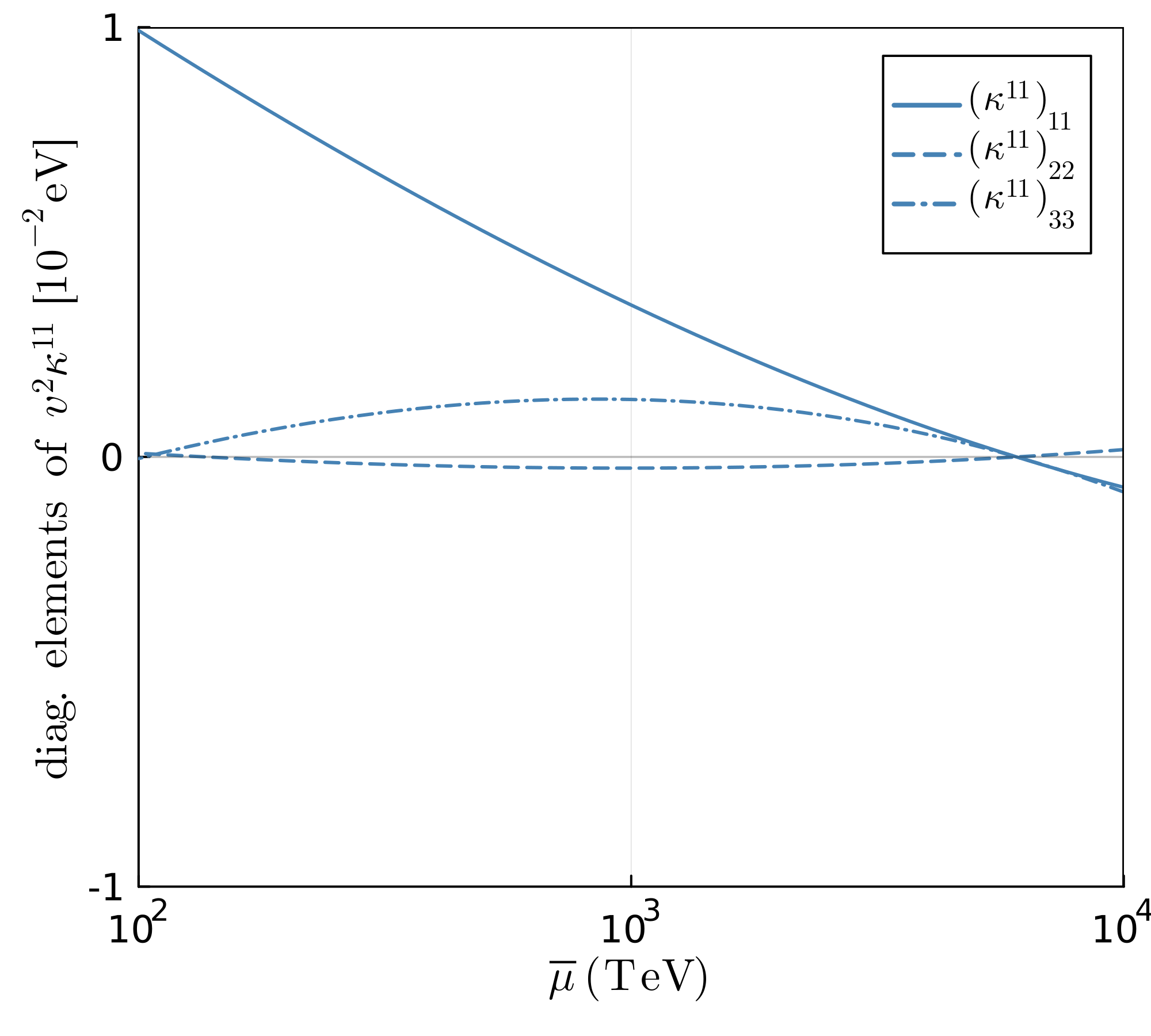}
    \caption{Running of $\kappa^{11}$.}
    \label{fig:inspectK11_B2}
  \end{subfigure}
  \hfill
  \begin{subfigure}[b]{0.45\textwidth}
    \centering
    \includegraphics[width=\textwidth]{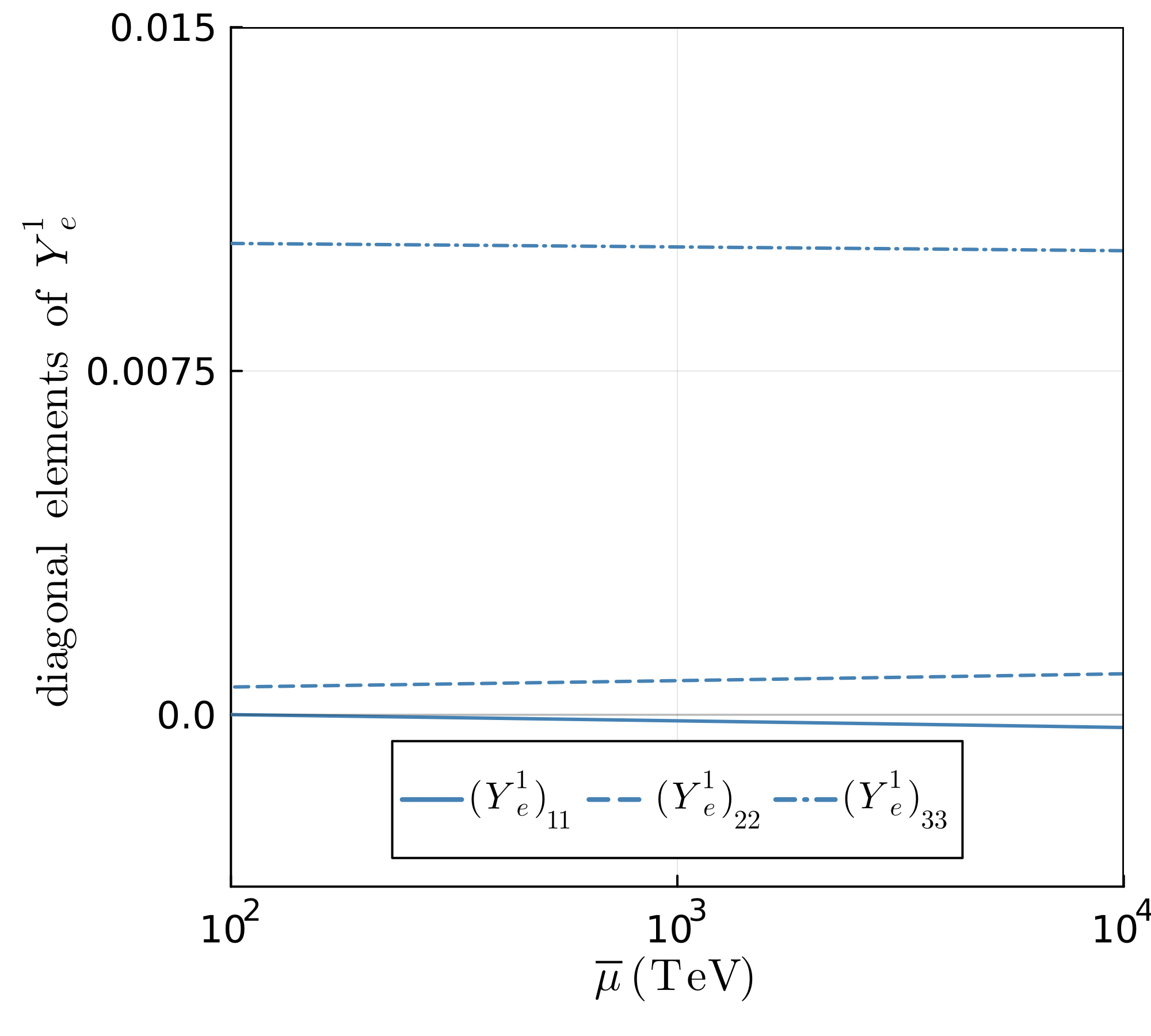}
    \caption{Running of $Y_e^1$.} 
    \label{fig:Ye1_B2}
  \end{subfigure}
  \caption{Running of $\kappa^{11}$ and $Y_e^1$ in the 2HDM EFT for the second set of high-scale benchmark parameters ($Y_e^1\sim Y_e^2$).
  The real component of the diagonal elements of each parameter matrix is plotted.}
  \label{fig:B2_running}
\end{figure}

To further understand the significance of the Yukawa couplings, let us consider benchmark 2 where we constrain the components of $Y_e^2$ such that they are roughly the same order of magnitude as the components of $Y_e^1$. All our choices for the mass scales, $\alpha$, and $\lambda_{1,\ldots,7}$ remain the same as in benchmark 1.  
Doing so results in the running for $\kappa^{11}$ and $Y_e^1$ shown in \cref{fig:B2_running}. 
We find that the much lessened decrease in the scale of $Y_e^1$ results in a much less drastic shift to the new gradient in $\beta_{\kappa^{11}}$.
We note that the running of $\kappa^{12}$ and $Y_e^2$ are very similar to the benchmark 1 case and not otherwise interesting here.

For these second benchmark values, the running of $Y_e^1$ is much less.
Although there still appears to be some non-linearity in the running of $\kappa^{11}$,
the relatively linear running of $Y_e^1$ suggests that it may be reasonable to consider a partial large log approximation where only the running of $\kappa^{11}$ is calculated from the full differential equation, while the running of all the other parameters is analytically described by a large log approximation.

An additional feature of this benchmark set is that the suppression of $Y_e^2$ to approximately the same order of magnitude as $Y_e^1$ allows the cLFV constraints in \cref{app:pheno} to be satisfied with a lower scale of new physics. To demonstrate the significance of these running effects even at scales much closer to the EW scale, we also generate a set of low scale benchmark parameters (also listed in \cref{app:benchmarks}) with $m_{H_2}=1$ TeV, $m_h=\mu_{\rm Zee}=100$ TeV and enforcing the same $Y_e^2\sim Y_e^1$ constraint on the elements of the Yukawa matrices.
We also take the opportunity of this lower scale to demonstrate a different choice of the quartic couplings
\begin{align}
    \lambda_1 = \lambda_2 = \lambda_6 = \lambda_7 &= 0.2, &
    \lambda_4 = \lambda_5 &= 0.4, &
    \lambda_3 &= 0,
\end{align}
which is consistent with the constraints in \cref{eq:lconstrainta,eq:lconstraintb} the general scale $\lambda_1=0.2$ is chosen to reproduce the correct SM Higgs mass at the low scale.
We see from the matching condition in \cref{eq:appMatchSMEFTmu} in combination with the matching condition for the quartic couplings between the 2HDM EFT and the SMEFT in \cref{eq:appMatchL}, that
\begin{align}
     \mu^{2} &= \mu_1^2 - m_{H_2}^2(2\lambda_3 + \lambda_4)\frac{1}{16\pi^{2}} \left(1+\log\left(\frac{\bar{\mu}^2}{m_{H_2}^2}\right)\right) 
\end{align}
allows for a non-zero $\lambda_3+\lambda_4$ if the mass scale $m_{H_2}$ is not too large compared to the Higgs vev, as is the case for $m_{H_2}=1$ TeV.

\begin{figure}[htb!]
  \centering
  \begin{subfigure}[b]{0.45\textwidth}
    \centering
    \includegraphics[width=\textwidth]{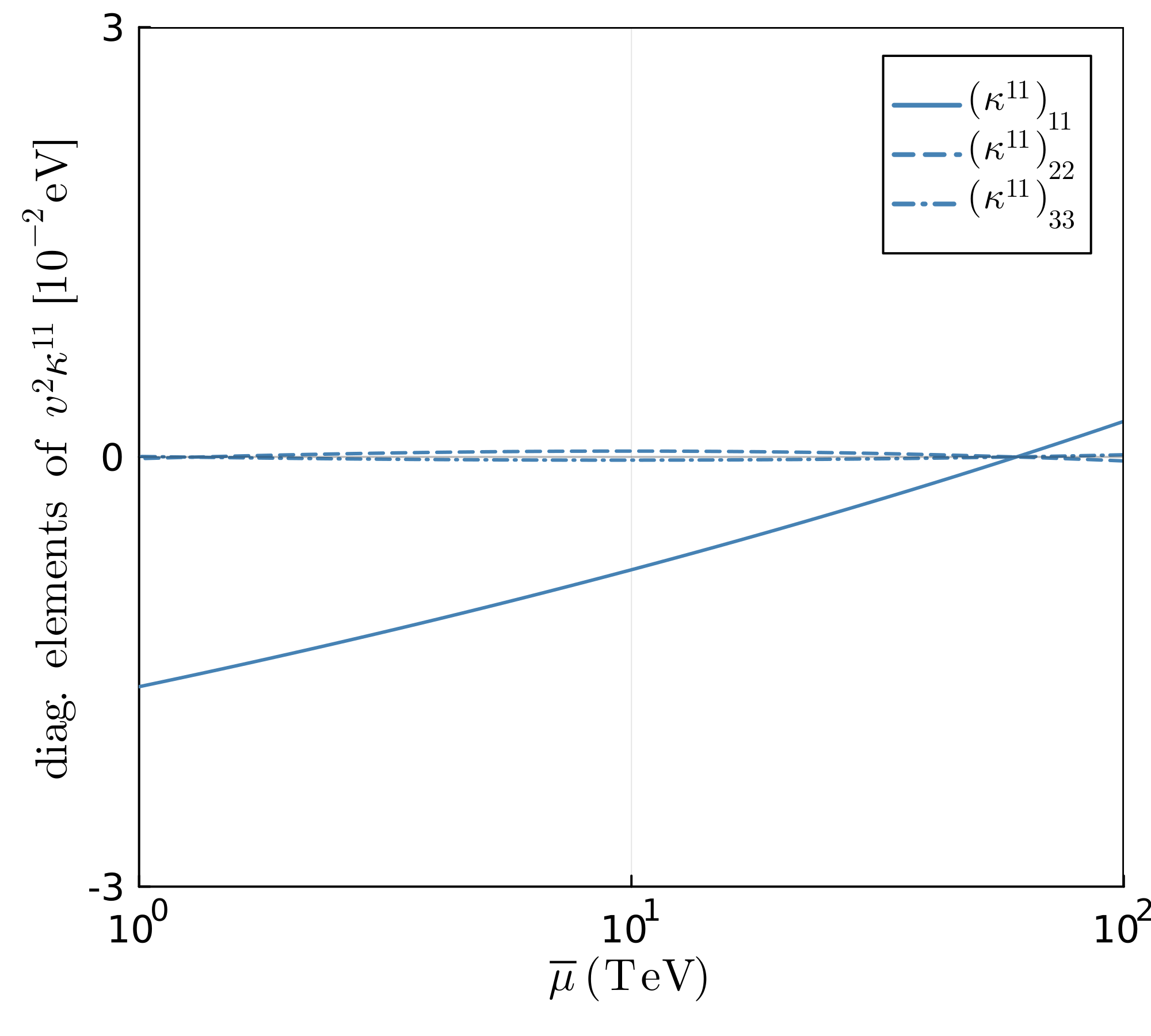}
    \caption{Running of $\kappa^{11}$.}
    \label{fig:inspectK11_LS}
  \end{subfigure}
  \hfill
  \begin{subfigure}[b]{0.45\textwidth}
    \centering
    \includegraphics[width=\textwidth]{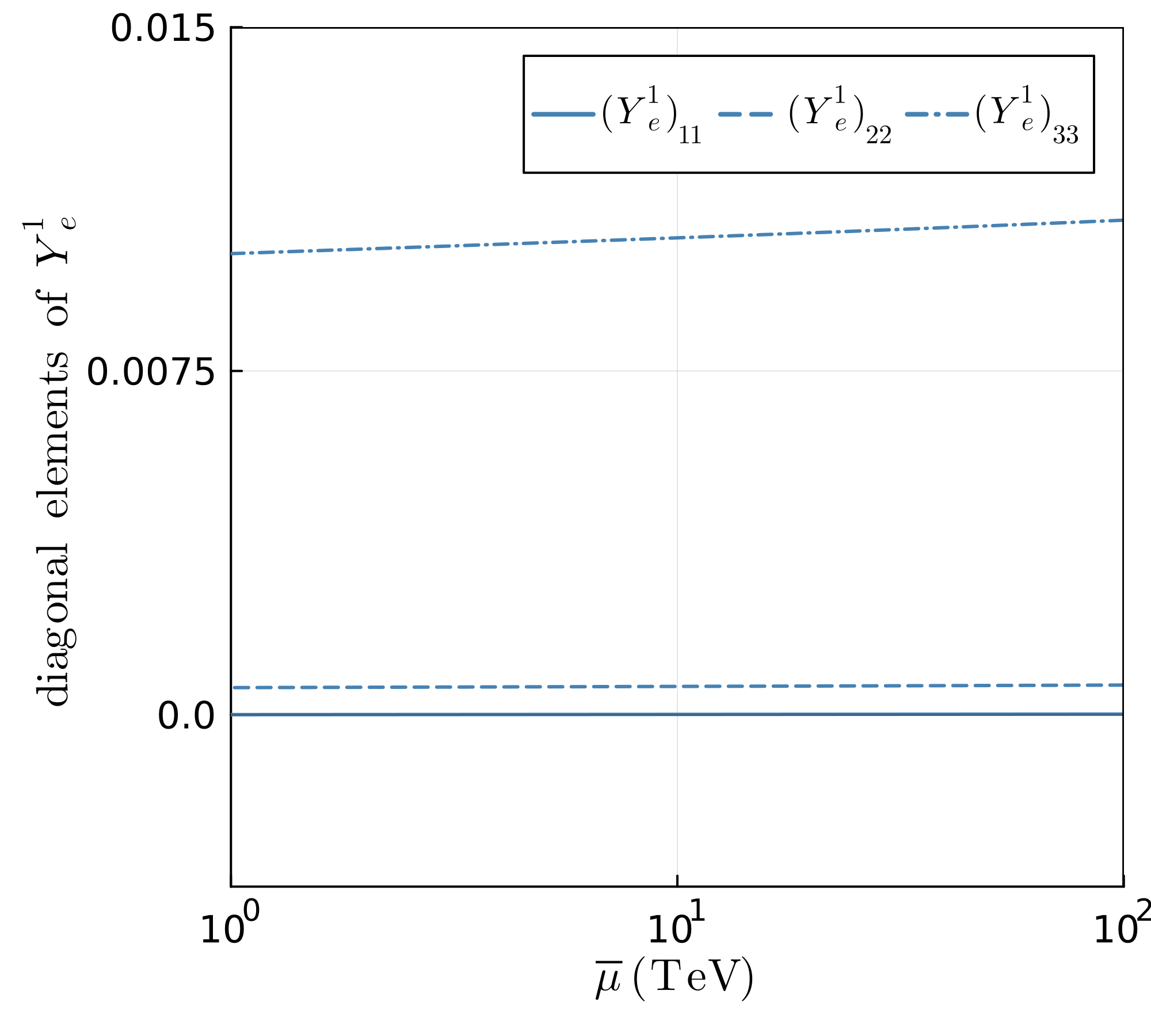}
    \caption{Running of $Y_e^1$.} 
    \label{fig:Ye1_LS}
  \end{subfigure}
  \caption{Running of $\kappa^{11}$ and $Y_e^1$ in the 2HDM EFT for the set of low-scale benchmark parameters with $m_{H_2}=1$TeV, $m_h=\mu_{\rm Zee}=100$TeV, and ($Y_e^1\sim Y_e^2$).
  The real component of the diagonal elements of each parameter matrix is plotted.}
  \label{fig:LS_running}
\end{figure}

The resulting running of $\kappa^{11}$ and $Y_e^{1}$ in the low scale 2HDM EFT is shown in \cref{fig:LS_running}.  

As predicted at the end of \cref{sec:framework},
we observe exactly the same qualitative behaviour with the lower scale benchmark parameters in \cref{fig:LS_running} as with the higher scale benchmark in \cref{fig:B2_running}.
This highlights the importance of considering the running effects for neutrino mass generating new physics across a broad range of scales. It also demonstrates that, at least for small $Y_e^2\sim Y_e^1$, the quartic couplings do not greatly influence the running.

\subsubsection{Benchmark 3: (\texorpdfstring{$\alpha=0$}{alpha = 0})}
\begin{figure}[!t]
  \centering
  \begin{subfigure}[b]{0.47\textwidth}
    \centering
    \includegraphics[width=\textwidth]{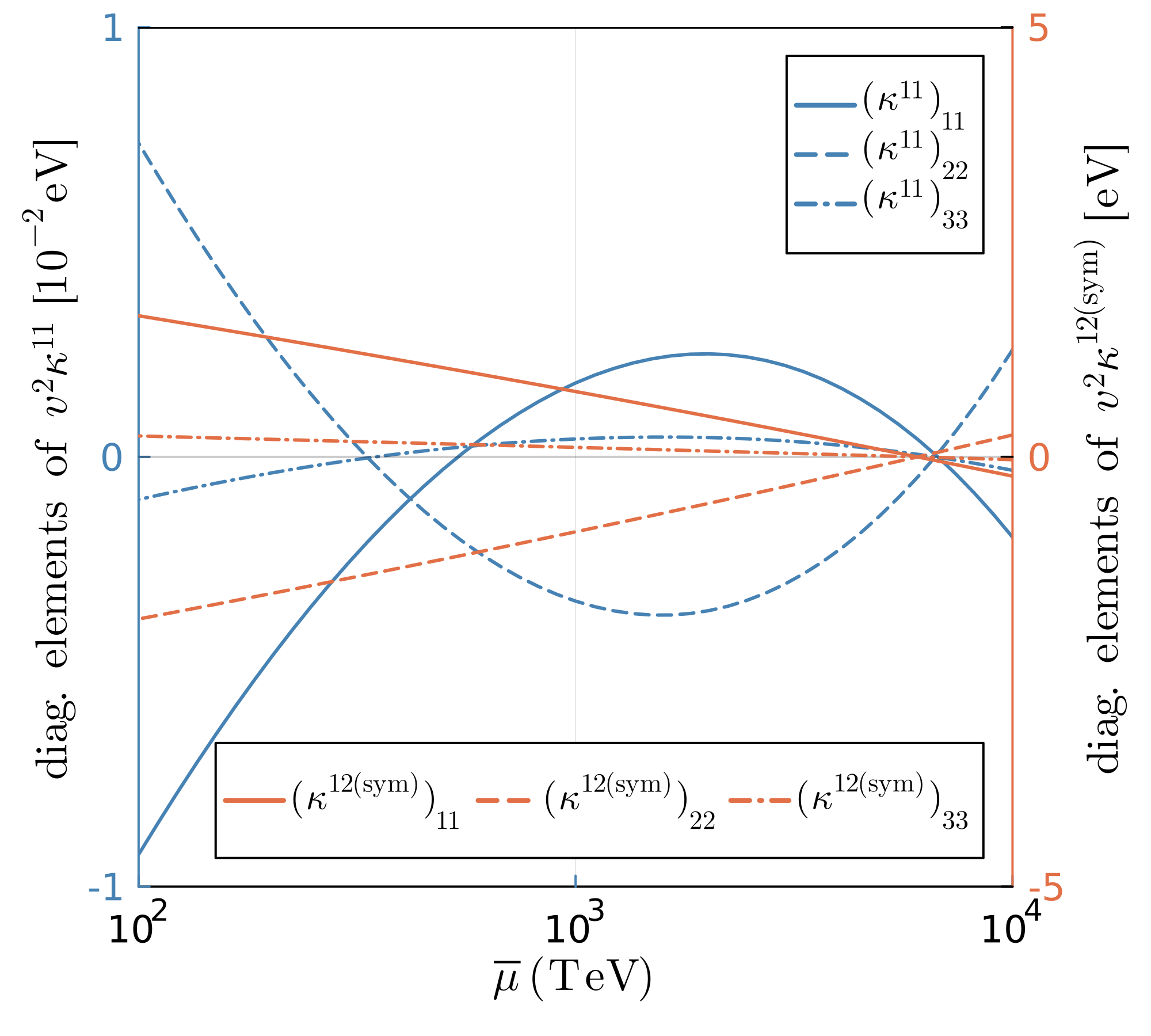}
    \caption{Running of $\kappa^{11}$.}
    \label{fig:inspectK11_B3}
  \end{subfigure}
  \hfill
  \begin{subfigure}[b]{0.47\textwidth}
    \centering
    \includegraphics[width=\textwidth]{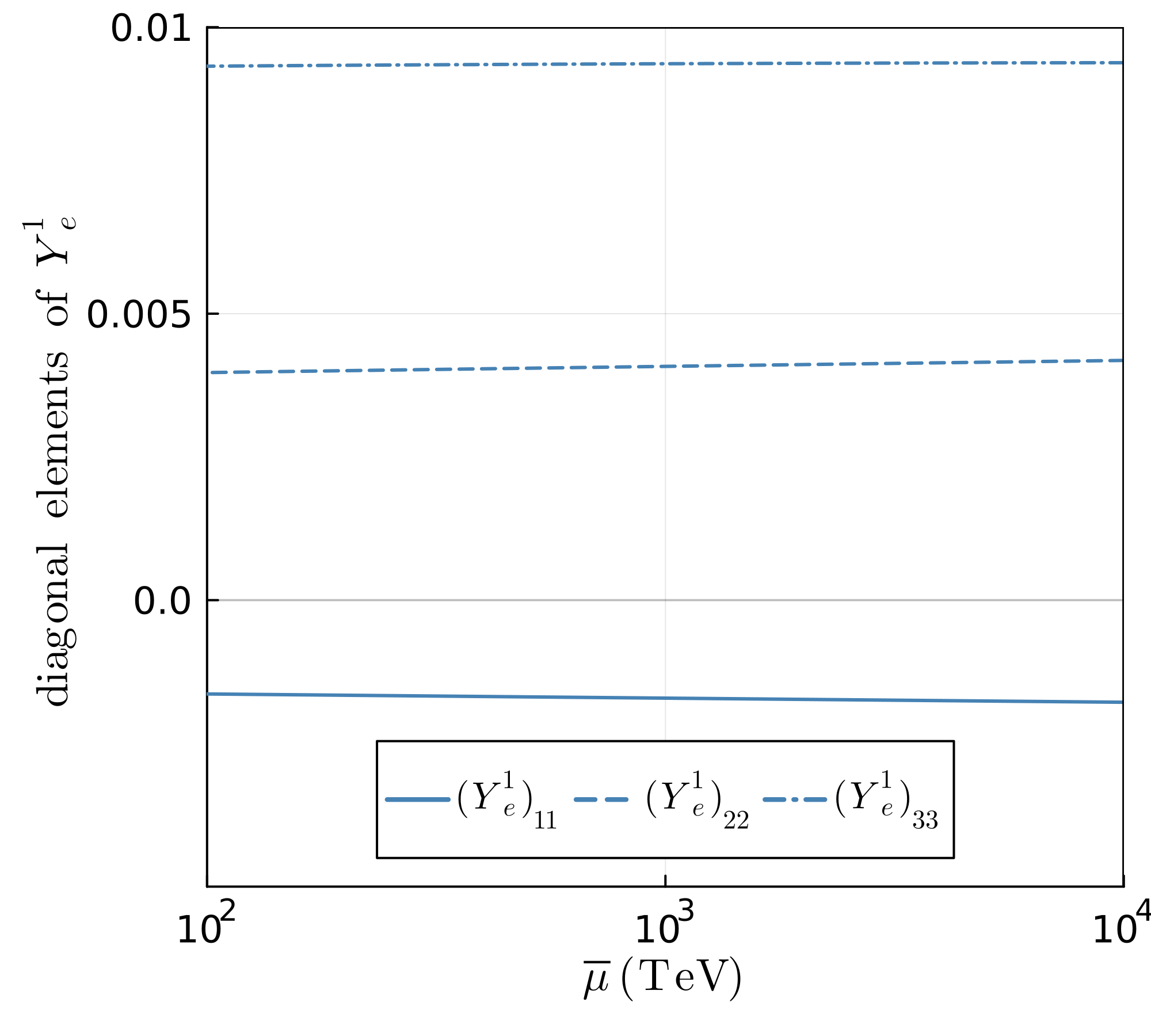}
    \caption{Running of $Y_e^1$.}
    \label{fig:Ye1_B3}
  \end{subfigure}
  \caption{Running of $\kappa^{11}$ and $Y_e^1$ in the 2HDM EFT for the third set of high-scale benchmark parameters ($\alpha=0$).
  The real component of the diagonal elements of each parameter matrix is plotted.}
  \label{fig:B3_running}
\end{figure}

It is clear that the scales of $Y_e^1$ and $Y_e^2$ play a significant role in the running in the 2HDM EFT.
We highlight here the $\beta$-function for $Y_e^1$ (also shown in \cref{app:betas}),
\begin{align}
16\pi^{2}\frac{d}{dt}Y_e^{1} = \sum_{k=1}^{2}\left( T_{1k}Y_e^{k}
+ Y_e^{k} Y_e^{k\dagger} Y_e^{1}  + \frac{1}{2}Y_e^{1}Y_e^{k\dagger} Y_e^{k} \right)
- \frac{9g_2^2 + 15g_1^2}{4}Y_e^{1} \ . \label{eq:betaYe1}
\end{align}
This beta function will be dominated by the terms with $\mathcal{O}(1)$ coefficients. These are the two terms proportional to the traces $T_{11}$ and $T_{12}$ that contain the quark Yukawa matrices, and the final term with the coefficient given by the gauge couplings. 
$Y_e^2$ contributes most strongly to the running of $Y_e^1$ through the term $T_{12}Y_e^2$.
This indicates that we can reduce the running of $Y_e^1$ by making $Y_e^2$ small, as we have seen in the second benchmark set, or alternatively by making $T_{12}$ small.
The easiest way to do so in our framework is by choosing $\alpha$ to be small.
This will have the additional effect of greatly reducing the strength of the term proportional to $\kappa^{12{\rm (sym)}}$ and make the scale of $\lambda_6$  important in the running of $\kappa^{11}$, as per \cref{eq:BetaK11}.
It is this extreme case of $\alpha = 0$ that is represented in our third set of benchmark parameters, resulting in the running shown in \cref{fig:B3_running}. Here, we return to a relatively large $Y_e^2\sim\mathcal{O}(0.1)$ and the quartic couplings set in the first two benchmarks, except for the choice $\lambda_6=-\lambda_7=\lambda_1(=0.25)$ which is required to ensure compatibility with both the low scale neutrino mass parameters in our fitting procedure and the bounds from the stability of the potential in \cref{eq:lconstrainta,eq:lconstraintb}.

Indeed, we see in \cref{fig:Ye1_B3} that there is extremely little running of $Y_e^1$. 
Nevertheless, the running of $\kappa^{11}$ shows a shift to the second gradient as per the first benchmark case.
This similar behaviour is not entirely surprising as $Y_e^2$ is large in both cases and a non-zero $\lambda_6$ in this case contributes in a similar way as a non-zero $\alpha$ contributes, through the trace term $T_{12}$, to the running of $\kappa^{11}$ given in \cref{eq:BetaK11}.
The shift from one dominating gradient to another must therefore manifest through the running of the $\kappa^{12}$ term. 
Comparing \cref{fig:inspectK11_B2,fig:inspectK11_B3}, we see that the running of $\kappa^{12{\rm (sym)}}$ is an order of magnitude larger in the third benchmark case, exaggerating the discrepancy in the running of $\kappa^{12}$ and $\kappa^{12{\rm (sym)}}$ which are the relevant terms in each gradient.

It is natural to ask what the running behaviour would be in a case where $\alpha$ is not a pure scale but includes a phase, or even, if $\alpha$ were some general matrix amounting to defining the quark Yukawas as general matrices themselves.
These cases differ very little as the quark Yukawa matrices only ever contribute to the matching and running of the charged-lepton Yukawa matrices via a trace term and so their shape is never significant. Furthermore, a phase in $\alpha$ would also contribute minimally as we take $Y_e^2$ itself to be a generic complex matrix that correctly reproduces the neutrino masses. The choice of $Y_e^2$ could effectively absorb any arbitrary phase in a component of its running. 


\subsection{Sensitivity of mixing parameters to UV model parameters}\label{sec:sensitivity}
While the results presented in the previous section demonstrate the running behaviour of the 2HDM EFT parameters directly as a function of the scale $\bar{\mu}$, these are not the physical neutrino mixing parameters that we wish to describe.
In our full EFT approach, the physical parameters are only well defined in the SMEFT at low energy, and thus their running in the region of the 2HDM EFT $m_{H_2} \leq \bar{\mu} \leq m_h$ is rather opaque.
Instead, we will convey the significance of the running on these mixing parameters by analysing their sensitivity to the chosen UV parameters.

The calculation of the neutrino mass matrix in the full theory, see \cref{eq:naiveMass}, is insensitive to changes in the individual scales of the three couplings $f$, $Y_e^1$ and $Y_e^2$ as well as the scales $\mu_{\rm Zee}$ and $m_h^2$, as long as the combined scale of these five terms remains invariant.
More concretely, let us choose a set of benchmark high scale parameters along with a scale factor $\gamma > 0$. Transforming 
\begin{align}
    f\to \gamma f \ , \ \  Y_e^1\to \gamma^{-1/2}Y_e^1 \ \text{ and } \ Y_e^2\to \gamma^{-1/2}Y_e^2 \ , \label{eq:paramStretchA}
\end{align}
we leave the neutrino mass matrix in the full theory unchanged.
Generically, this scaling will slightly spoil the connection to other measured parameters in the SMEFT, in particular the charged lepton masses
\footnote{Note that there is freedom in the choice of both $Y_e^1$ and $Y_e^2$ with respect to the measured charged lepton masses. The freedom in $Y_e^2$ can be interpreted as freedom in $Y_e^1$ because $Y_e^2$ contributes in a highly non-trivial way to the running of $Y_e^1$. Therefore, one should not be concerned that scaling $Y_e^1$ is any more egregious for the connection to the charged lepton masses than scaling $Y_e^2$.}.
Indeed, we find that depending on the characteristics of the benchmark parameters used, the charged lepton masses in particular can vary up to $\mathcal{O}(10\%)$ when $\gamma$ scales the high energy parameters by $\sim \mathcal{O}(1\%)$.
The amplification of this scaling through the running itself suggests that the total running is considerably non-linear.
For the particular analysis in this subsection, we have chosen to forego strict connections to these peripheral SMEFT parameters so that we can simply and clearly demonstrate the qualitative behaviour of the running of the neutrino mass parameters. 

Importantly, although inconsequential in the full theory, the scaling in \cref{eq:paramStretchA} is significant in the EFT calculation of the neutrino mass matrix as each coupling parameter runs differently.
Therefore in the full EFT approach, the relative shift in the final value for a given mixing parameter due to a scaling of the high scale model parameters that leaves the calculation in the full theory invariant provides a measure of the significance of the running effects.
If we compare this measure across different choices of scaling of the UV parameters, we can also study the relevance of different parameter choices to the running.

\begin{figure}[!tb]
    \centering
    \begin{subfigure}[b]{0.32\textwidth}
        \centering
        \includegraphics[width=\textwidth]{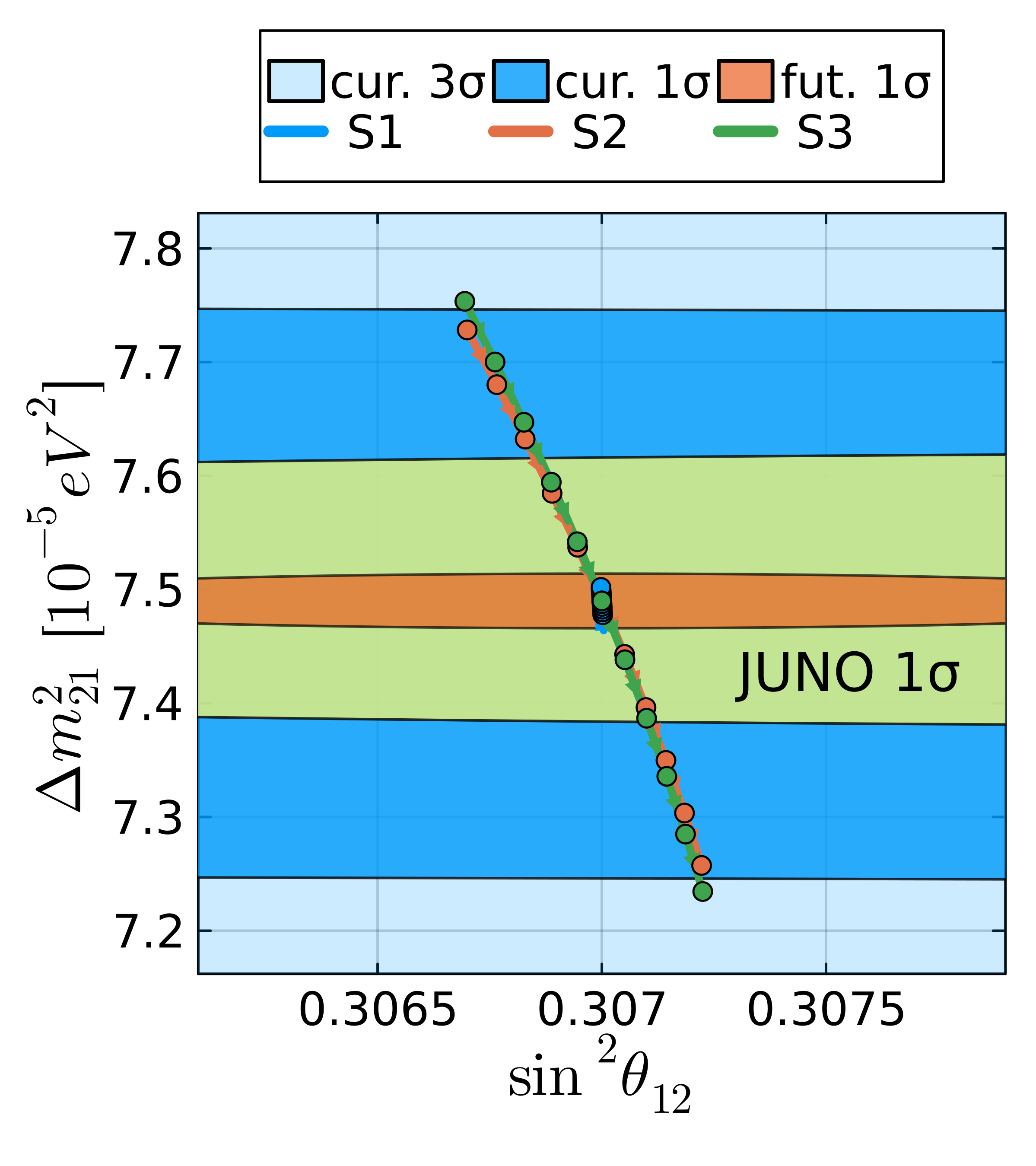}
        \caption{BM 1 ($Y_e^2\gg Y_e^1$ and $\alpha=1$)}
        \label{fig:paramStretchAB1}
    \end{subfigure}
    \begin{subfigure}[b]{0.32\textwidth}
        \centering
        \includegraphics[width=\textwidth]{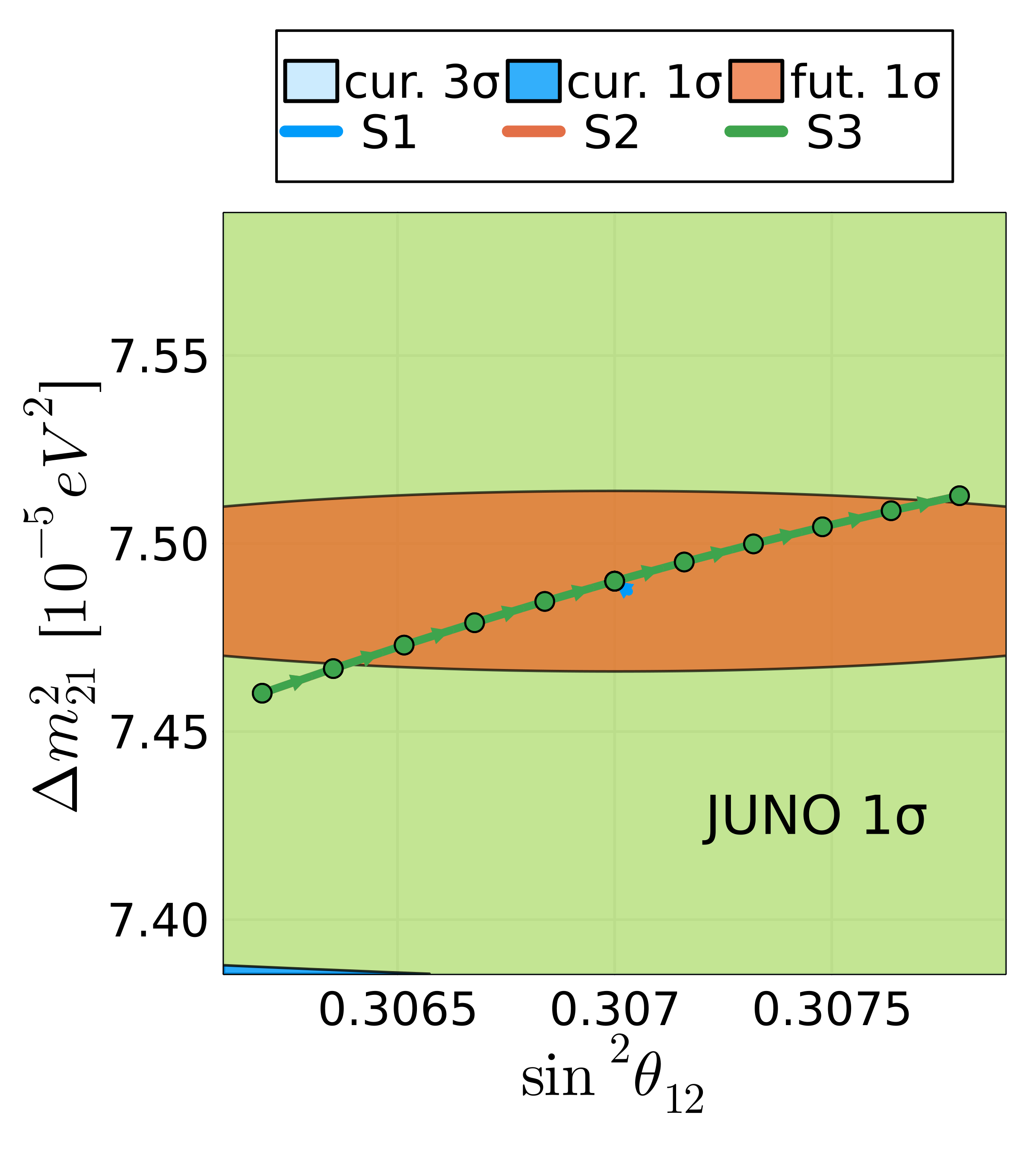}
        \caption{BM 2 ($Y_e^2\sim Y_e^1$ and $\alpha=1$)}
        \label{fig:paramStretchAB2}
    \end{subfigure}
    \begin{subfigure}[b]{0.32\textwidth}
        \centering
        \includegraphics[width=\textwidth]{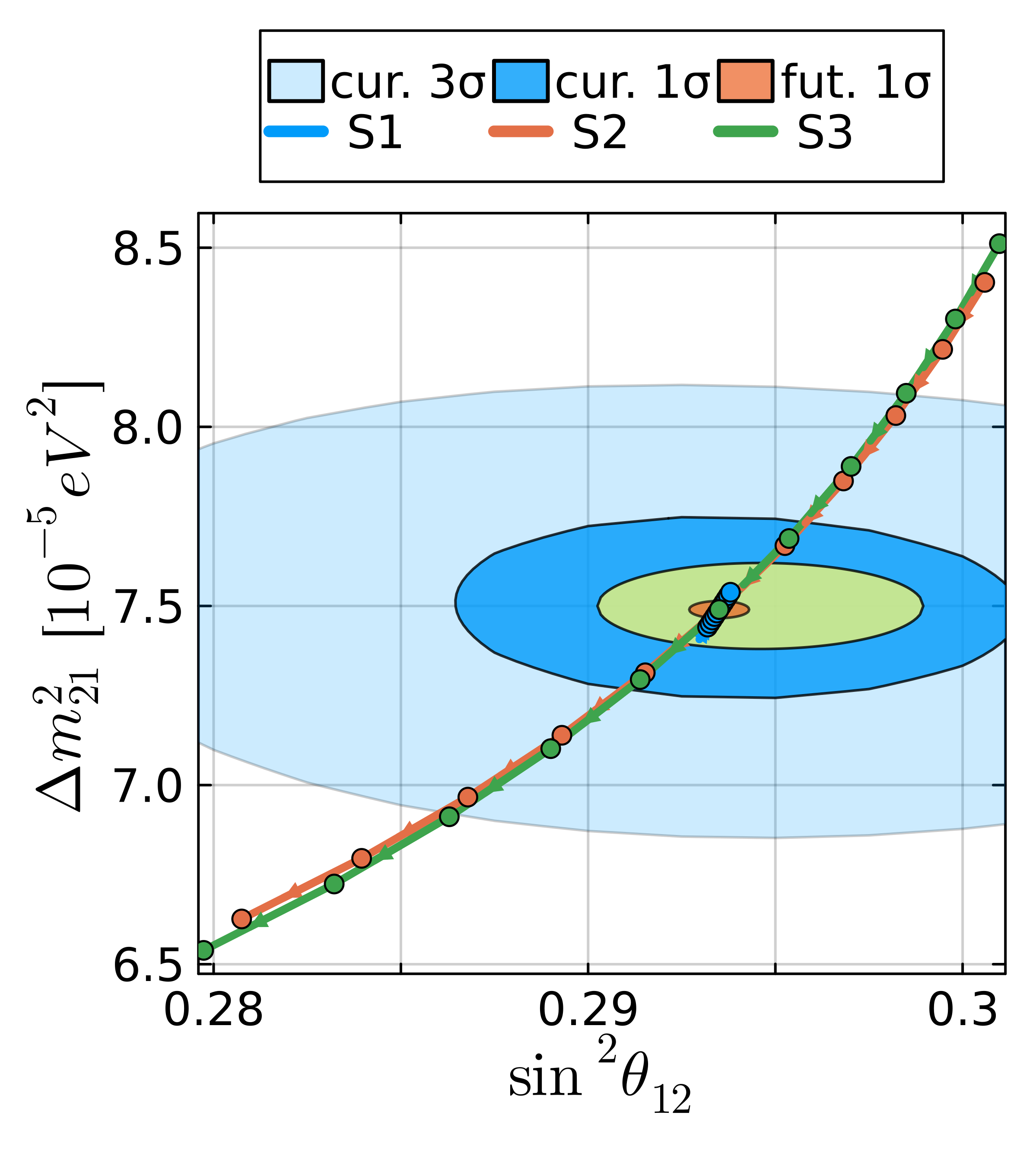}
        \caption{BM 3 ($Y_e^2\gg Y_e^1$ and $\alpha=0$)}
        \label{fig:paramStretchAB3}
    \end{subfigure}
    \caption{
    Sensitivity of the solar neutrino mixing parameters to the high-scale model parameters in the EFT framework for each set of benchmark parameters with $m_{h}=10^4$TeV. 
    The high-scale parameters are scaled to preserve the neutrino mass matrix in the full theory with: S1 $\implies \gamma^{-1/2}Y_e^{1,2}$, $\gamma f$; S2 $\implies \gamma^{-1}Y_e^{2}$, $\gamma f$; S3 $\implies \gamma^{-1}f$, $\gamma Y_e^1$; and with $0.99\leq\gamma\leq1.01$ increasing in increments of $0.002$ in the direction of the arrows marked.
    The shaded regions represent current and future estimated experimental uncertainties including the latest $1\sigma$ result of the JUNO collaboration~\cite{JUNO:2025gmd} (future projected to 6 years by the JUNO collaboration \cite{JUNO:2022mxj}), using the current central values from a global fit \cite{Esteban:2024eli}.
    Note that S2 and S3 overlap in figure (b). 
    }
    \label{fig:paramStretchA}
\end{figure}

For each of our three regular benchmark (BM) sets (c.f. \cref{tbl:benchmarks}), we plot the final neutrino mass parameters due to the scaling in \cref{eq:paramStretchA} as well as the scalings
\begin{align}
    f\to \gamma f \ , \ \  Y_e^2\to \gamma^{-1}Y_e^2\label{eq:paramStretchB}\\
    \text{and, } \ f\to \gamma^{-1} f \ , \ \  Y_e^1\to \gamma Y_e^1 \label{eq:paramStretchC}
\end{align}
in \cref{fig:paramStretchA}.
We choose $\gamma$ to scale the high scale couplings by $\pm 1\%$ and compare the size of the corrections to the current and future experimental uncertainty, as projected by the JUNO collaboration \cite{JUNO:2022mxj}.

If we were calculating the neutrino mass matrix in the full theory, each line in the plots in \cref{fig:paramStretchA} would remain fixed at the central point for all values of $\gamma$.
The visible displacement from this point immediately demonstrates that the running contributes non-trivially to the calculation. 
The larger this displacement, the greater the effect of the running.
We see that the largest of the effects shown are always for the scalings S2 and S3 (these two scalings overlap for the second benchmark in \cref{fig:paramStretchAB2}). 
In each of these cases, we are scaling one of $Y_e^1$ or $Y_e^2$ independently of the other. 
We have already demonstrated that the relative scale of $Y_e^1$ and $Y_e^2$ is a crucial factor in the running.

We can understand the qualitative behaviour in each of these plots analytically by considering the running of $\kappa^{11}$.
The $\beta$-function for $\kappa^{11}$ (see \cref{eq:BetaK11}) contains terms with a $\kappa^{ij}$ containing matching contributions from either tree- or loop-level or both.
The only term that contains a $\kappa^{ij}$ with tree-level matching contributions is the term that contains $\kappa^{12}$. This term also contains one of each Yukawa coupling $Y_e^{1,2}$ and thus the tree-level component in $\kappa^{12}$ is suppressed by a product of the two Yukawa couplings.
Many terms in $\beta_{\kappa^{11}}$ therefore contain the same $f{Y_e^1}^\dagger Y_e^2$ structure found in the full theory calculation and thus will also not be sensitive to the scalings S1, S2, and S3.
However, there are three exceptions. 

The first is the term $\kappa^{12{\rm (sym)}}$, which contains a loop-level matching contribution of the form $f{Y_e^i}^\dagger Y_e^i$ with two Yukawa matrices of the same `Higgs index' $i$.
These terms will be sensitive to the scalings S2 and S3 that scale $Y_e^1$ and $Y_e^2$ independently of each other.
Being the only such terms with two Yukawa matrices, the variation in these $\kappa^{12{\rm (sym)}}$ terms with respect to $\gamma$ will dominate the variation in the running.

The second exception is the terms containing two of the same Yukawa couplings and $\kappa^{11}$ itself. Considering the matching of $\kappa^{11}$ at the UV scale, we can consider these terms to effectively contain four Yukawa couplings, and the Yukawa-like coupling $f$ at the UV scale. These terms will be sensitive to all three scalings as there will always be some remaining factor of $\gamma$ in the combined scaling.

The final exception is the term containing the full $\kappa^{12}{Y_e^1}^\dagger Y_e^2$. The part combining with the tree-level contribution to $\kappa^{12}$ will be insensitive to the scalings, as indicated previously. However, the loop-level contribution will be similar to the previous exceptional case and contains four Yukawa couplings, rendering it also sensitive to the scalings. In this way the scaling S1 allows us to probe the significance of the Yukawa-suppressed contributions to the running as the dominant terms with only two Yukawa couplings are far less sensitive to the scaling S1.\footnote{It should be noted that decomposing the $\kappa^{ij}$ in $\beta_{\kappa^{11}}$ in terms of their matching conditions is only valid at the UV scale. As each $\kappa^{ij}$ runs, they do not strictly maintain the same functional form in terms of the UV parameters. However, this approximation is sufficient to gain a qualitative understanding of the running behaviour that we discuss here.}
We then see these effects in the plots of \cref{fig:paramStretchA}.

Across the three benchmarks, S1 always results in the smallest variation in the final neutrino mass as it is only felt through the terms with extra Yukawa suppression.
In \cref{fig:paramStretchAB2}, the second set of benchmark parameters enforces the smallness of $Y_e^2$, and therefore the Yukawa suppression is maximised, leading to negligible variation under S1.
The scalings S2 and S3 result in a much larger variation in the final neutrino mass as they act through the dominant $\kappa^{12{(\rm sym)}}$ term in $\beta_{\kappa^{11}}$. 
These two scalings produce similar behaviour due to the highly symmetric way $Y_e^1$ and $Y_e^2$ contribute in the 2HDM EFT.
The second set of benchmark parameters also creates a special case for these scalings as when $Y_e^1$ and $Y_e^2$ are enforced to be of roughly the same scale, they become almost entirely interchangeable parameters and thus the effect of scaling either one is indistinguishable, as seen by the overlap of the S2 and S3 lines in \cref{fig:paramStretchAB2}.

\begin{figure}[!tb]
    \centering
    \begin{subfigure}[b]{0.32\textwidth}
        \centering
        \includegraphics[width=\textwidth]{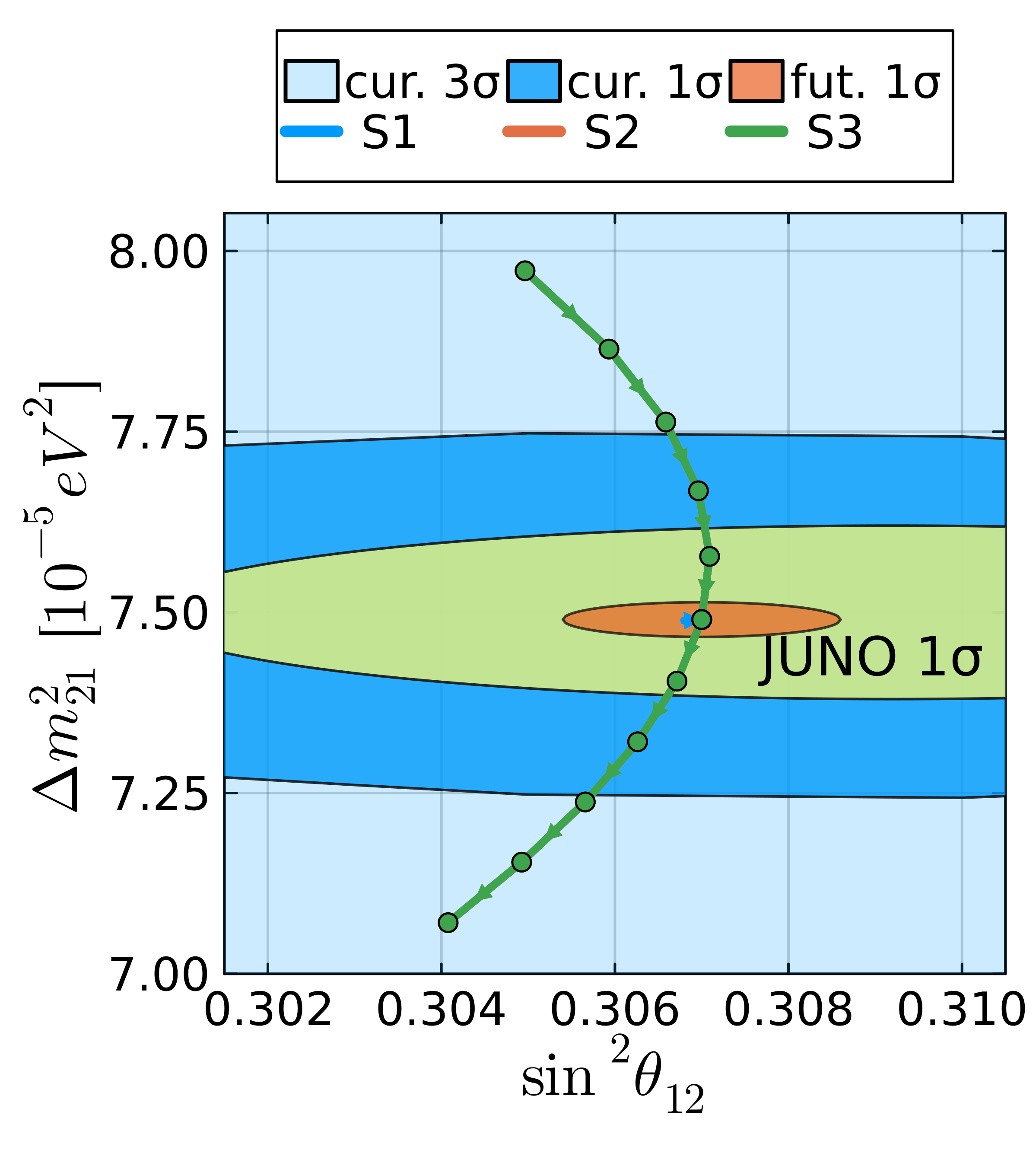}
        \caption{Solar parameters.}
        \label{fig:paramStretchALS}
    \end{subfigure}
    \begin{subfigure}[b]{0.32\textwidth}
        \centering
        \includegraphics[width=\textwidth]{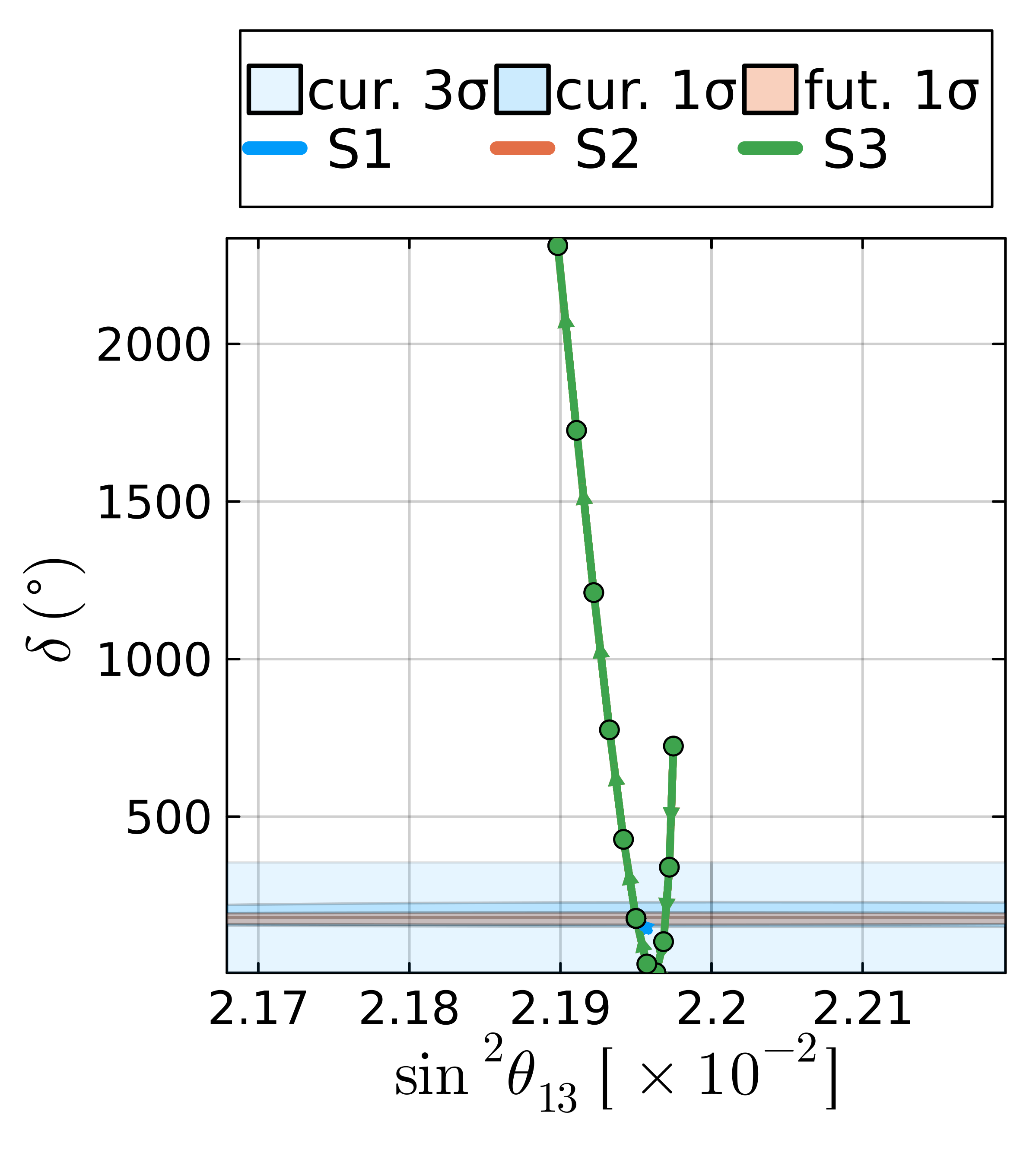}
        \caption{CP violating phase.}
        \label{fig:paramStretchDLS}
    \end{subfigure}
    \\
    \begin{subfigure}[b]{0.64\textwidth}
        \centering
        \includegraphics[width=0.5\textwidth]{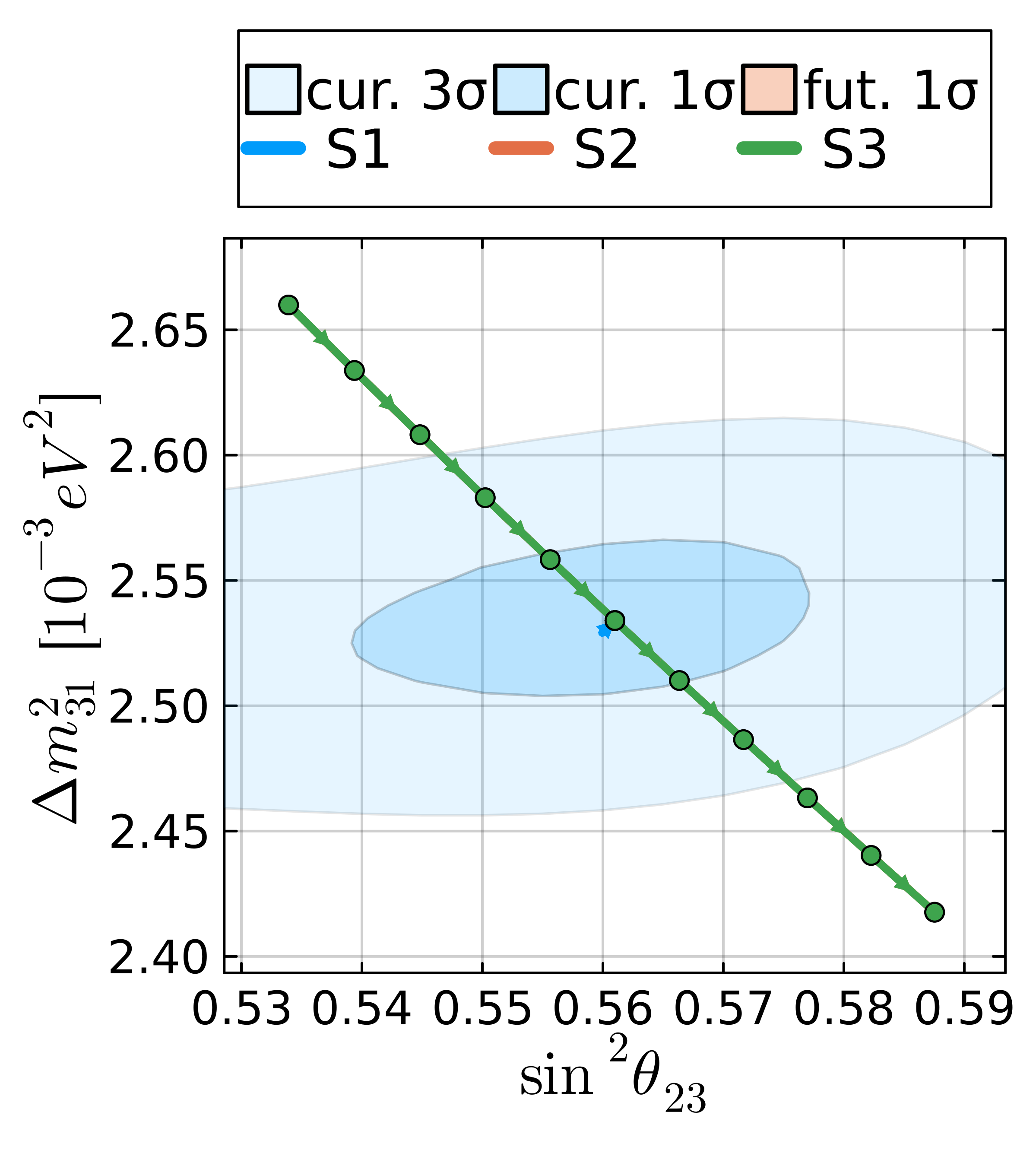}%
        \includegraphics[width=0.5\textwidth]{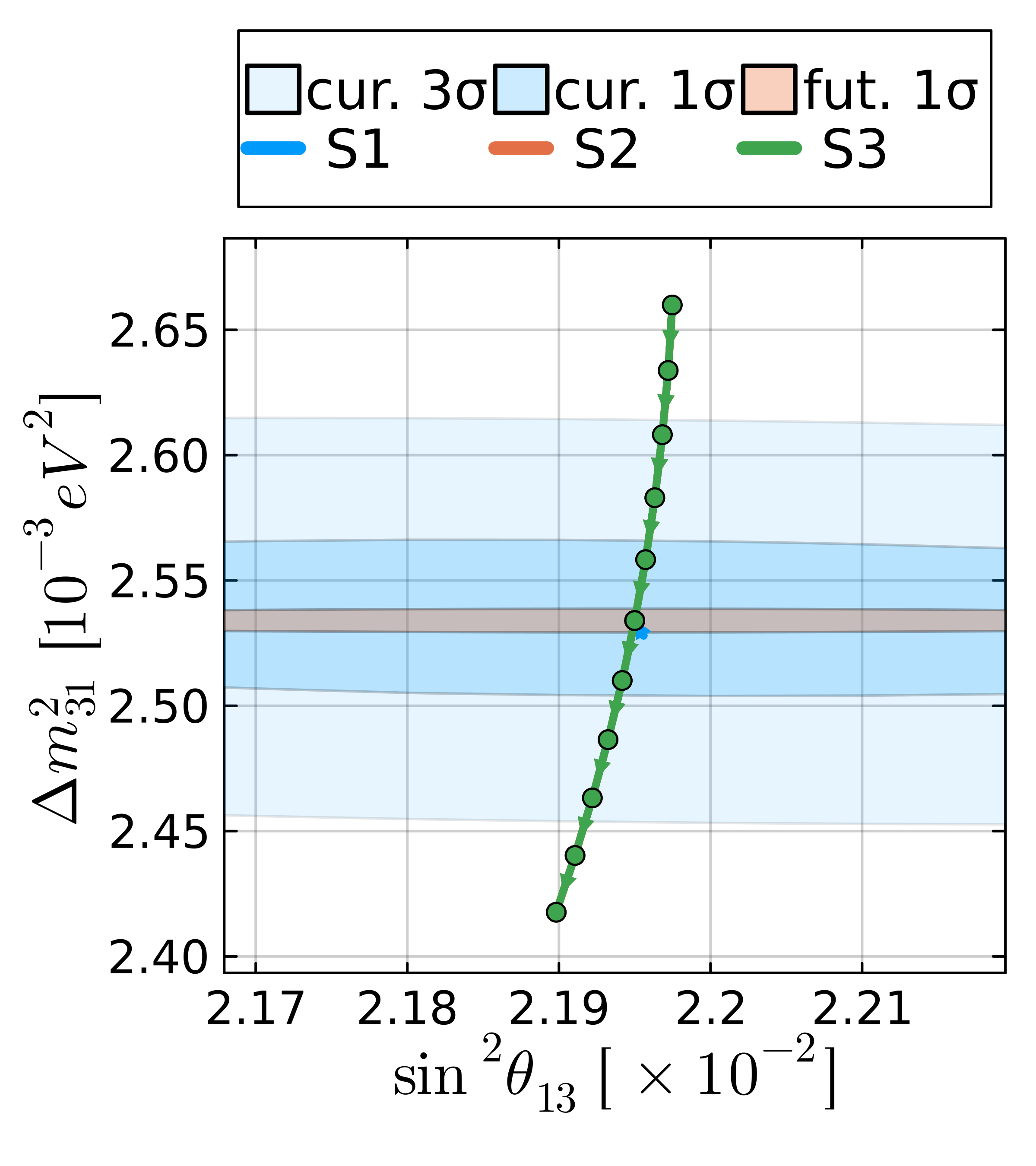}
        \caption{Atmospheric parameters.}
        \label{fig:paramStretchBLS}
    \end{subfigure}
    
    \caption{
    Sensitivity of the neutrino mixing parameters to the UV model parameters of the low scale benchmark set with $m_h=100$ TeV.
    Details of the plots are the same as in \cref{fig:paramStretchA} with, in this instance, $0.95\leq\gamma\leq1.05$ increasing in increments of $0.01$.
    The CP violating phase $\delta$ is experimentally measured as $\cos\delta$ and should be interpreted as cyclic over the range $0$ to $360^{\rm o}$.
    \Cref{fig:paramStretchDLS} extends this range to best show the continuous motion of $\delta$ under the scaling of the UV parameters.}
    \label{fig:paramStretchLS}
\end{figure}

We show the equivalent plots for the case of the low scale benchmark values in \cref{fig:paramStretchLS}. In this instance, $\gamma$ scales the parameters by $\pm5\%$. The sensitivity is slightly weakened due to $m_h=100$ TeV being closer to the experimental scale. However, the sensitivity is still significant, and we observe a qualitatively similar result to the regular benchmark cases, albeit over a slightly enlarged scaling by $\gamma$. This again demonstrates that the qualitative behaviour of the running is mostly independent of a shift in the energy scale.

One can consider the plotted lines of varying $\gamma$ as representing a subspace of high-scale model parameters that the calculation in the full theory does not constrain.
Each line in the plots of \cref{fig:paramStretchA,fig:paramStretchLS} represents a variance in the UV parameters by $\approx \pm 1\%$ in the regular benchmark cases, or $\pm 5\%$ in the low scale case (precisely, this depends on the choice of scaling, e.g. for S1 (blue line) $f$ is varied by $\sim\pm 5\%$ but the Yukawa couplings are scaled by $\sim \pm \sqrt{5\%}$).
We then see from the plots that this results in a displacement from the central value in the direction of the mass scale by up to $\sim 10\%$ in both cases. 
Most importantly, as seen in both \cref{fig:paramStretchA,fig:paramStretchALS}, even this $1(5)\%$ deviation results in corrections to the neutrino mass parameters that exceed future and even current experimental precisions.
A similar story applies to the other mixing parameters; their corresponding plots for the regular benchmark cases are shown in \cref{app:sensitivity}.
We note here the exception of the S1 scaling, particularly in the second but also in the third benchmark set, \cref{fig:paramStretchAB2,fig:paramStretchAB3}, which does not meaningfully extend from the central value. We have discussed this point in previous paragraphs. 

These results demonstrate that the EFT calculation can constrain the subspace of UV parameters not constrained by the full theory to the level of the measurement, or better in cases where the variance in the measured parameters is larger than that in the UV parameters.
Such a high sensitivity to the running can be expected, as we have previously seen in \cref{sec:theory} that the neutrino mass matrix is generated entirely by the running in the EFT calculation.
This also applies more broadly across both the alternate hierarchy within the Zee model, discussed in \cref{sec:matching}, and other radiative neutrino mass models where the logarithm of scales in the neutrino mass matrix in the full theory is generated by running effects in the EFT calculation. 
Similar large running effects have also been identified in the scotogenic model~\cite{Merle:2015ica}.
It is clearly imperative to consider the running effects in such cases.

\section{Summary and outlook}
\label{sec:summ}
We considered the Zee model with a hierarchical scalar mass spectrum, where the singly-charged scalar is heavier than the electroweak doublet scalars. Within this framework, a sequence of effective field theories enables precise predictions of the neutrino mass parameters that avoid the potentially large logarithm term in the calculation of the neutrino mass matrix in the full theory. 
We calculated the 1-loop matching of the Zee model onto the general 2HDM EFT and the 1-loop matching of the general 2HDM EFT onto the SMEFT, which we combined with the existing 1-loop RG equations~\cite{Grimus:2004yh,Li:2016} in the general 2HDM to demonstrate an EFT treatment.

By carefully analysing the different contributions to neutrino masses, discussing their relative importance, and comparing them to the direct calculation in the Zee model, we demonstrated the importance of renormalisation group corrections (and of carrying out a proper EFT treatment) in the Zee model, a typical radiative mass model. 
This is particularly important to achieve a theoretical uncertainty at the same level as the experimental sensitivity. A 1\% correction to parameter combinations in the fitted model, which leave the full theory prediction of the neutrino mass parameters invariant, can lead to corrections that exceed the uncertainties of mass squared differences and mixing angles at the next generation of neutrino experiments. RG corrections are particularly important for sizeable quartic scalar coupling $\lambda_6$, sizeable Yukawa couplings of the second Higgs doublet $Y_{e,u,d}^2$, as well as a large hierarchy in the scalar mass spectrum. Hence, it will be important to include RG corrections in phenomenological studies to fully exploit the information gained from the next-generation neutrino oscillation experiments, which motivates carrying out a new phenomenological study of the Zee model.  

We expect similarly-sized renormalisation group corrections in other radiative neutrino mass models. Although the exact mechanisms that contribute to the behaviour of the running will differ across models, the running effects commonly enter at the same order as the radiative masses are generated (1-loop).
The running effects, therefore, contribute significantly to the neutrino mass parameters.

With the increased precision of JUNO, Hyper-Kamiokande, and DUNE, statistical surveys of the parameter space of new physics models should consider an EFT framework to match the precision of these experiments. 

\section*{Acknowledgements}
J.V. is supported by an Australian Government Research Training Program (RTP) Scholarship.

\appendix
\section*{Appendix}

\section{Matching}
\label{app:matching}
We list here the matching conditions between the Zee model and the 2HDM EFT, as well as those between the 2HDM EFT and the SMEFT.

\subsection{Zee Model to 2HDM EFT}
As we define our initial quartic and Yukawa couplings at $\bar{\mu}=m_h$ in the 2HDM EFT, the only matching condition that we use between the Zee model and the 2HDM EFT is that for the Weinberg like coefficients $\kappa^{ij}$,
\begin{align}
        \kappa^{ij}(\bar{\mu}) &= -\frac{\mu_{\rm Zee}}{m_{h}^2}\epsilon_{ij}f -\frac{1}{16\pi^{2}}\frac{\mu_{\rm Zee}}{m_{h}^2}\left(1+\log\left[\frac{\bar{\mu}^2}{m_h^2}\right]\right)\times \nonumber\\
                    &\phantom{=}\, \left[\delta_{ik}\epsilon_{lj}f{Y_e^k}^{\dagger}Y_e^l + \delta_{lj}\epsilon_{ik}{Y_e^k}^{\rm T}{Y_e^l}^{*}f - \epsilon_{ij}f(\lambda_8+\lambda_9-4\lambda_h)\right] \ .\label{eq:appMatchK}
\end{align}
Strictly, the Yukawa and quartic couplings appearing in \cref{eq:appMatchK} refer to the respective couplings in the Zee model. As discussed in \cref{sec:matching}, we assume them to be equivalent to the couplings defined in the 2HDM EFT at the scale $\bar{\mu}=m_h$ in our numeric computations.

Although we do not use the matching for any of the quartic couplings between these theories in the computation, we include here for reference the matching conditions of $\lambda_3$ and $\lambda_4$;
\begin{align}
    \lambda_3 &= \lambda_3' - \frac{\mu_{\rm Zee}^2}{m_h^2} - \frac{1}{16\pi^2} \left[
    \lambda_8'\lambda_9' \log\left(\frac{\bar{\mu}^2}{m_h^2}\right)
    + \frac{\mu_{\rm Zee}^2}{m_h^2}\bigg[-\lambda_8'-\lambda_9' \right. \nonumber\\
    & \left.+ 2(\lambda_1'+\lambda_2' +\lambda_3'-\lambda_8'-\lambda_9'+2\lambda_h')\left(1+\log\left(\frac{\bar{\mu}^2}{m_h^2}\right)\right) \bigg]
    + \frac{\mu_{\rm Zee}^4}{m_h^4}\left(2+\log\left(\frac{\bar{\mu}^2}{m_h^2}\right)\right)
    \right] \ , \\
    \lambda_4 &= \lambda_4' + \frac{\mu_{\rm Zee}^2}{m_h^2} + \frac{1}{16\pi^2}|\lambda_{10}'|^2\log\left(\frac{\bar{\mu}^2}{m_h^2}\right)\nonumber\\
    &- \frac{2}{16\pi^2}\frac{\mu_{\rm Zee}^2}{m_h^2}\left(\lambda_3'+\lambda_4'-\lambda_8'-\lambda_9'+2\lambda_h\right)\left(1+\log\left(\frac{\bar{\mu}^2}{m_h^2}\right)\right)\ .
\end{align}
In the above, $\lambda_i'$ should be understood as the quartic couplings in the Zee model, while $\lambda_3$ and $\lambda_4$ are in the 2HDM EFT.

\subsection{2HDM EFT to SMEFT}
We then have the matching of the 2HDM to the SMEFT, at 1-loop and up to operators of dimension $5$. 
In the following, we denote the SMEFT Higgs field as $H$ and refer to the Higgs fields in the 2HDM EFT as $H_1$ and $H_2$, as in the main text.
The matching conditions are:
\begin{itemize}
\item Weinberg coefficient ($\mathcal{L}_{\rm SMEFT} \supset C_{\alpha\beta}(\bar{\tilde{L}}_{\alpha}H)(\tilde{H}^{\dagger}L_{\beta})$), 
\begin{align}
    C &= \kappa^{11} - \frac{1}{16\pi^{2}}\kappa^{22} \lambda_5 \log\left(\frac{\bar{\mu}^2}{m_{H_2}^2}\right)\nonumber\\
    &\phantom{=}\ - \frac{1}{16\pi^{2}}\left(1+\log\left(\frac{\bar{\mu}^2}{m_{H_2}^2}\right)\right)\bigg[(\lambda_{6}^{*} - 3\lambda_{7}^{*})(\kappa^{12} + \kappa^{21}) 
    - \kappa^{12}{Y_e^1}^{\dagger} {Y_e}^{2} - {{Y_e}^{2}}^{\rm T} {{Y_e}^{1}}^*\kappa^{21} \bigg]\nonumber\\
    &\phantom{=}\  -\frac{1}{16\pi^{2}}\frac{1}{8}\left(1+2\log\left(\frac{\bar{\mu}^{2}}{m_{H_2}^{2}}\right)\right)
    \bigg[\kappa^{11}{Y_e^{2}}^\dagger Y_e^{2}
    + {Y_e^{2}}^{\rm T}{Y_e^{2}}^*\kappa^{11}\bigg] \ . \label{eq:appMatchC}
\end{align}    

\item  Quartic coupling ($\mathcal{L}_{\rm SMEFT} \supset - \lambda (H^{\dagger}H)^{2}$),
    \begin{align}
        \lambda &= \frac{1}{2}\lambda_{1} - \left(\lambda_{3}^{2} + \lambda_{3}\lambda_{4} + \frac{1}{2}\lambda_{4}^{2} + \frac{1}{2}\lambda_{5}^{*}\lambda_{5}\right)\frac{1}{16\pi^{2}}\log\left(\frac{\bar{\mu}^{2}}{m_{H_2}^{2}}\right)\nonumber\\
                &\phantom{=}\qquad \qquad + \left(3\lambda_{6}\lambda_{7}^{*} -6\lambda_{6}\lambda_{6}^{*} + 3\lambda_{6}^{*}\lambda_{7}\right)\frac{1}{16\pi^{2}}\left[1+\log\left(\frac{\bar{\mu}^{2}}{m_{H_2}^{2}}\right)\right] \ . \label{eq:appMatchL}
    \end{align}

    \item Yukawa couplings ($\mathcal{L}_{\rm SMEFT} \supset \bar{L}Y^{\dagger}_{e}H{e_{R}}$),
    \begin{align}
        Y_{e} &= Y^{1}_{e} + \frac{1}{16\pi^{2}}
        \bigg(3Y_{e}^{2}\lambda_{7}\left[1 + \log\left(\frac{\bar{\mu}^{2}}{m_{H_2}^{2}}\right)\right]\nonumber\\
              &\qquad - \frac{1}{8}\left(2Y_{e}^{2}{Y_{e}^{2}}^{\dagger}Y_{e}^{1} + Y_{e}^{1}{Y_{e}^{2}}^{\dagger}Y_{e}^{2}\right)\left[1 + 2\log\left(\frac{\bar{\mu}^{2}}{m_{H_2}^{2}}\right)\right]\bigg) \ , \label{eq:appMatchYe} \\
        Y_{u} &= Y^{1}_{u} + \frac{1}{16\pi^{2}}
        \left(\left(3Y_u^2\lambda_7^* + Y_u^2{Y_d^1}^\dagger Y_d^2\right)\left[1+\log\left(\frac{\bar{\mu}^2}{m_{H_2}^2}\right)\right] \right.\nonumber\\
        &\phantom{=}\left.- \frac{1}{8}\left(Y_u^1{Y_d^2}^\dagger Y_d^2 + Y_u^{1}{Y_u^2}^\dagger Y_u^2 + 2Y_u^2{Y_u^2}^\dagger Y_u^1\right)\left[1+2\log\left(\frac{\bar{\mu}^2}{m_{H_2}^2}\right)\right]
        \right) \ , \label{eq:appMatchYu} \\
        Y_{d} &= Y^{1}_{d} + \frac{1}{16\pi^{2}}
        \left(\left(3Y_d^2\lambda_7 + Y_d^2{Y_u^1}^\dagger Y_u^2\right)\left[1+\log\left(\frac{\bar{\mu}^2}{m_{H_2}^2}\right)\right] \right.\nonumber\\
        &\phantom{=}\left.- \frac{1}{8}\left(Y_d^1{Y_u^2}^\dagger Y_u^2 + Y_d^{1}{Y_d^2}^\dagger Y_d^2 + 2Y_d^2{Y_d^2}^\dagger Y_d^1\right)\left[1+2\log\left(\frac{\bar{\mu}^2}{m_{H_2}^2}\right)\right]
        \right) \ . \label{eq:appMatchYd}
    \end{align}

\item Gauge couplings ($g_2$ and $g_1$) 
        \begin{align}
            g^{2}_{2 \ \rm SMEFT} &= \frac{g_2^{2}}{1 + \frac{1}{6}\frac{g_2^2}{16\pi^{2}}\log\left(\frac{\bar{\mu}^{2}}{m_{H_2}^{2}}\right)}
            \qquad \to \qquad 
            g_2^{2} = \frac{g_{2 \ \rm SMEFT}^{2}}{1 - \frac{1}{6}\frac{g_{2 \ \rm SMEFT}^{2}}{16\pi^{2}}\log\left(\frac{\bar{\mu}^{2}}{m_{H_2}^{2}}\right)} \\
            g^{2}_{1 \ \rm SMEFT} &= \frac{g_1^{2}}{1 + \frac{1}{6}\frac{g_1^2}{16\pi^{2}}\log\left(\frac{\bar{\mu}^{2}}{m_{H_2}^{2}}\right)}
            \qquad \to \qquad 
            g_1^{2} = \frac{g_{1\ \rm SMEFT}^{2}}{1 - \frac{1}{6}\frac{g_{1\ \rm SMEFT}^{2}}{16\pi^{2}}\log\left(\frac{\bar{\mu}^{2}}{m_{H_2}^{2}}\right)}
        \end{align}

\item Higgs quadratic ($\mathcal{L}_{\rm SMEFT} \supset -\mu^{2}H^{\dagger}H$)
    \begin{align}
    \mu^{2} &= \mu_1^2 - m_{H_2}^2(2\lambda_3 + \lambda_4)\frac{1}{16\pi^{2}} \left(1+\log\left(\frac{\bar{\mu}^2}{m_{H_2}^2}\right)\right) \label{eq:appMatchSMEFTmu}
    \end{align}
\end{itemize}
Of note in this calculation is the need for gauge field renormalisation. After using equations of motion to simplify the result in \texttt{Matchete} (using the routine EOMSimplify) \cite{Fuentes-Martin:2022jrf}, we are left with the following terms in the EFT (SMEFT) Lagrangian:
\begin{align}
        \mathcal{L}_{\rm SMEFT} \supset -\frac{1}{4}(G^{\mu\nu A})^2
    -\frac{1}{4}\left(1 + \frac{1}{6}\frac{g_2^2}{16\pi^{2}}\log\left(\frac{\bar{\mu}^2}{m_{H_2}^{2}}\right)\right)(W^{\mu\nu I})^2\nonumber\\
    -\frac{1}{4} \left(1+ \frac{1}{6}\frac{g_1^2}{16\pi^{2}}\log\left(\frac{\bar{\mu}^2}{m_{H_2}^{2}}\right)\right)(B^{\mu\nu})^2 \ .
\end{align}
This normalisation can be interpreted as a redefinition of the gauge coupling. We relabel
\begin{align}
        W^{\mu}_{a} \to \frac{W^{\mu}_{a}}{g} \text{ and } D^{\mu} \to \partial^{\mu} + \frac{i}{2}W_{b}^{\mu}\tau_{b} + \ldots\ ,
\end{align}
meaning we have
\begin{align}
        \mathcal{L}_{\rm SMEFT} \supset -\frac{1}{4}\left(1 + \frac{1}{6}\frac{g_2^2}{16\pi^{2}}\log\left(\frac{\bar{\mu}^{2}}{m_{H_2}^{2}}\right)\right)\frac{(W_{a}^{\mu\nu I})^2}{g_2^{2}} 
        =
        -\frac{1}{4}\frac{(W_{a}^{\mu\nu I})^2}{g_{2\ \rm SMEFT}^{2}} \ ,
\end{align}
where
\begin{align}
        g^{2}_{2\ \rm SMEFT} &= \frac{g_2^{2}}{1 + \frac{1}{6}\frac{g_2^2}{16\pi^{2}}\log\left(\frac{\bar{\mu}^{2}}{m_{H_2}^{2}}\right)} \ .
\end{align}
Then we make a second relabelling
\begin{align}
        \frac{W_{a}^{\mu}}{g_{2\ \rm SMEFT}} \to W_{a}^{\mu} \ ,
\end{align}
such that
\begin{align}
        \mathcal{L}_{\rm SMEFT} \supset -\frac{1}{4}\frac{(W_{a}^{\mu\nu I})^2}{g_{2\ \rm SMEFT}^{2}} \to 
        -\frac{1}{4}(W_{a}^{\mu\nu I})^2 \ .
\end{align}
This amounts to the renormalisation
\begin{align}
        {W^{\mu}_{a}}_{\rm SMEFT} &= \frac{g_{2\ \rm SMEFT}}{g_2}W^{\mu}_{a} \ .
\end{align}
A similar rescaling is required for the hypercharge field $B_\mu$. This renormalisation of the gauge couplings is used for their matching condition between the 2HDM EFT and the SMEFT.

\section{Renormalisation group equations}
\label{app:betas}
We calculated all $\beta$-functions using the \texttt{RGBeta}~\cite{Thomsen:2021ncy} package for Mathematica.
For those in the 2HDM EFT, we verified our calculated $\beta$-functions with the calculations for the n-HDM presented in \cite{Grimus:2004yh} and the calculation of the $\beta$-function for $\kappa^{ij}$ in \cite{Li:2016}.
The $\beta$-functions of the SMEFT parameters are well known across the literature and have been taken from~\cite{Antusch:2005gp}.
Recall $t=\ln{\bar{\mu}}$.

To keep the notation compact in the following, we define a new indexing of the quartic couplings, as used in~\cite{Davidson:2005cw}
\begin{equation} 
\begin{aligned}
\lambda_{1111} &= \frac{\lambda_1}{2}\,, & 
\lambda_{1122}  & = \lambda_{2211} = \frac{\lambda_3}{2}\,, & 
\lambda_{1112} &= \lambda_{1211} = \lambda_{1121}^* = \lambda_{2111}^* = \frac{\lambda_6}{2}\,,
&
\lambda_{1212} &= \lambda_{2121}^* = \frac{\lambda_5}{2}\,, 
\\
\lambda_{2222} &= \frac{\lambda_2}{2}\,, &
\lambda_{1221}  &= \lambda_{2112} = \frac{\lambda_4}{2}\,, & 
\lambda_{2212} &= \lambda_{1222} = \lambda_{2221}^* = \lambda_{2122}^* = \frac{\lambda_7}{2} \,. 
\end{aligned}
\end{equation}

\subsection{2HDM EFT}
All of the renormalisation group $\beta$-functions for the parameters of the 2HDM EFT relevant in our calculations are listed below.
\begin{itemize}
\item Weinberg-like $\kappa^{ij}$ couplings,
\begin{align}
        16\pi^{2}\frac{d}{dt}\kappa^{ij} &=
-3g^2(2\kappa^{ij}-\kappa^{ji}) + 4\sum_{k,l=1}^2\kappa^{kl}\lambda_{kilj}\nonumber\\
                               &\phantom{=}\ + 2\sum_{k=1}^2\bigg[\kappa^{kj}{Y_e^{i\dagger}}{Y_e}^k 
                                       - \left(\kappa^{ik} + \kappa^{ki}\right){Y_e^{j\dagger}}Y_e^{k} \nonumber\\
&\phantom{=}\qquad\qquad\qquad
+ {Y_e^k}^{\rm T}{Y_e^j}^*\kappa^{ik} 
- {Y_e^k}^{\rm T}{Y_e^i}^*(\kappa^{kj} + \kappa^{jk})\bigg]\nonumber\\
&\phantom{=}\ + \sum_{k=1}^2[T_{ki}\kappa^{kj} + T_{kj}\kappa^{ik}] + \kappa^{ij}S + (S^T)\kappa^{ij} \ , \label{eq:appBetaK2HDM}\\
        \text{where, }\ S &= \frac{1}{2}\sum_{k=1}^2{Y_e^k}^{\dagger} {Y_e^k} \ \text{ and, } \ T^{ij} = \text{Tr}\left(Y_e^{i\dagger} Y_e^{j} + 3Y_u^{i\dagger} Y_u^{j} + 3Y_d^{i\dagger} Y_d^{j}\right) \ .
\end{align}

\item Quartic couplings,
\begin{align}
16\pi^{2}\frac{d}{dt}\lambda_{ijkl} &=
4\sum_{m,n=1}^2\left(2\lambda_{ijmn}\lambda_{nmkl} + \lambda_{ijmn}\lambda_{kmnl} + \lambda_{imnj}\lambda_{mnkl} + \lambda_{imkn}\lambda_{mjnl} + \lambda_{mjkn}\lambda_{imnl}\right)\nonumber\\
&\phantom{=}\ - (9g_2^2 + 3g_1^2)\lambda_{ijkl} + \frac{9g_2^4 + 3g_1^4}{8}\delta_{ij}\delta_{kl} + \frac{3g_2^2g_1^2}{4}(2\delta_{il}\delta_{kj} - \delta_{ij}\delta_{kl})\nonumber\\
&\phantom{=}\ + \sum_{m=1}^2(T_{mj}\lambda_{imkl} + T_{ml}\lambda_{ijkm} + T_{im}\lambda_{mjkl} + T_{km}\lambda_{ijml}) -2 \ \text{tr}(Y_iY_j^{\dagger} Y_kY_l^{\dagger}) \; .\label{eq:appBetaL2HDM}
\end{align}

\item Yukawa couplings,
\begin{align}
16\pi^{2}\frac{d}{dt}Y_e^{i} &= \sum_{k=1}^{2}\left( T_{ik}Y_e^{k}
    + Y_e^{k} Y_e^{k\dagger} Y_e^{i}  + \frac{1}{2}Y_e^{i}Y_e^{k\dagger} Y_e^{k} \right) - \frac{9g_2^2 + 15g_1^2}{4}Y_e^{i} \ , \label{eq:appBetaYe2HDM} \\
16\pi^2\frac{d}{dt}Y^i_u &= -\frac{17}{12}g_1^2Y^i_u - \frac{9}{4}g_2^2Y^i_u -8g_3^2Y_u^i \nonumber\\
    &\phantom{=} + \sum_{k=1}^2 \left(T^{ik}Y^k_u + Y^k_uY_d^{k\dagger} Y^i_d + \frac{1}{2}Y^i_uY_d^{k\dagger} Y^k_d
    + Y^k_uY_u^{k\dagger} Y^i_u + \frac{1}{2}Y^i_uY_u^{k\dagger} Y^k_u\right) \ ,\label{eq:appBetaYu2HDM} \\
16\pi^2\frac{d}{dt}Y^i_d &= -\frac{5}{12}g_1^2Y^i_d - \frac{9}{4}g_2^2Y^i_d -8g_3^2Y_d^i \nonumber\\
    &\phantom{=} + \sum_{k=1}^2 \left(T^{ik}Y_d^k + Y^k_dY_u^{k\dagger} Y^i_u + \frac{1}{2}Y^i_dY_u^{k\dagger} Y^k_u
    + Y^k_dY_d^{k\dagger} Y^i_d + \frac{1}{2}Y^i_dY_d^{k\dagger} Y^k_d\right) \ . \label{eq:appBetaYd2HDM}
\end{align}

\item Gauge couplings,
\begin{align}
\frac{d}{dt}g_1 = \frac{1}{16\pi^{2}}7g_1^3 \qquad
&\implies \qquad g_1(\mu) = \left(\frac{g_1(\mu_0)^2}{1 - \frac{14g_1(\mu_0)^2}{16\pi^2}\log\left(\frac{\mu}{\mu_0}\right)}\right)^{1/2} \ , \label{eq:appBetag12HDM}
\end{align}

\begin{align}
\frac{d}{dt}g_2 = -\frac{1}{16\pi^{2}}3g_2^3 \qquad 
&\implies \qquad g_2(\mu) = \left(\frac{g_2(\mu_0)^2}{1+\frac{6g_2(\mu_0)^2}{16\pi^2}\log\left(\frac{\mu}{\mu_0}\right)}\right)^{1/2} \ , \label{eq:appBetag22HDM}
\end{align}

\begin{align}
\frac{d}{dt}g_3 = -\frac{1}{16\pi^{2}}7g_3^3 \qquad 
&\implies \qquad g_3(\mu) = \left(\frac{g_3(\mu_{0})^2}{1 + \frac{14g_3(\mu_{0})^2}{16\pi^2}\log\left(\frac{\mu}{\mu_0}\right)}\right)^{1/2} \ . \label{eq:appBetag32HDM}
\end{align}

\item Quadratic couplings (with $\mu_{1,2,3}$ defined in the 2HDM EFT as per \cref{eq:ZeePot}),
    \begin{align}
        16\pi^{2}\frac{d}{dt}(\mu_{1}^{2}) &= \left(-\frac{3}{2}g_{1}^{2} - \frac{9}{2}g_{2}^{2} + 6\lambda_{1} + 2T_{11}\right) \mu_{1}^{2} + (4\lambda_{3} + 2\lambda_{4})\mu_{2}^{2}\nonumber\\ 
                                &\phantom{=} - \left(\left(6\lambda_{6}^{*} + T_{21}\right)\mu_{3}^{2} + {\rm h.c.}\right) \ , \\
        16\pi^{2}\frac{d}{dt}(\mu_{2}^{2}) &= \left(-\frac{3}{2}g_{1}^{2} - \frac{9}{2}g_{2}^{2} + 6\lambda_{2} + 2T_{22}\right) \mu_{2}^{2} + (4\lambda_{3} + 2\lambda_{4})\mu_{1}^{2}\nonumber\\ 
                                &\phantom{=} - \left(\left(6\lambda_{7}^{*} + T_{12}\right)\mu_{3}^{2} + {\rm h.c.}\right) \ ,\\
        16\pi^{2}\frac{d}{dt}(\mu_{3}^{2}) &= \left(-\frac{3}{2}g_{1}^{2} - \frac{9}{2}g_{2}^{2} + 2\lambda_{3} + 4\lambda_{4} + T_{11} + T_{22}\right) \mu_{3}^{2}\nonumber\\ 
                                &\phantom{=} + 6\lambda_{5}(\mu_{3}^{*})^{2}  + \left(6\lambda_{6} + T_{12}\right)\mu_{1}^{2} + \left(6\lambda_{7} + T_{12}\right)\mu_{2}^{2} \ .
    \end{align}
    We only use these $\beta$-functions for the analysis in \cref{sec:constraints}, they are not used in the computation. See the discussion in \cref{sec:ZeeModelScales} for the connection of these quadratic couplings to the physical mass states in the 2HDM and Zee models.
\end{itemize}

\subsection{SMEFT}
All of the $\beta$-functions for the parameters of the SMEFT relevant to our calculations are listed below. For the running of the Weinberg coefficient, we use the result in \cite{Antusch:2001ck}.
\begin{itemize}
\item Weinberg coefficient,
\begin{align}
    16\pi^{2}\frac{d}{dt}C &= -\frac{3}{2}\left[C(Y_{e}^{\dagger}Y_{e}) + (Y_{e}^{\dagger}Y_{e})^{T}C\right] + 4\lambda C - 3g_{2}^{2}C + 2TC \label{eq:appBetaCSMEFT} \ ,
\end{align}
    where $T \equiv \Tr(3Y_{u}^{\dagger}Y_{u} + 3Y_{d}^{\dagger}Y_{d} + Y_{e}^{\dagger}Y_{e})$ .
\item Quartic coupling,
\begin{align}
        16\pi^{2}\frac{d}{dt}\lambda &= \frac{3}{8}g_{1}^{4} + \frac{3}{4}g_{1}^{2}g_{2}^{2} + \frac{9}{8}g_{2}^{4} + (4T - 3g_{1}^{2} - 9g_{2}^{2})\lambda  + 24\lambda^{2}\nonumber\\
                                 &\phantom{=}\ - 2\Tr(Y_{e}Y_{e}^{\dagger}Y_{e}Y_{e}^{\dagger}) - 6\Tr(Y_{u}Y_{u}^{\dagger}Y_{u}Y_{u}^{\dagger}) - 6\Tr(Y_{d}Y_{d}^{\dagger}Y_{d}Y_{d}^{\dagger}) 
                                 \nonumber\\
                                 &\phantom{=}\ - 12\Tr(Y_{d}Y_{d}^{\dagger}Y_{u}Y_{u}^{\dagger}) \ .\label{eq:appBetaLSMEFT}
\end{align}

\item Yukawa couplings,
\begin{align}
        16\pi^{2}\frac{d}{dt}Y_{e} &= -\frac{15}{4}g_{1}^{2}Y_{e} - \frac{9}{4}g_{2}^{2}Y_{e} + TY_{e} + \frac{3}{2}Y_{e}Y_{e}^{\dagger}Y_{e} \ , \label{eq:appBetaYeSMEFT} \\
    16\pi^2 \frac{d}{dt}Y_u &= -\frac{17}{12}g_1^2Y_u - \frac{9}{4}g_2^2Y_u - 8g_3^2Y_u + TY_u + \frac{3}{2}Y_uY_d^\dagger Y_d + \frac{3}{2}Y_u Y_u^\dagger Y_u \ , \label{eq:appBetaYuSMEFT} \\
    16\pi^2 \frac{d}{dt}Y_d &= -\frac{5}{12}g_1^2Y_d - \frac{9}{4}g_2^2Y_d - 8g_3^2Y_d + TY_d + \frac{3}{2}Y_dY_u^\dagger Y_u + \frac{3}{2}Y_d Y_d^\dagger Y_d \ . \label{eq:appBetaYdSMEFT}
\end{align}

\item Gauge couplings,
 \begin{align}
        16\pi^{2}\frac{d}{dt}g_{i} = - b_i g_{i}^{3}
        \qquad \implies \qquad
        g_{i}(\mu) = \left(\frac{g_{i}(\mu_{0})^{2}}{1 + \frac{b}{16\pi^{2}} g_{i}(\mu_{0})^{2}\ln\left(\frac{\mu}{\mu_{0}}\right)}\right)^{1/2} \label{eq:appBetagSMEFT}
    \end{align}
    for $(b_1,b_2,b_3) = (-\frac{41}{6},\frac{19}{6},7)$.
    \item Quadratic coupling, 
    \begin{align}
        16\pi^{2}\frac{d}{dt}(\mu^{2}) &= \left(-\frac{3}{2}g_{1}^{2} - \frac{9}{2}g_{2}^{2} + 6\lambda + 2T\right) \mu^{2} \;.
    \end{align}
 
\end{itemize}

\section{Constraints on parameter space for neutrino mass running}
\label{app:pheno}
In this appendix, we outline the calculation of the branching ratios of cLFV processes that we use to estimate bounds on the values of second Yukawa couplings $Y_e^2$ in particular.

\begin{figure}[b!]
    \centering
    \begin{subfigure}[c]{0.32\textwidth}
        \centering
        \includegraphics[width=\textwidth]{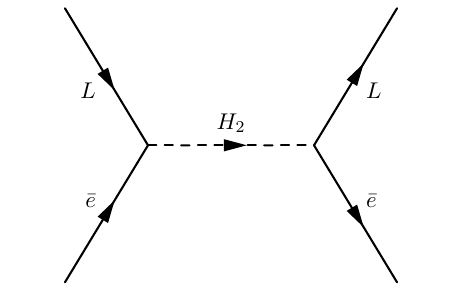}
        \caption{}
        \label{fig:feyncLFVa}
    \end{subfigure}
    \begin{subfigure}[c]{0.32\textwidth}
        \centering
        \includegraphics[width=\textwidth]{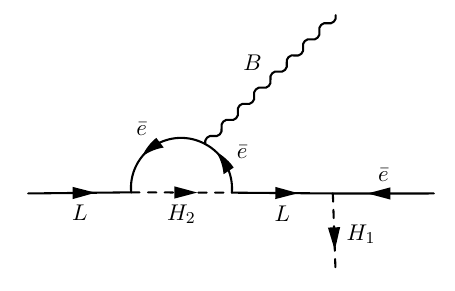}
        \caption{}
        \label{fig:feyncLFVb}
    \end{subfigure}
    \hfill
    \begin{subfigure}[c]{0.32\textwidth}
        \centering
        \includegraphics[width=\textwidth]{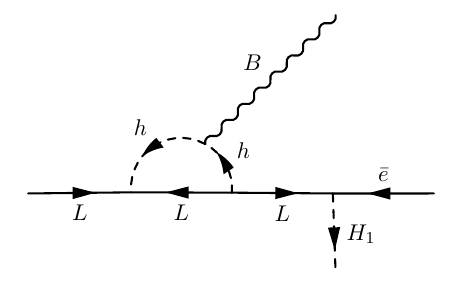}
        \caption{}
        \label{fig:feyncLFVc}
    \end{subfigure}
    \caption{Representative contributions to cLFV processes from the high scale Higgs fields. }
    \label{fig:feyncLFV}
\end{figure}

Recall from \cref{sec:leptonsector} that in the Higgs basis, $Y_e^2$ is a general complex matrix. 
Subsequently, cLFV processes in the SMEFT can be induced in the Zee model at both tree and loop level through the off-diagonal components of $Y_e^2$, as depicted in \cref{fig:feyncLFVa,fig:feyncLFVb}. 
In principle, there is also a tree-level contribution with the charged singlet scalar $h$ propagator. 
However, because $h$ is charged, the structure of the coupling $h\bar{\tilde{L}} f L$ necessitates that $f$ always couples a charged lepton to a neutrino.
The resulting 4-lepton cLFV processes will be those that include two neutrinos, which have experimental bounds that are far less constraining and can be disregarded here.
The charged singlet will however result in a contribution to the loop-level cLFV radiative decays via the Feynman diagram depicted in \cref{fig:feyncLFVc} with a neutrino in the loop.
cLFV processes are generally well constrained experimentally and allow us to estimate bounds on the scale of the elements of $Y_e^2$.

In line with our EFT approach, we match these processes in the 2HDM EFT to their respective cLFV effective operators in the SMEFT using \texttt{Matchete}~\cite{Fuentes-Martin:2022jrf}.
The tree-level diagram in \cref{fig:feyncLFV} generates the effective four-lepton SMEFT operator
\begin{align}
        \mathcal{L}_{\rm SMEFT} \supset C^{(Le)}_{ijkl} (\bar{L}_i\gamma^{\mu}L_j)(\bar{e}_k\gamma^{\mu}e_l)
\end{align}
with dimensionful Wilson coefficient
\begin{align}
        C_{ijkl}^{(Le)} &= -\frac{1}{2m_{H_2}^2}({Y_e^2}^\dagger)_{li}{(Y_e^2)}_{kj} \ . \label{eq:3lWC}
\end{align}
Using the formulas collected in \cite{Calibbi:2021pyh}, we calculate the leptonic 3-body decay branching ratio
\begin{align}
    {\rm BR}(\ell_i\to\ell_j\ell_k\bar{\ell}_l)
    &= \frac{2m_{\ell_i}^5}{3(16\pi)^3\Gamma_{\ell_i}m_{H_2}^4}|(Y_e^2)_{jl} (Y_e^2)_{ki}|^2 \;,
    \label{eq:BR3l}
\end{align}
where $\Gamma_{\ell_i}$ is the total decay width of the decaying lepton. 
The strongest bound for leptonic 3-body decays is that on the branching ratio ${\rm BR}(\mu\to ee\bar{e}) \leq 1.0\times10^{-12}$ \cite{SINDRUM:1987nra}.
The branching ratios of the $\tau\to 3\ell$ decays are approximately bound by $\text{BR}(\tau\to3\ell) \lesssim 2.0\times10^{-8}$ across all of the different flavour combinations \cite{Hayasaka:2010np,ParticleDataGroup:2020ssz}.
Evaluating \cref{eq:BR3l} in each case, we find the approximate bounds
\begin{align}
    |(Y_e^2)_{ee}(Y_e^2)_{e\mu}| &\lesssim 1.8 \times 10^{-4}
    \left(\frac{m_{H_2}}{\rm TeV}\right)^2\ , \label{eq:Yboundee}\\
    |(Y_e^2)_{ij}(Y_e^2)_{k\tau}| &\lesssim 0.06
    \left(\frac{m_{H_2}}{\rm TeV}\right)^2 \ \text{ where $i,j,k \in \{e,\mu\}$} \ . \label{eq:Yboundgen}
\end{align}
In the above, we use the particle decay widths $\Gamma_\tau \simeq 2.3\times10^{-3}$ eV and $\Gamma_\mu \simeq 3\times10^{-10}$ eV.
For the typical scale $m_{H_2}=10^2$ TeV used in this paper, \cref{eq:Yboundgen} only constrains the elements of $Y_e^2$ to generally be of the scale $|Y_e^2| \lesssim 20$ and \cref{eq:Yboundee} slightly strengthens the constraint for the elements $(Y_e^2)_{ee},(Y_e^2)_{e\mu} \lesssim 1$.

The diagram in \cref{fig:feyncLFVb} generates a dipole operator in the SMEFT.
Although generated at loop level, these radiative decays provide a competitive bound on the elements of $Y_e^2$ as the radiative cLFV processes are much more experimentally constrained than the 3-lepton decays.
After enacting the Higgs mechanism, the 2HDM EFT induces the SMEFT dipole operators
\begin{align}
        \mathcal{L}_{\rm SMEFT} &= \frac{v}{\sqrt{2}}C_{e\gamma}^{ij}\left(\overline{L_{i}}\sigma^{\mu\nu}e_{j}\right) F_{\mu\nu} \ .
\end{align}
with Wilson coefficients
\begin{align}
    C_{e\gamma}^{ij} &= \frac{e}{24m_{H_2}^2}\frac{1}{16\pi^2}\left(2({Y_e^2}^\dagger Y_e^2{Y_e^1}^\dagger) + (Y_1^\dagger Y_e^2{Y_e^2}^\dagger) \right)_{ij}\\
    &= \frac{e}{24m_{H_2}^2}\frac{1}{16\pi^2}\frac{\sqrt{2}}{v}\bigg(2m_{\ell_j}({Y_e^2}^\dagger Y_e^2)_{ij}+ m_{\ell_i}({Y_e^2}{Y_e^2}^\dagger)_{ij}
    \bigg) \ ,
\end{align}
where the second equality is true in the Higgs basis of the full theory with $Y_e^{1} = (\sqrt{2}/v)\times{\rm diag}(m_{e},m_\mu,m_\tau)$\footnote{Here we are assuming all terms are at their low energy values, such that we can consider $Y_e^1$ diagonal without having considered running effects.}.
We then calculate the branching ratio of the cLFV decays in terms of the Wilson coefficient $C_{e\gamma}^{ij}$, again using the formulas collected in \cite{Calibbi:2021pyh},
\begin{align}
    {\rm BR}(\ell_i\to\ell_j\gamma) &= \frac{m_{\ell_i}^3v^2}{8\pi \Gamma_{\ell_i}}\left(|C_{e\gamma}^{ji}|^2 + |C_{e\gamma}^{ij*}|^2\right) \ . \label{eq:BRdipole}
\end{align}
Energy conservation ensures the lepton in the final state will always be a lighter flavour than the decaying lepton, and the difference in mass scales of the lepton flavours allows us to neglect the mass of the final lepton ($m_\tau/m_\mu \sim 17$, $m_\mu/m_e\sim 200$).
In this scenario, we calculate 
\begin{align}
    |C_{e\gamma}^{ij}|^2 &= -\frac{1}{576}\frac{e^2}{256\pi^2}\frac{2}{v^2}
    \left[4m_j^2|A_{ij}|^2 + m_i^2|A_{ij}|^2 + 4m_im_j\left({\rm Re}(A_{ij})^2 - {\rm Im}(A_{ij})^2\right)\right]\\
    &\simeq -\frac{1}{576}\frac{e^2}{256\pi^2}\frac{2}{v^2} 
    \times\begin{cases}
        m_i^2|A_{ij}|^2 & (m_{\ell_i}\gg m_{\ell_j})\\
        4m_j^2|A_{ij}|^2 & (m_{\ell_j}\gg m_{\ell_i})
    \end{cases}
    \label{eq:CegammaCalc}
\end{align}
where
\begin{align}
A_{ij} &= \sum_{k=e,\mu,\tau} (Y_e^2)^\dagger_{ik}(Y_e^2)_{kj}
\end{align}
and the expression for the branching ratio in \cref{eq:BRdipole} shows that we will require each scenario of \cref{eq:CegammaCalc}. 

The most constrained cLFV radiative decay is ${\rm BR}(\mu\to e\gamma) < 1.5\times10^{-13}$ \cite{MEGII:2025gzr} which, using \cref{eq:BRdipole,eq:CegammaCalc}, implies the bound
\begin{align}
\left(|A_{\mu e}|^2 + 4|A_{e\mu}|^2\right) \lesssim 6\times10^{-7}\left(\frac{m_{H_2}}{\rm TeV}\right)^4 \ ,
\end{align}
where we use $e=\sqrt{4\pi\alpha}\simeq0.303$ for the electric charge.
At an order of magnitude estimate, this bounds the scale of $|(Y_e^2)_{e\mu}|<
0.01
\left(\frac{m_{H_2}}{\rm TeV}\right)$. For $m_{H_2}=100$ TeV, this is roughly the same order of magnitude as the bound from the 3-lepton decays. 

For a TeV scale $m_{H_2}$, each of these bounds constrains the Yukawa couplings to within the numeric range considered in the analysis of \cite{Herrero-Garcia:2017xdu}. For larger $m_{H_2}$, this mass scale suppression allows for slightly larger values in $Y_e^2$, outside of the range considered in \cite{Herrero-Garcia:2017xdu}, however, only by one order of magnitude before $Y_e^2$ is bound by perturbativity constraints.

The contribution to the radiative decay due to the charged singlet can be calculated in a similar fashion, with Wilson coefficient given by
\begin{align}
   C_{e\gamma} &=- \frac{1}{6}\frac{e}{16\pi^2}\frac{1}{m_h^2}f^\dagger f Y_e^{1\dagger} \ .
\end{align}
This simplifies to the expression for the branching ratio of the radiative decay given in \cite{Felkl:2021qdn} from which we find the bound
\begin{align}
    {\rm BR}(\ell_i\to\ell_j\gamma) 
    &\simeq 3\times10^{-16} 
    |f_{jk}f_{ik}|^2 \leq 10^{-13} \text{ for } f_{ij}\lesssim \mathcal{O}(1)
\end{align}
for even the lowest $m_h=10^5\, \mathrm{GeV} = 100\, \mathrm{TeV}$ we consider. For our perturbative maximum $f_{ij}\leq \mathcal{O}(1)$, we see that this renders the Higgs singlet contribution to the branching ratio much less than any of the current bounds ${\rm BR}(\ell_i\to\ell_j\gamma) < 10^{-13}$. Therefore, the cLFV constraints place no further bound on the Yukawa like coupling $f$ to the charged singlet.

\section{Input parameters for numerics}\label{app:benchmarks}
We use the SM running parameters calculated in \cite{Antusch:2025fpm}
as inputs at the scale of the top quark pole mass, listed in \cref{tbl:inputs}. The calculation of RGEs in \cite{Antusch:2025fpm} are all at 2-loop order or greater, ensuring ample precision for our 1-loop calculation.
\begin{table}[tb!]
\centering
\begin{tabular}{@{}clcc@{}}
\toprule
\multicolumn{3}{c}{Higgs parameters} & quark masses \\ \midrule
\multicolumn{3}{c}{$m_{h_{\rm SM}\rm (pole)} = 125.25\ {\rm GeV} 
$} 
& \multirow{6}{*}{
$\begin{aligned}
m_u &= 1.23 \text{ MeV}\\[-0.1em]
m_c &= 0.621 \text{ GeV}\\[-0.1em]
m_t &= 168.62 \text{ GeV}\\[-0.1em]
m_d &= 2.69 \text{ MeV}\\[-0.1em]
m_s &= 53 \text{ MeV}\\[-0.1em]
m_b &= 2.842 \text{ GeV}
\end{aligned}$
} \\
\multicolumn{3}{c}{$\lambda = 0.139825 \qquad v = 246.604$ GeV} &  \\ \cmidrule{1-3}
\multicolumn{2}{c}{lepton masses} & gauge couplings &  \\ \cmidrule{1-3}
\multicolumn{2}{c}{\multirow{3}{*}{
$\begin{aligned}
m_e &= 0.483246 \text{ MeV}\\[-0.3em]
m_\mu &= 0.102017 \text{ GeV}\\[-0.3em]
m_\tau &= 1.7329 \text{ GeV}
\end{aligned}$
}} & \multirow{3}{*}{
$\begin{aligned}
g_1 &= 0.357266\\[-0.3em]
g_2 &= 0.65096\\[-0.3em]
g_3 &= 1.2123
\end{aligned}$
} &  \\
\multicolumn{2}{c}{} &  &  \\
\multicolumn{2}{c}{} &  &  \\ \bottomrule
\end{tabular}
\caption{SM running parameters in the $\overline{\rm MS}$ scheme used as inputs for our numeric calculations. All quantities are at the scale of the top quark pole mass $m_t = 172.4$ GeV as calculated in \cite{Antusch:2025fpm} unless specified otherwise.}
\label{tbl:inputs}
\end{table}

Unless otherwise stated, we use the values listed in \cref{tbl:UVscales} for the various remaining scales and choices of non-crucial parameters in the Zee and 2HDM EFTs. Review \cref{sec:constraints,sec:BMs} for the discussion of these choices. 

\begin{table}[tb!]
    \centering
    \begin{tabular}{cc}\toprule
        mass scales & coupling strengths \\\midrule
        $\mu_{\rm Zee} = m_h = 10^4$ TeV & $\lambda^{\rm (Zee)}_{8,9,10} = 0 \ , \ \ \lambda_h=10^{-4}$ \\
        $m_{H_2} = 10^2$ TeV & $\lambda_{1,...,7}$ (see \cref{tbl:benchmarks})\\\bottomrule
    \end{tabular}
    \caption{Choice of UV scales and scalar potential couplings used in numeric calculations. The quartic coupling $\lambda_h$ is chosen small but non-zero to guarantee stability of the potential for large field values of $h$ despite having very little effect on the overall computation.}
    \label{tbl:UVscales}
\end{table}

Finally, we also have the benchmark values of the parameters defined at the UV scale $\bar{\mu} = m_h$. Recall from the calculation framework discussed in \cref{sec:framework} that Zee model parameters such as $f$ and $\mu_{\rm Zee}$ are defined in the Zee model, while parameters shared between the Zee and 2HDM EFT, such as the quartic couplings $\lambda_{1,\ldots,7}$ and the Yukawa matrices $Y_{e}^{1,2}$ are defined in the 2HDM EFT at the same $\bar{\mu}=m_h$ scale. The three sets of benchmark parameters discussed in the main text are listed explicitly in \cref{tbl:benchmarks} below.
\begin{table}[tbp!]
    \centering
    \begin{tabular}{c}\toprule
         Benchmark Set 1 \\ \midrule
         $Y_e^2\gg Y_e^1$, $\alpha=1$, $m_{H_2}=100$ TeV, $m_h=10^4$ TeV  \\
         $\lambda_{1,2}=0.25$, $\lambda_{3,\ldots,7}=0$\\[1em]
         $\begin{aligned}
         f &\simeq \begin{pmatrix}
         0 & 7.2\times10^{-6} & 1.5\times10^{-5}\\
         -7.2\times10^{-6} & 0 & 1.5\times10^{-9}\\
         -1.5\times10^{-5} & -1.5\times10^{-9} & 0
         \end{pmatrix}\\
         Y_e^2(m_{h}) & \simeq \begin{pmatrix}
         0.36 -\ i0.24 & -0.11 +\ i0.089 & -0.080 -\ i0.052\\
         -0.099 -\ i0.11 & 0.21 +\ i0.42 & 0.11 -\ i0.026\\
         -0.11 +\ i0.11 & -0.0467 +\ i0.11 & -0.55 -\ i0.066
         \end{pmatrix}\\
         \end{aligned}$ \\\midrule\midrule
         Benchmark Set 2 \\ \midrule
         $Y_e^2 \sim Y_e^1$, $\alpha=1$, $m_{H_2}=100$ TeV, $m_h=10^4$ TeV \\
         $\lambda_{1,2}=0.25$, $\lambda_{3,\ldots,7}=0$\\[1em]
         $\begin{aligned}
         f &\simeq \begin{pmatrix}
         0 & 0.0093 & 0.0017\\
         -0.0093 & 0 & 9.3\times10^{-7}\\
         -0.0017 & -9.3\times10^{-7} & 0
         \end{pmatrix}\\
         Y_e^2(m_{h}) & \simeq \begin{pmatrix}
         -0.0053 +\ i0.0053 & 0.0053 +\ i0.0053 & -0.0053 +\ i0.0053\\
         0.0050 +\ i0.0053 & 0.0054 -\ i0.0050 & 0.0053 -\ i0.00050\\
         -0.0026 -\ i0.0053 & 0.00020 +\ i0.00080 & -0.0048 -\ i0.0049
         \end{pmatrix}\\
         \end{aligned}$ \\\midrule\midrule
         Benchmark Set 3 \\\midrule
         $Y_e^2\gg Y_e^1$,  $\alpha=0$, $m_{H_2}=100$ TeV, $m_h=10^4$ TeV  \\
         $\lambda_{1,2}=0.25$, $\lambda_{3,4,5}=0$, $\lambda_6=-\lambda_7=0.25$\\[1em]
         $\begin{aligned}
         f &\simeq \begin{pmatrix}
         0 & -1.4\times10^{-4} & 7.4\times10^{-6}\\
         1.4\times10^{-4} & 0 & 1.4\times10^{-8}\\
         -7.4\times10^{-6} & -1.4\times10^{-8} & 0
         \end{pmatrix}\\
         Y_e^2(m_{h}) & \simeq \begin{pmatrix}
         -0.35 -\ i0.70 & 0.035 -\ i0.10 & -0.10 -\ i0.10\\
         0.0096 +\ i0.032 & 0.72 -\ i0.12 & -0.027 -\ i0.056\\
         0.072 -\ i0.10 & -0.10 +\ i0.10 & -0.20 -\ i0.19
         \end{pmatrix}\\
         \end{aligned}$ \\\midrule\midrule
         Low Scale Benchmark Set \\ \midrule
         $Y_e^2 \sim Y_e^1$, $\alpha=1$, $m_{H_2}=1$TeV, $m_h=100$TeV \\
         $\lambda_{1,2,6,7}=0.2$, $\lambda_{4,5}=0.4$, $\lambda_{3}=0$\\[1em]
         $\begin{aligned}
         f &\simeq \begin{pmatrix}
         0 & 0.00014 & -2.8\times10^{-6}\\
         -0.00014 & 0 & 1.4\times10^{-8}\\
         2.8\times10^{-6} & -1.4\times10^{-8} & 0 
         \end{pmatrix}\\
         Y_e^2(m_{h}) & \simeq \begin{pmatrix}
         0.00016 +\ i4.9\times10^{-5} & -0.0056 +\ i0.0046 & 0.0026 -\ i0.0024\\
         0.0056 -\ i0.00018 & 0.00055 +\ i0.0019 & 0.0055 -\ i0.0056\\
         0.0056 +\ i0.0056 & 0.0042 -\ i0.0038 & 0.0064 +\ i0.0056
         \end{pmatrix}\\
         \end{aligned}$ \\\bottomrule
    \end{tabular}
    \caption{Benchmark values for $f$, $Y_e^1$ and $Y_e^2$ at $\bar{\mu}=m_h$, rounded to 2 significant figures and chosen to reproduce the experimentally observed neutrino mixing parameters via a direct calculation in the full theory. See \cref{sec:results} for a full discussion. Note that the first three benchmark sets are at $m_h=10^4$TeV and the low scale benchmark set is at $m_h=100$TeV.}
    \label{tbl:benchmarks}
\end{table}

\section{UV sensitivity of remaining mixing parameters}\label{app:sensitivity}
\begin{figure}[!p]
    \centering
    \begin{subfigure}[b]{\textwidth}
        \includegraphics[width=0.32\textwidth]{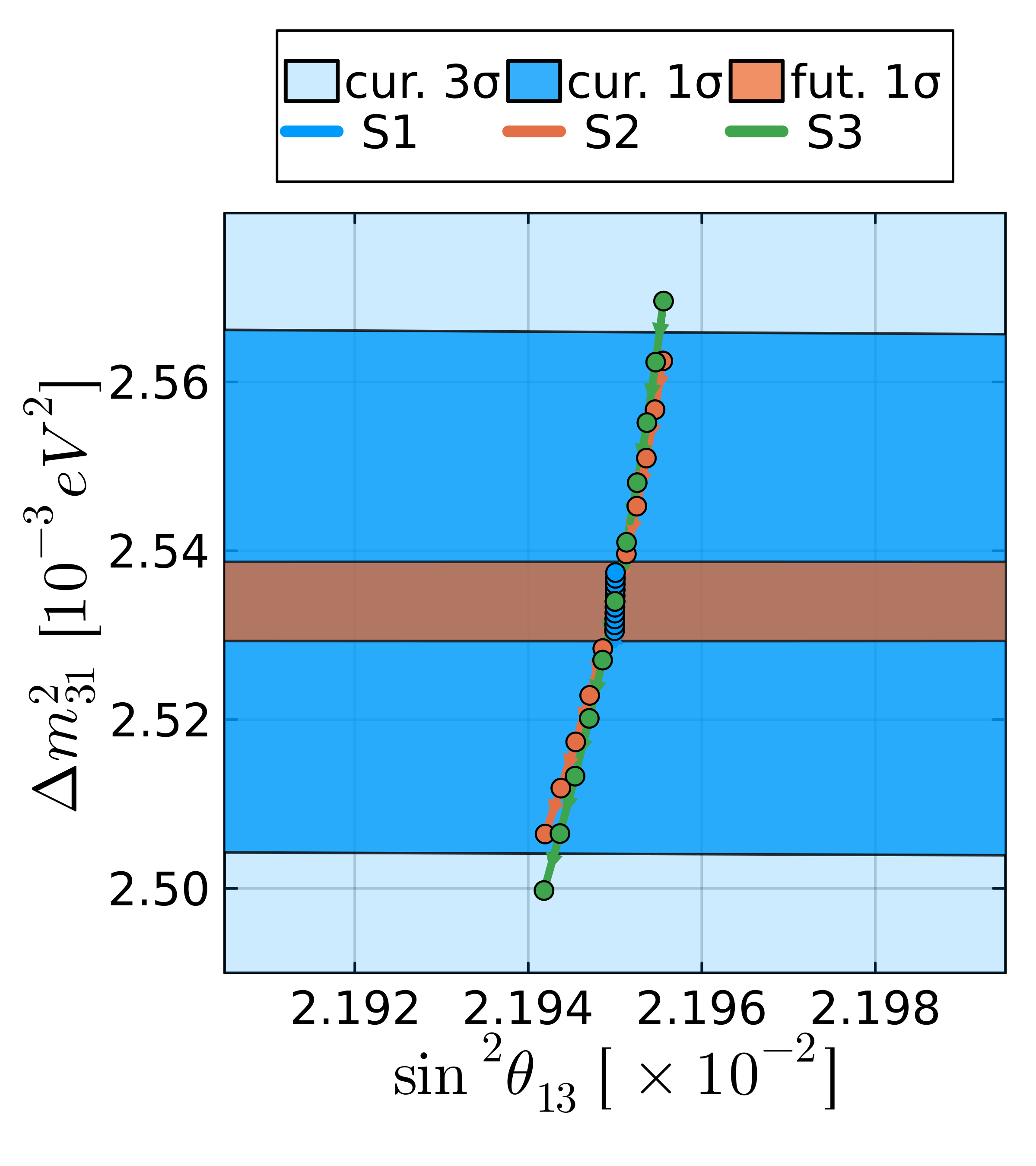}%
        \includegraphics[width=0.32\textwidth]{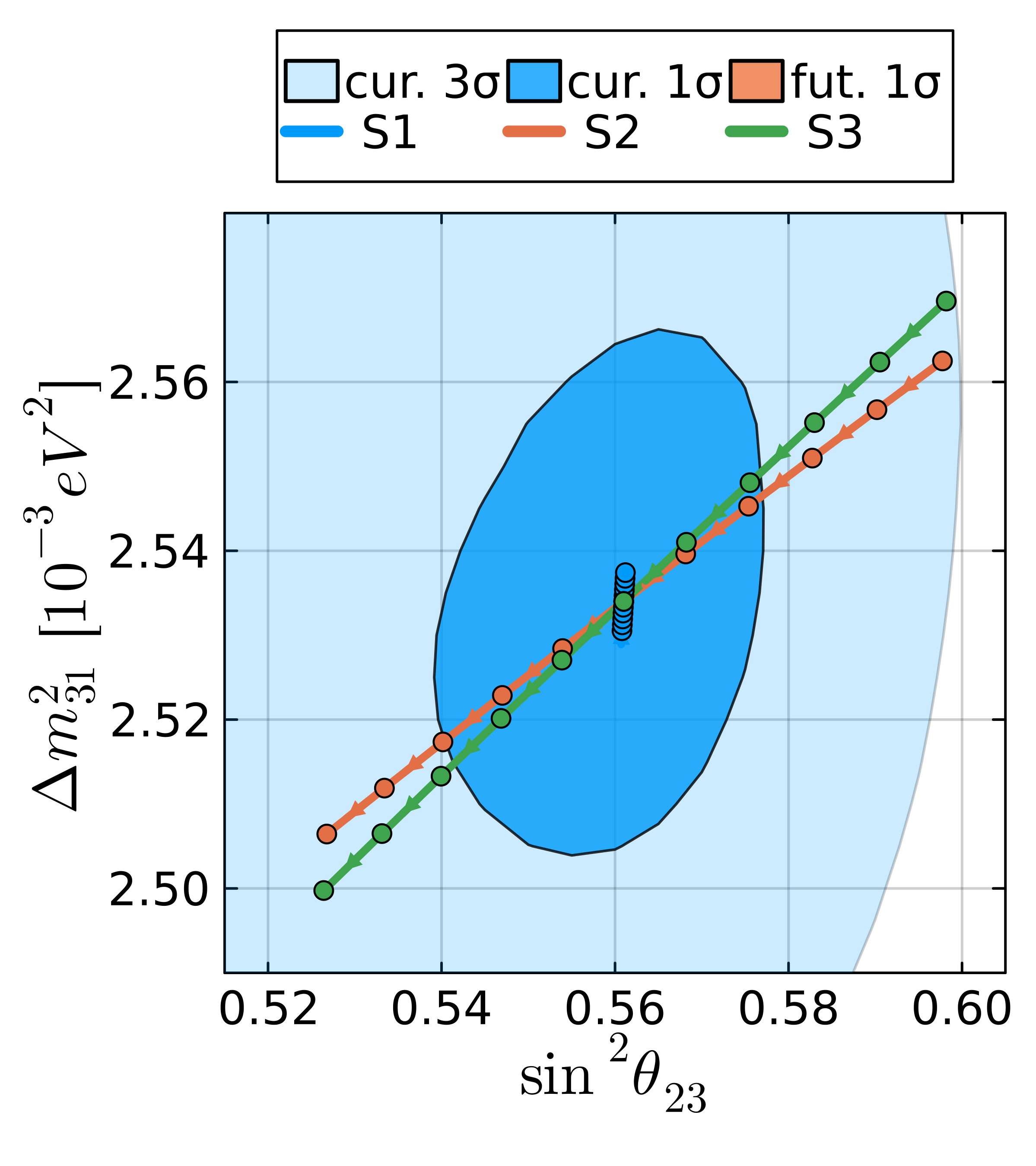}%
        \includegraphics[width=0.32\textwidth]{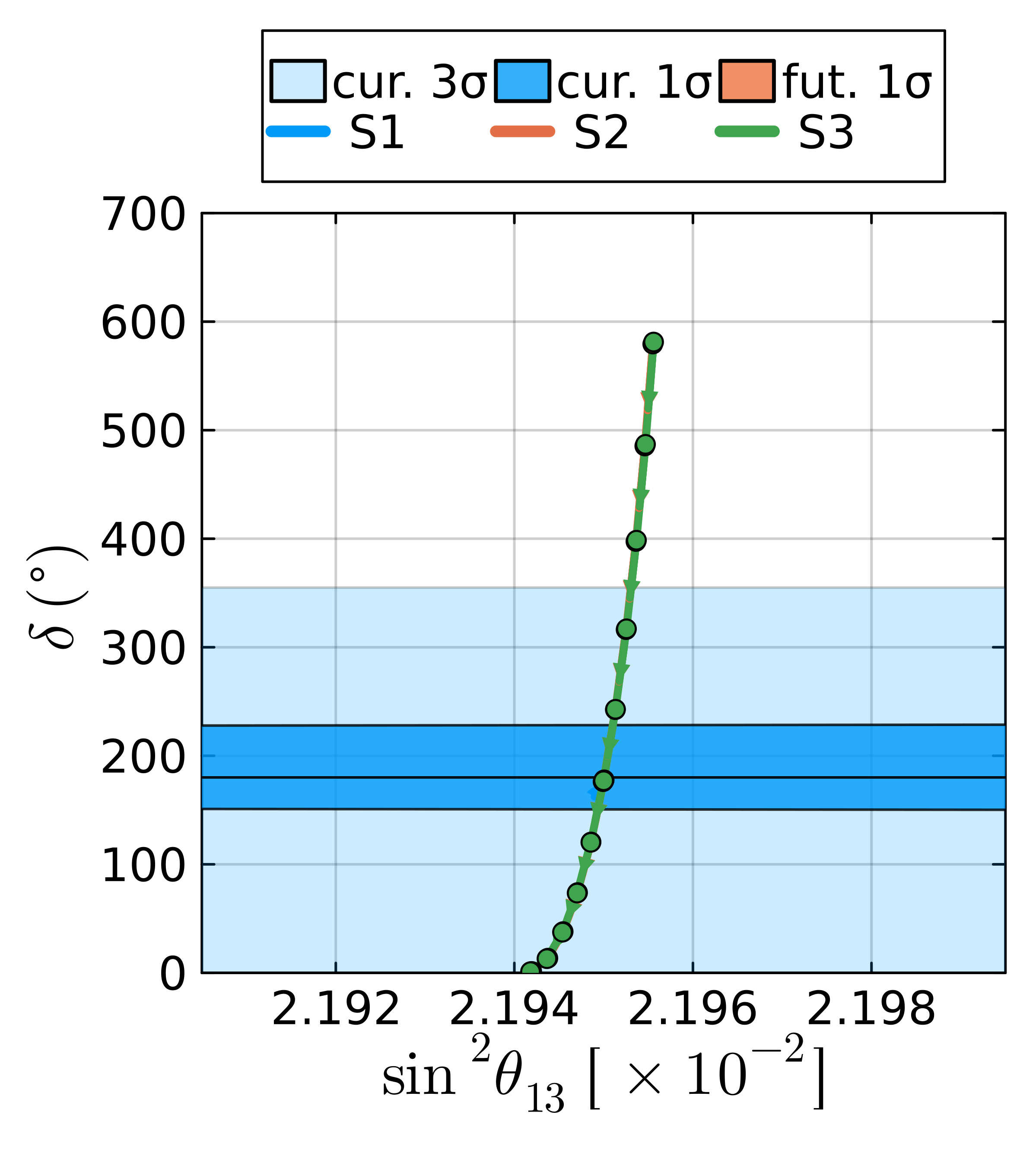}
        \caption{BM 1 ($Y_e^2\gg Y_e^1$ and $\alpha=1$)}
        \label{fig:paramStretchBB1}
    \end{subfigure}
    
   \vspace{2ex}
   
    \begin{subfigure}[b]{\textwidth}
        \includegraphics[width=0.32\textwidth]{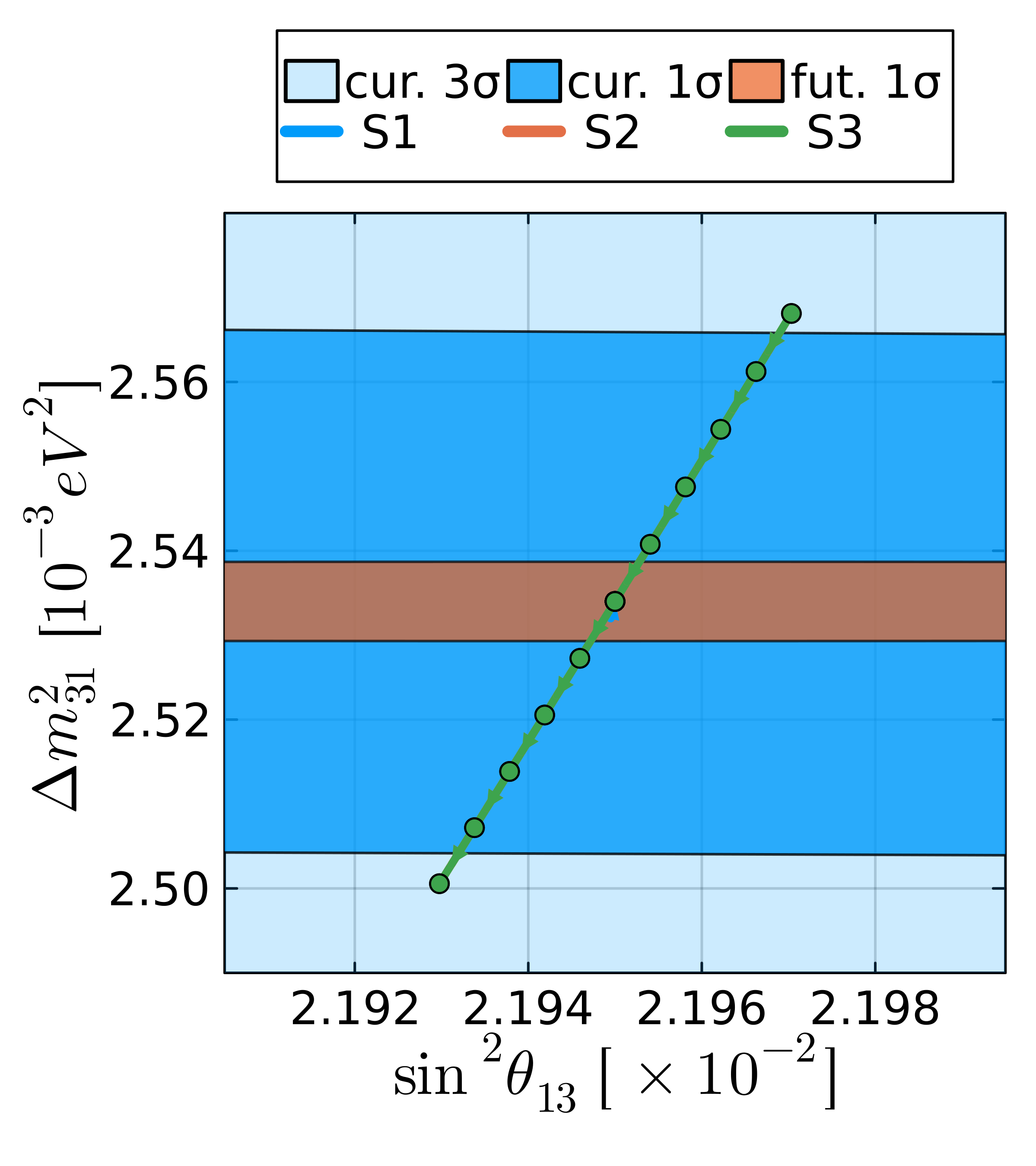}%
        \includegraphics[width=0.32\textwidth]{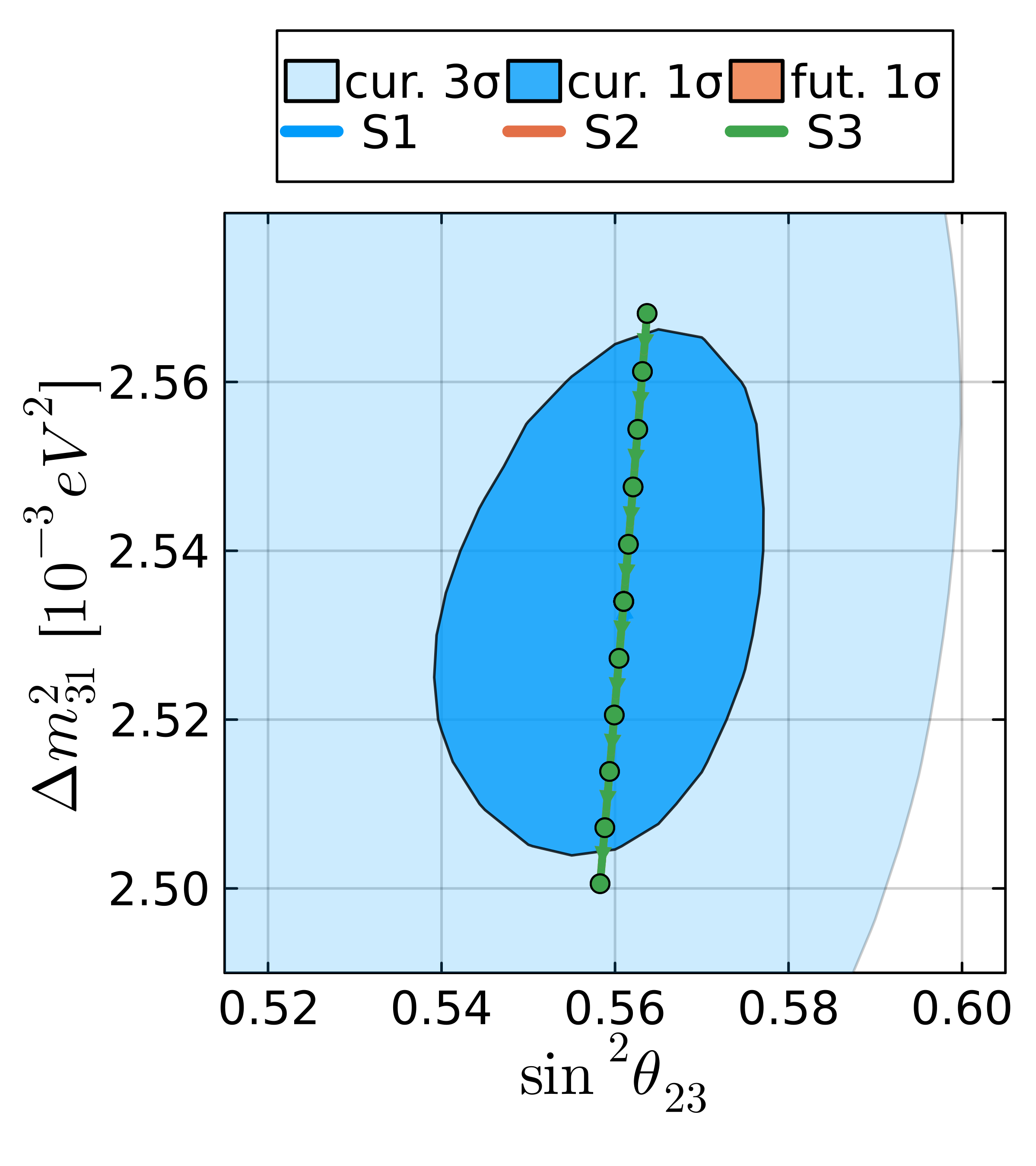}%
        \includegraphics[width=0.32\textwidth]{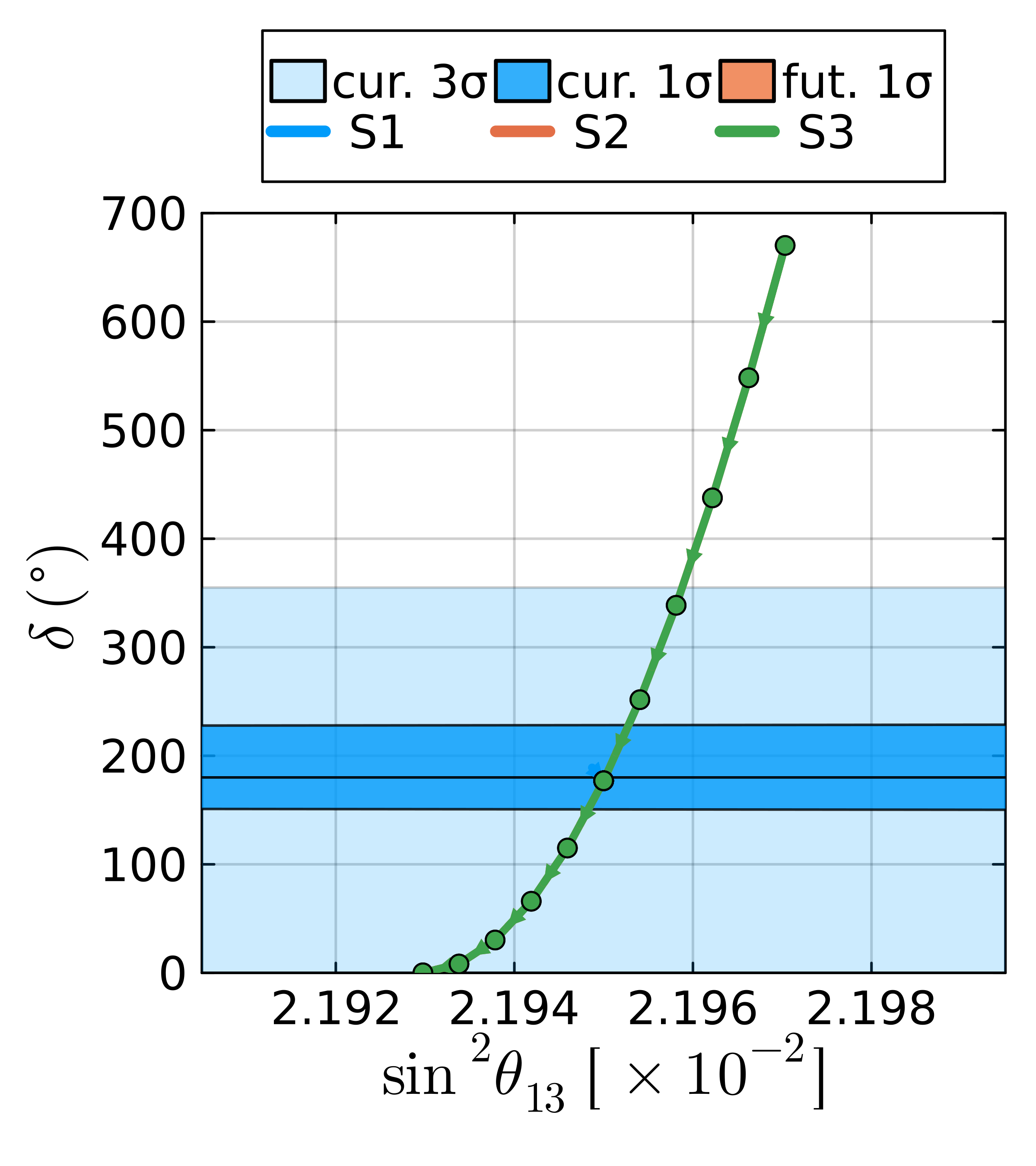}
        \caption{BM 2 ($Y_e^2\sim Y_e^1$ and $\alpha=1$)}
        \label{fig:paramStretchBB2}
    \end{subfigure}

   \vspace{2ex}
    
    \begin{subfigure}[b]{\textwidth}
        \includegraphics[width=0.32\textwidth]{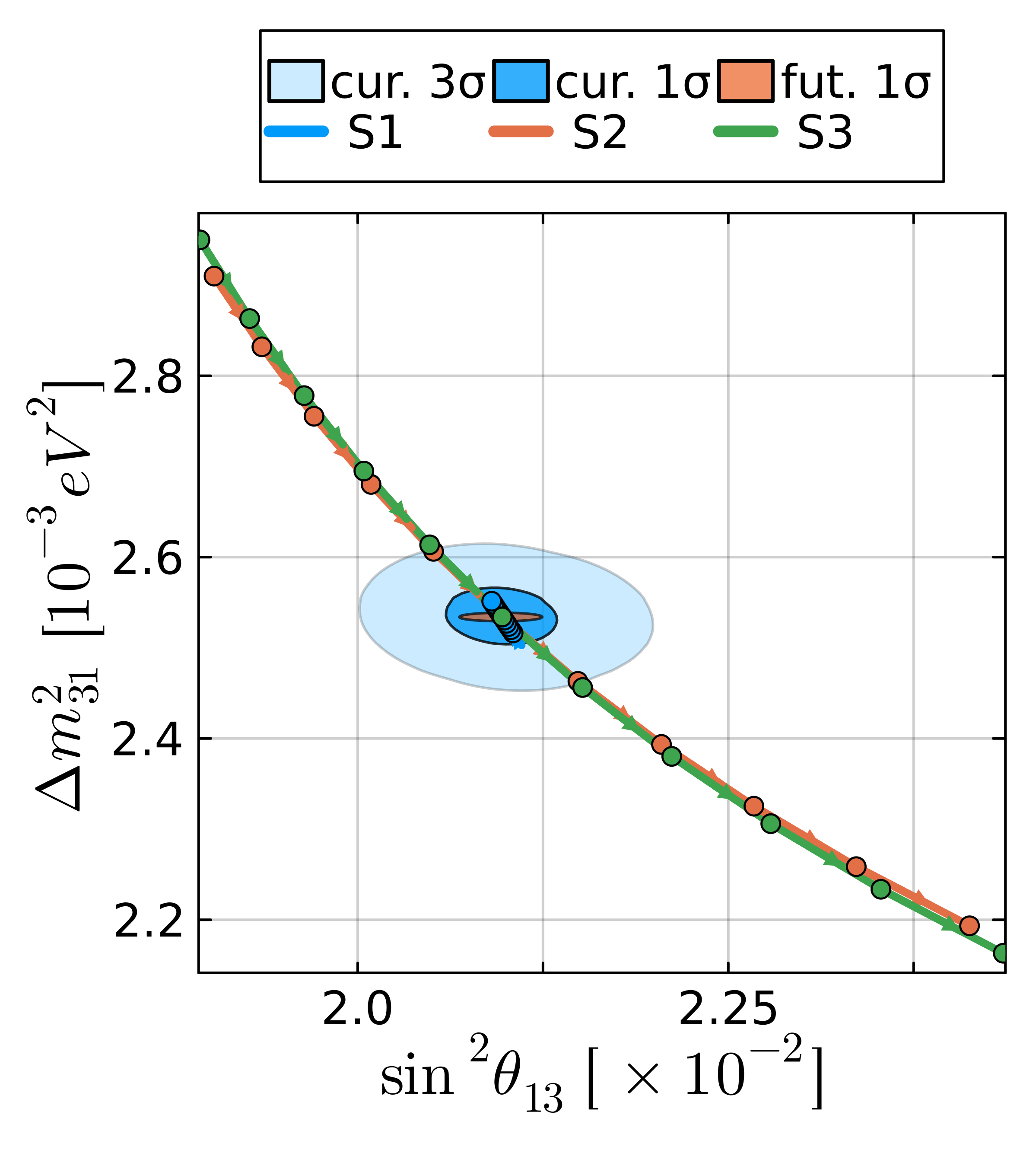}%
        \includegraphics[width=0.32\textwidth]{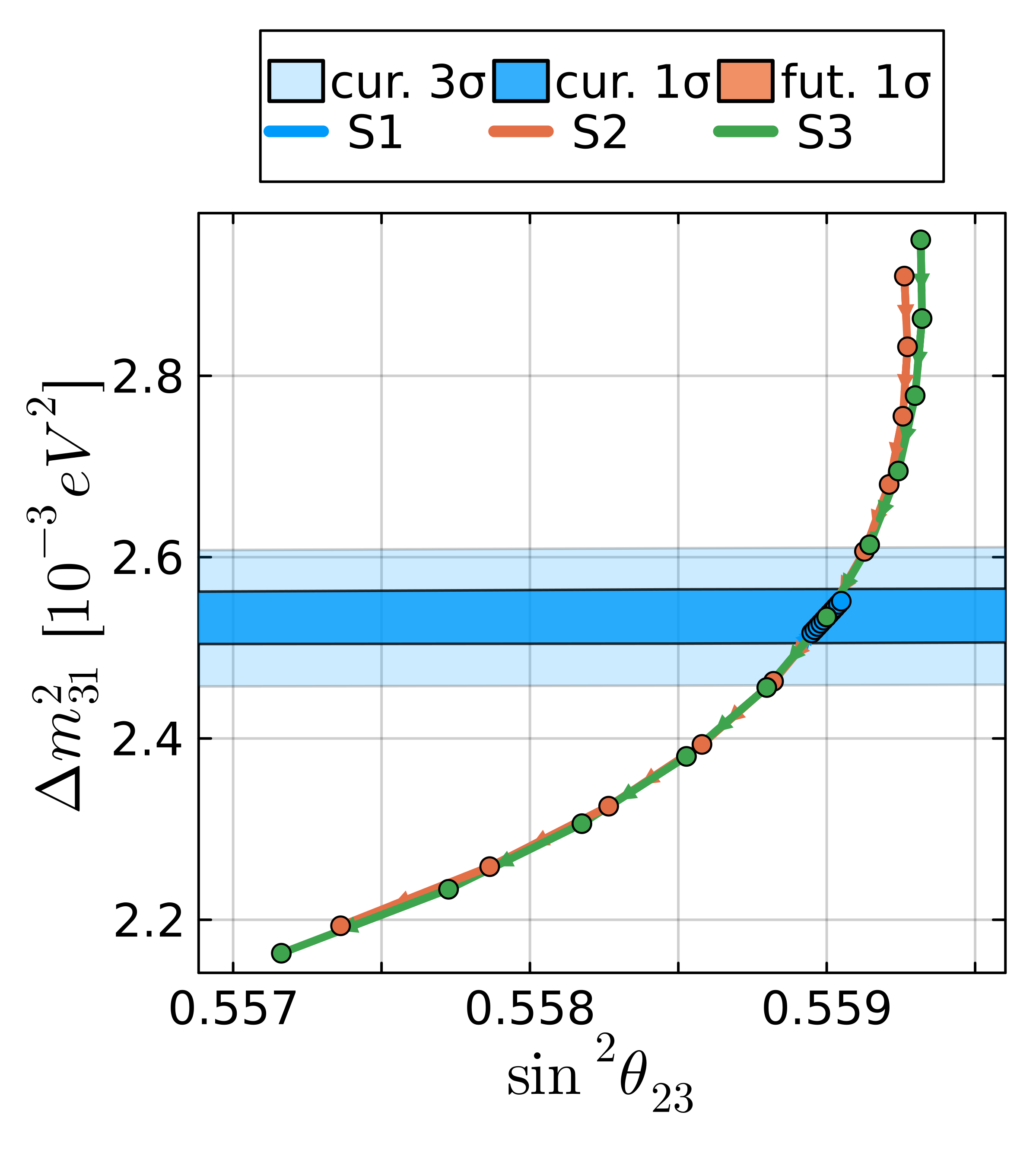}%
        \includegraphics[width=0.32\textwidth]{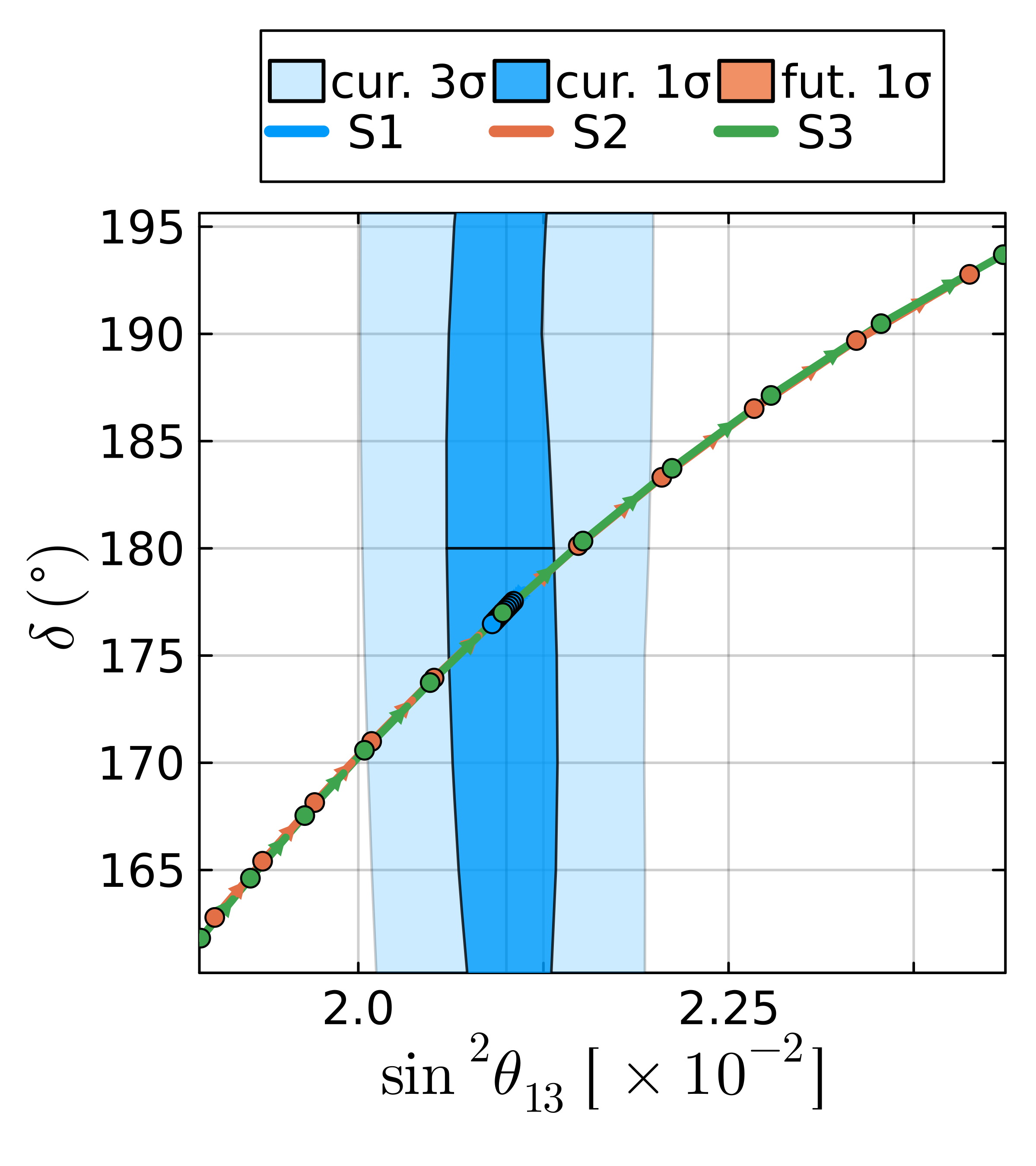}
        \caption{BM 3 ($\alpha=0$)}
        \label{fig:paramStretchBB3}
    \end{subfigure}

    \caption{
    Sensitivity of the normal-ordered atmospheric neutrino mixing parameters to the UV model parameters in the EFT framework for each set of benchmark (BM) parameters with $m_h=10^4$TeV. 
    As per \cref{fig:paramStretchA},
    the UV parameters are scaled to preserve the neutrino mass matrix in the full theory with: S1 = $\gamma^{-1/2}Y_e^{1,2}$, $\gamma f$; S2 = $\gamma^{-1}Y_e^{2}$, $\gamma f$; S3 = $\gamma^{-1}f$, $\gamma Y_e^1$; and with $0.95\leq\gamma\leq1.05$ increasing in the direction of the arrows marked.
    The shaded regions represent current and future estimated experimental uncertainties (future projected to 6 years by the JUNO collaboration \cite{JUNO:2022mxj}), using the current central values.
    }
    \label{fig:paramStretch}
\end{figure}

As per the discussion in \cref{sec:results}, \cref{fig:paramStretch} 
contains plots of the sensitivity of the neutrino mixing parameters in the EFT framework, other than the solar mixing parameters that are shown in \cref{fig:paramStretchA}, to the UV parameters in the three regular benchmark sets.

The sensitivity of the mixing parameters shown in these figures tells much the same story as those discussed in \cref{sec:sensitivity} in the main text. The sensitivity of the CP violating phase $\delta$ shown in the right column of \cref{fig:paramStretch} is particularly large; $\delta$ is also the least precisely measured mixing parameter.

\clearpage

\bibliographystyle{JHEP}
\bibliography{refs.bib}
\end{document}